\documentclass[%
 reprint,preprintnumbers,
 amsmath,amssymb,
 aps,nofootinbib,
 prd,superscriptaddress
 ]{revtex4-2}

\usepackage{hyperref}

\usepackage[utf8]{inputenc}
\usepackage{comment}

\usepackage{graphicx}% Include figure files
\usepackage{dcolumn}% Align table columns on decimal point
\usepackage{bm}% bold math
\usepackage{multibib}
\usepackage{placeins}
\usepackage{xcolor}
\usepackage{orcidlink}

\usepackage{siunitx} % pretty formating of units
\usepackage{cleveref} % \cref for standard refences, \Cref for the beginning of a sentence
\crefname{figure}{Figure}{Figures}
\crefname{table}{Table}{Tables}
\crefname{section}{section}{sections} % subsection and lower inherit
\crefname{appendix}{appendix}{appendixes}
\crefname{equation}{Eq.}{Eqs.}
\Crefname{figure}{Figure}{Figures}
\Crefname{table}{Table}{Tables}
\Crefname{section}{Section}{Sections} % subsection and lower inherit
\Crefname{appendix}{Appendix}{Appendixes}
\Crefname{equation}{Eq.}{Eqs.}
\AddToHook{cmd/appendix/before}{\crefalias{section}{appendix}}
\newcommand{\ket}[1]{\ensuremath{| {#1} \rangle }}
\newcommand{\bra}[1]{\ensuremath{\langle {#1} |}}

\renewcommand{\vec}[1]{\bm{#1}}

\newcites{SM}{Supplementary Material Bibliography}

\begin{document}

\setlength\abovedisplayskip{10pt}
\setlength\belowdisplayskip{10pt}

\setlength{\parskip}{14pt}
\setlength{\parindent}{0pt}

\preprint{HIP-2025-9/TH}

\title{Inclusive semileptonic decays of the \texorpdfstring{$D_s$}{Ds} meson:\\ A first-principles lattice QCD calculation}

\author{Alessandro~\surname{De~Santis}~\orcidlink{0000-0002-2674-4222}}
\affiliation{Helmholtz-Institut Mainz, Johannes Gutenberg-Universit{\"a}t Mainz, 55099 Mainz, Germany}
\affiliation{GSI Helmholtz Centre for Heavy Ion Research, 64291 Darmstadt, Germany}

\author{Antonio~\surname{Evangelista}~\orcidlink{0000-0002-3320-3176}}
\affiliation{Dipartimento di Fisica \& INFN, Universit\`a di Roma ``Tor Vergata'', Via della Ricerca Scientifica 1, I-00133 Rome, Italy}

\author{Roberto~\surname{Frezzotti}~\orcidlink{0000-0001-5746-0065}}
\affiliation{Dipartimento di Fisica \& INFN, Universit\`a di Roma ``Tor Vergata'', Via della Ricerca Scientifica 1, I-00133 Rome, Italy}

\author{Giuseppe~\surname{Gagliardi}~\orcidlink{0000-0002-4572-864X}}
\affiliation{Dipartimento di Matematica e Fisica, Universit\`a Roma Tre, Via della Vasca Navale 84, I-00146 Rome, Italy}
\affiliation{INFN, Sezione di Roma Tre, Via della Vasca Navale 84, I-00146 Rome, Italy}

\author{Paolo~\surname{Gambino}~\orcidlink{0000-0002-7433-4914}}
\affiliation{Dipartimento di Fisica, Universit\`a di Torino \& INFN, Sezione di Torino, Via Pietro Giuria 1, I-10125 Turin, Italy}

\author{Marco~\surname{Garofalo}~\orcidlink{0000-0002-4508-6421}}
\affiliation{HISKP (Theory) \& Bethe Centre for Theoretical Physics, Rheinische Friedrich-Wilhelms-Universit\"at Bonn, Nussallee 14-16, D-53115 Bonn, Germany}

\author{Christiane~Franziska~\surname{Gro\texorpdfstring{\ss}}~\orcidlink{0009-0009-5876-1455}}
\affiliation{HISKP (Theory) \& Bethe Centre for Theoretical Physics, Rheinische Friedrich-Wilhelms-Universit\"at Bonn, Nussallee 14-16, D-53115 Bonn, Germany}

\author{Bartosz~\surname{Kostrzewa}~\orcidlink{0000-0003-4434-6022}}
\affiliation{HISKP (Theory) \& Bethe Centre for Theoretical Physics, Rheinische Friedrich-Wilhelms-Universit\"at Bonn, Nussallee 14-16, D-53115 Bonn, Germany}

\author{Vittorio~\surname{Lubicz}~\orcidlink{0000-0002-4565-9680}}
\affiliation{Dipartimento di Matematica e Fisica, Universit\`a Roma Tre, Via della Vasca Navale 84, I-00146 Rome, Italy}
\affiliation{INFN, Sezione di Roma Tre, Via della Vasca Navale 84, I-00146 Rome, Italy}

\author{Francesca~\surname{Margari}~\orcidlink{0000-0003-2155-7679}}
\affiliation{Dipartimento di Fisica \& INFN, Universit\`a di Roma ``Tor Vergata'', Via della Ricerca Scientifica 1, I-00133 Rome, Italy}

\author{Marco~\surname{Panero}~\orcidlink{0000-0001-9477-3749}}
\affiliation{Dipartimento di Fisica, Universit\`a di Torino \& INFN, Sezione di Torino, Via Pietro Giuria 1, I-10125 Turin, Italy}
\affiliation{Department of Physics \& Helsinki Institute of Physics, University of Helsinki, PL 64, FIN-00014 Helsinki, Finland}

\author{Francesco~\surname{Sanfilippo}~\orcidlink{0000-0002-1333-745X}}
\affiliation{INFN, Sezione di Roma Tre, Via della Vasca Navale 84, I-00146 Rome, Italy}

\author{Silvano~\surname{Simula}~\orcidlink{0000-0002-5533-6746}}
\affiliation{INFN, Sezione di Roma Tre, Via della Vasca Navale 84, I-00146 Rome, Italy}

\author{Antonio~\surname{Smecca}~\orcidlink{0000-0002-8887-5826}}
\affiliation{Department of Physics, Faculty of Science and Engineering, Swansea University (Singleton Park Campus), Singleton Park, SA2 8PP Swansea, Wales, United Kingdom}

\author{Nazario~\surname{Tantalo}~\orcidlink{0000-0001-5571-7971}}
\affiliation{Dipartimento di Fisica \& INFN, Universit\`a di Roma ``Tor Vergata'', Via della Ricerca Scientifica 1, I-00133 Rome, Italy}

\author{Carsten~\surname{Urbach}~\orcidlink{0000-0003-1412-7582}}
\affiliation{HISKP (Theory) \& Bethe Centre for Theoretical Physics, Rheinische Friedrich-Wilhelms-Universit\"at Bonn, Nussallee 14-16, D-53115 Bonn, Germany}

\begin{abstract}
\centerline{\includegraphics[height=4.2cm]{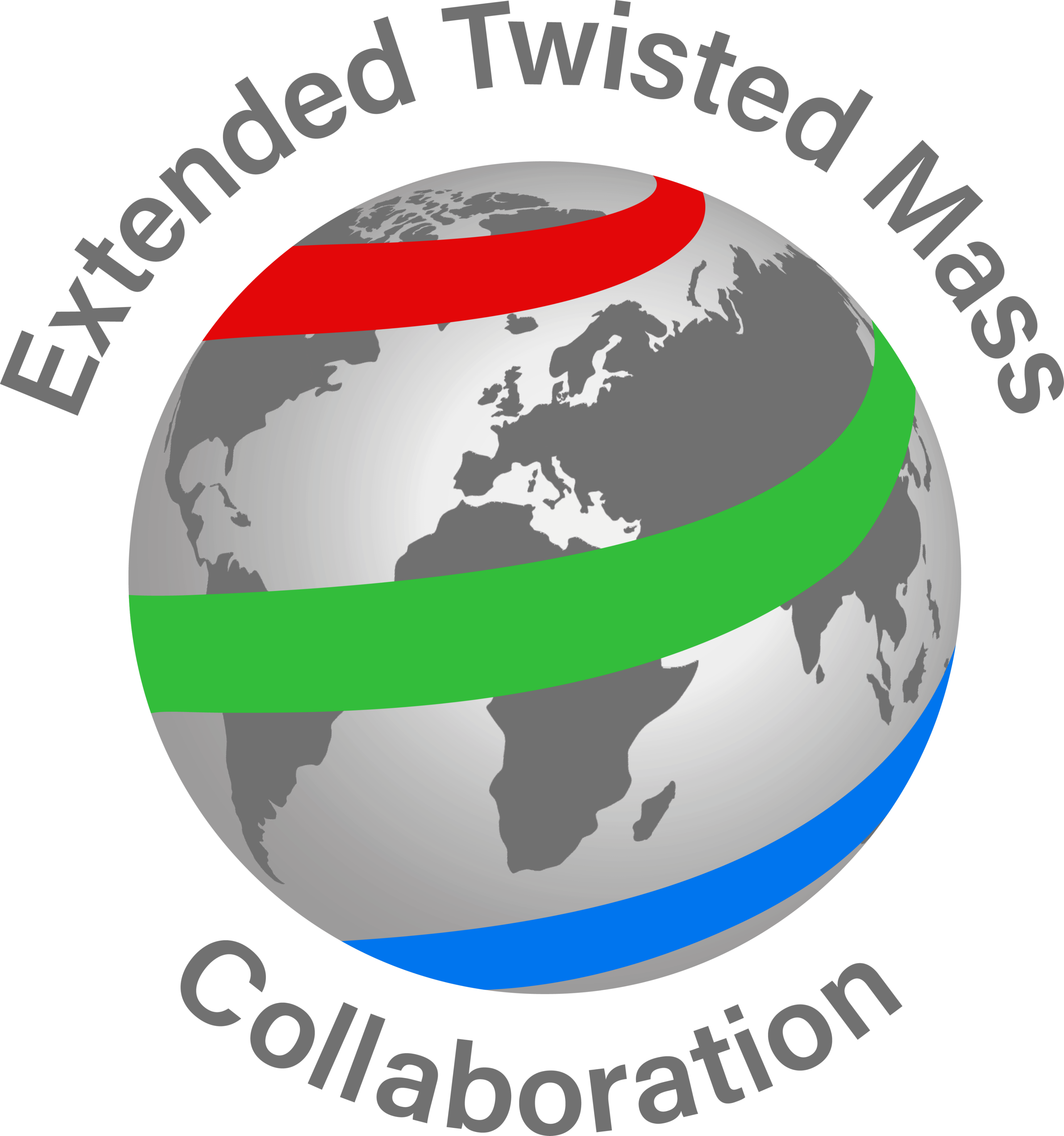}}
\vspace{0.22cm}
We present the results of a first-principles theoretical study of the inclusive semileptonic decays of the \texorpdfstring{$D_s$}{Ds} meson. We performed a state-of-the-art lattice QCD calculation using the gauge ensembles produced by the Extended Twisted Mass Collaboration (ETMC) with dynamical light, strange and charm quarks with physical masses and employed the so-called Hansen-Lupo-Tantalo (HLT) method to extract the decay rate and the first two lepton-energy moments from the relevant Euclidean correlators. We have carefully taken into account all sources of systematic errors, including the ones associated with the continuum and infinite-volume extrapolations and with the HLT spectral reconstruction method. We obtained results in very good agreement with the currently available experimental determinations and with a total accuracy at the few-percent level, of the same order of magnitude of the experimental error.
%reported by the BES-III collaboration. 
Our total error is dominated by the lattice QCD simulations statistical uncertainties and is certainly improvable. From 
the results presented and thoroughly discussed in this paper we conclude that it is nowadays possible to study heavy mesons inclusive semileptonic decays on the lattice at a phenomenologically relevant level of accuracy. The phenomenological implications of our physical results are the subject of a companion letter ~\cite{DeSantis:2025yfm}.
\end{abstract}

\maketitle

\section{
\label{sec:introduction}
Introduction
}

Understanding the origin and the structure of the flavor sector of the Standard Model (SM) is one of the main open challenges of particle physics. After many years of tireless experimental and theoretical efforts, advancing our knowledge on flavor physics requires performing very accurate (at the sub-percent level) studies of weak-interaction processes involving hadrons and leptons. Among the many interesting processes, a very important r\^ole is played by the semileptonic decays of QCD-stable pseudoscalar mesons, that couple the leptonic and the hadronic flavor sectors and give access to the matrix elements of the Cabibbo--Kobayashi--Maskawa (CKM) matrix. 

On the theoretical side, the exclusive semileptonic decays of kaons and heavy ($D_{(s)}$, $B_{(s)}$) pseudoscalar mesons have been extensively studied by performing lattice QCD simulations. An updated picture of the level of theoretical accuracy currently reached on different interesting processes can be found in the latest edition of the FLAG review~\cite{FlavourLatticeAveragingGroupFLAG:2024oxs}. In some cases, e.g.\ $K\mapsto \pi\ell \bar \nu_\ell$ decays, the sub-percent accuracy level has already been achieved, by relying though on the isospin-symmetric approximation of QCD (isoQCD), and further progress can only be made by performing challenging lattice QCD$+$QED calculations.

In the present work and in the companion paper~\cite{DeSantis:2025yfm}, we face another long-standing challenge in the theoretical study of flavor physics, namely the non-perturbative calculation of \emph{inclusive} semileptonic decay rates. 
In particular, by performing state-of-the-art isoQCD lattice simulations, we have calculated the decay rate and the first two lepton-energy moments for the inclusive process $D_s\mapsto X \ell\bar \nu_\ell$, in which a negatively-charged $D_s$ meson decays into all possible (kinematically and flavor allowed) hadronic states $X$, a lepton $\ell$ (in the approximation in which it is massless) and the corresponding anti-neutrino $\bar \nu_\ell$.   

On the experimental side, depending upon the specific process and the experimental setup, inclusive semileptonic decay rates can be obtained by summing the decay rates of all possible exclusive channels or measured directly by using tailored techniques. The latter is the case of $D_s\mapsto X \ell\bar \nu_\ell$ processes (see Refs.~\cite{CLEO:2009uah,BESIII:2021duu} for more details) that, therefore, provide independent information and different control on the experimental systematics w.r.t.\ that provided by the corresponding exclusive channels.

From a phenomenological perspective, our first-principles lattice results are important because they allow one to use the experimental information of Refs.~\cite{CLEO:2009uah,BESIII:2021duu} to constrain the CKM matrix elements $V_{cs}$, $V_{cd}$. The study of the phenomenological implications of our results is the subject of the companion paper~\cite{DeSantis:2025yfm}.

From a theoretical perspective, our results are important because they show that inclusive semileptonic decays can nowadays be studied from first-principles on the lattice. This is a non-trivial result. Indeed, while the hadronic form-factors parametrizing the decay rates of exclusive processes involving QCD-stable hadrons in the external states can be extracted by studying the asymptotic behavior at large Euclidean times of lattice correlators, the lattice calculation of inclusive decay rates requires radically different theoretical and numerical techniques. Although the key ingredients were already present in the more general, mathematically-oriented and forward-looking Ref.~\cite{Barata:1990rn} (see also Ref.~\cite{Patella:2024cto} for a recent generalization), these techniques have been developed only recently~\cite{Hansen:2017mnd,Hashimoto:2017wqo,Hansen:2019idp,Gambino:2020crt,Gambino:2022dvu}.

Together with other collaborators, some of us made a first important step toward the demonstration of the numerical feasibility of lattice calculations of inclusive semileptonic decay rates in Ref.~\cite{Gambino:2022dvu}. In that work, by using the methods of Refs.~\cite{Hashimoto:2017wqo,Hansen:2019idp,Gambino:2020crt}, we studied the inclusive processes $H\mapsto X \ell\bar \nu_\ell$ at unphysical values of the heavy-quark mass of the decaying pseudoscalar meson $H$ and compared the lattice results, obtained at fixed lattice spacing and fixed volume, with the analytical results obtained by relying on quark-hadron duality and the Operator Product Expansion (OPE). In fact, in the absence of first-principles approaches, OPE techniques~\cite{Manohar:1993qn,Blok:1993va,Bigi:1992su,Bigi:1993fe,Chay:1990da}, that are particularly well motivated in the case of the phenomenologically very relevant $B_{(s)}$ inclusive decays, have been for many years the only viable theoretical approach to heavy meson inclusive semileptonic decays. Ref.~\cite{Gambino:2022dvu} has shown that in the regions of the parameters space where the OPE was expected to be reliable, the lattice results were in fairly nice agreement (at the level of 1 standard deviation) with the analytical predictions. This preliminary study has thus highlighted the  necessity of a detailed investigation aiming at establishing whether lattice calculations can now provide phenomenologically relevant information on inclusive processes. This is the main subject of the present work.

The problem of the lattice determination of inclusive observables has already been addressed in the case of other phenomenologically relevant processes, namely the (energy-smeared) $R$-ratio~\cite{ExtendedTwistedMassCollaborationETMC:2022sta} and the inclusive hadronic decays of the $\tau$ lepton~\cite{Evangelista:2023fmt,ExtendedTwistedMass:2024myu}, by producing first-principles isoQCD lattice results at a level of accuracy that can only be improved by including the neglected isospin breaking effects. At the same time, other lattice groups~\cite{Barone:2022gkn,Kellermann:2022mms,Barone:2023tbl,Kellermann:2023yec,Barone:2023iat,Kellermann:2024zfy,Hashimoto:2024pnd} have started to face the challenge of providing phenomenologically relevant lattice results for heavy mesons inclusive semileptonic decay rates\footnote{See also~\cite{Kellermann:2025pzt} which appeared on the arXiv when this work and the companion paper~\cite{DeSantis:2025yfm} were already finalized. The authors of Ref.~\cite{Kellermann:2025pzt}  perform a lattice QCD analysis of the $D_s\mapsto X \ell\bar \nu_\ell$ inclusive decays and focus on the systematic errors associated with the chosen spectral reconstruction technique and with finite volume effects,  performing simulations at fixed lattice spacing, fixed physical volume, and unphysical  pion mass ($m_\pi=300$~MeV), neglecting the $\bar c d$ and $\bar u s$ flavour channels as well as the so-called weak-annihilation contribution.}.

In this work we computed the differential decay rate and the first two lepton-energy moments for the inclusive process $D_s\mapsto X \ell\bar \nu_\ell$. We have carefully investigated and quantified all sources of systematic errors, including the ones associated with the necessary continuum and infinite-volume extrapolations. 
As shown in \cref{sec:conclusions}, and discussed in more details in the companion paper~\cite{DeSantis:2025yfm}, our first-principles theoretical results have a total accuracy of $O(5\%)$ and are in very good agreement with the corresponding experimental results of Refs.~\cite{CLEO:2009uah,BESIII:2021duu}. 

The plan of the paper is as follows. In \cref{sec:contratemoments,sec:contrate} we set our notation and derive the formulae for the decay rate. In \cref{sec:contmoments} we present the formulae for the lepton-energy moments. In \cref{sec:sigmato0} we derive the asymptotic formulae that we will use to extrapolate our results, obtained with the HLT algorithm of Ref.~\cite{Hansen:2019idp} at increasingly smaller values of a smearing parameter $\sigma$, down to the $\sigma\mapsto 0$ limit. In \cref{sec:correlators} we define the lattice Euclidean correlators from which we extract our physical results. In \cref{sec:hlt} we discuss the details of the implementation of the HLT algorithm  used in this work. In the \cref{sec:cs_DGammaDq2,sec:cd_DGammaDq2,,sec:su_DGammaDq2} we present our results for the different flavor contributions to the decay rate, while the results for the lepton-energy moments are presented in \cref{sec:anamom}. In \cref{sec:conclusions} we summarize our results and present our conclusions. In \cref{sec:exp} we explain how we obtained the experimental result for the decay rate by starting from the currently available measurements of the branching-ratios. 

%%%%%%%%%%%%%%%%%%%%%%%%%%%%%%%%%%%%%%%%%%%%%%%%%%%%%%%%%%%%%%%%%%%%%%%%%%%%%%%%%%%%%%%%%%%%%%%%
\section{
\label{sec:contratemoments}
The differential decay rate 
}
\begin{figure}
\includegraphics[width=0.9\columnwidth]{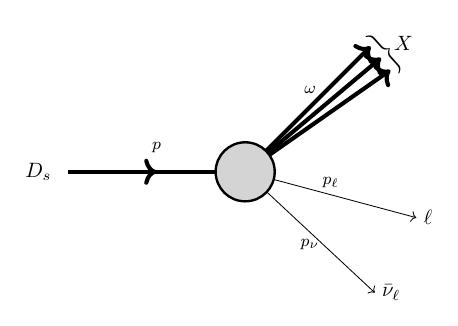}
\caption{\label{fig:kinematic} The kinematics of the inclusive $D_s\mapsto X\ell \bar \nu_\ell$ semileptonic decay. The incoming $D_s$ meson carries momentum $p$, and the outgoing lepton, neutrino and generic hadron state carry momentum $p_\ell$, $p_\nu$ and $\omega$, respectively.} 
\end{figure}

%%%%%%%%%%%%%%%%%%%%%%%%%%%%%%%%%%%%%%%%%%%%%%%%%%%%%%%%%%%%%%%%%%%%%%%%%%%%%%%%%%%%%%%%%%%%%%%%
We work in the rest-frame of the decaying $D_s$ meson and call
\begin{flalign}
&
p=m_{D_s}(1,\vec 0)\;,
\qquad\quad
\omega=m_{D_s}(\omega_0,\vec \omega)\;,
\nonumber \\[8pt]
&
p_\ell=m_{D_s}(e_\ell,\vec k_\ell)\;,
\qquad
p_\nu=m_{D_s}(e_\nu,\vec k_\nu)\;,
\end{flalign}
the four-momenta of the $D_s$, of the generic hadronic state $X$, of the lepton and of the neutrino, so that the energy-momentum conservation relation $p=p_\ell+p_\nu+\omega$ (see \cref{fig:kinematic}) implies
\begin{flalign}
\omega_0=1-e_\ell-e_\nu\;,
\qquad
\vec \omega=-\vec k_\ell -\vec k_\nu\;.
\end{flalign}
We work in the approximation in which the charged lepton is massless and therefore we have $\vec k_\ell^2=e_\ell^2$ as well as $\vec k_\nu^2=e_\nu^2$.

The fully inclusive process $D_s\mapsto X\ell \bar \nu_\ell$ can be separated into three different flavor channels. These are mediated by the flavor components $J^\mu_{\bar cs}$, $J^\mu_{\bar cd}$  and $J^\mu_{\bar us}$ of the hadronic weak current, given by
\begin{flalign}
&
J^\mu_{\bar fg}(x)
=\bar \psi_{\bar f}(x) \gamma^\mu(1-\gamma^5) \psi_g(x)\;.
\end{flalign}
When the flavor indexes $\bar f g$ are omitted, we refer to the fully inclusive process that is mediated by the sum of the three different flavor contributions weighted by the corresponding CKM matrix elements, 
\begin{flalign}
J^\mu(x) = 
V_{cs} J^\mu_{\bar cs}(x) +
V_{cd} J^\mu_{\bar cd}(x) +
V_{us} J^\mu_{\bar us}(x)\;.
\end{flalign}
Taking into account that the $D_s$ meson has $\bar s c$ flavor, in the channel mediated by $J^\mu_{\bar cd}$ the final hadrons are $\bar s d$-flavored and, therefore, denoted as $X_{\bar s d}$. Analogously, in the channel mediated by $J^\mu_{\bar us}$ the final hadrons are $\bar u c$-flavored and denoted as $X_{\bar u c}$. In the channel mediated by $J^\mu_{\bar cs}$ the final hadrons are flavorless and, in this case, we denote them as $X_{\bar ss}$. In the following, we shall call $X_{\bar F G}$ the hadronic states in the channel mediated by the current $J^\mu_{\bar fg}$. We thus have the correspondence
\begin{flalign}
\bar fg=\{\bar c s, \bar c d,\bar u s\}
\quad
\longleftrightarrow
\quad
\bar FG=\{\bar s s, \bar s d,\bar u c\}
\end{flalign}
between the flavor indexes of the currents and of the states.

The currently available experimental results~\cite{CLEO:2009uah,BESIII:2021duu} provide the fully-inclusive decay rate $\Gamma\equiv \Gamma[D_s\mapsto X \ell \bar \nu_\ell]$ which is the sum,
\begin{flalign}
\Gamma=
\vert V_{cs}\vert^2 \Gamma_{\bar cs} +
\vert V_{cd}\vert^2 \Gamma_{\bar cd} +
\vert V_{us}\vert^2 \Gamma_{\bar us} \;,
\end{flalign}
of the contributions corresponding to the different flavor channels. In this work, however, we also provide separate results for all of the contributions, i.e.\ for the dominant channel $\Gamma_{\bar cs}$ as well as for the Cabibbo-suppressed channels $\Gamma_{\bar cd}$  and $\Gamma_{\bar us}$. 

Each contribution to the decay rate can be written as
\begin{flalign}
&
\Gamma_{\bar fg}=
\nonumber \\[8pt]
&
G_F^2 S_\mathrm{EW} \int \frac{d^3 p_\nu}{(2\pi)^3}\, 
\frac{d^3 p_\ell}{(2\pi)^3}\,
\frac{
L_{\mu\nu}(p_\ell,p_\nu) H_{\bar fg}^{\mu\nu}(p,p-p_\ell-p_\nu) }
{4 m_{D_s}^2 e_\ell e_\nu}\;,
\label{eq:gammfgstart}
\end{flalign}
where $G_F$ is the Fermi constant and $S_\mathrm{EW}=1.013$  accounts for the logarithmic electroweak correction~\cite{Sirlin:1981ie} and for the QED threshold corrections\footnote{
In \cref{sec:su_DGammaDq2} we provide quantitative evidence  that $\Gamma_{\bar u s}$ is negligible w.r.t.\ the dominant $\Gamma_{\bar c s}$ and the Cabibbo-suppressed $\Gamma_{\bar c d}$ contributions. This allows us to ignore the fact that  QED threshold corrections are different in the $\bar u s$ channel and, therefore, to use the same $S_\mathrm{EW}$ factor for all channels.} that have been computed  in Ref.~\cite{Bigi:2023cbv}.
In this work, we perform an isoQCD calculation and neglect long-distance isospin breaking effects. We define the leptonic tensor as
\begin{flalign}
&
L^{\alpha\beta}(p_\ell,p_\nu) = 
\nonumber \\[8pt]
&
4\left\{
p_\ell^\alpha p_\nu^\beta + p_\ell^\beta p_\nu^\alpha - g^{\alpha\beta} p_\ell\cdot p_\nu +i\epsilon^{\alpha\beta\gamma\delta} (p_\ell)_\gamma (p_\nu)_\delta 
\right\}\;,
\label{eq:leptens}
\end{flalign}
where $\epsilon^{\alpha\beta\gamma\delta}$ is the totally antisymmetric four-index Levi--Civita symbol, with $\epsilon^{0123}=1$. The so-called hadronic tensor, which is in fact an hadronic spectral density, is expressed by 
\begin{flalign}
H_{\mu\nu}(p,\omega)
=\frac{(2\pi)^4}{2m_{D_s}}
\bra{D_s(p)}J_\mu^\dagger(0)\, \delta^4(\mathbb{P}-\omega)\, J_\nu(0)\ket{D_s(p)}\;,
\end{flalign}
where $\mathbb{P}=(H,\vec P)$ is the QCD four-momentum operator. Based on Lorentz and time-reversal covariance, $H^{\mu\nu}(p,\omega)$ can be decomposed into five form-factors,
\begin{flalign}
&
m_{D_s}^3 H^{\mu\nu}(p,\omega) 
=
g^{\mu\nu} m_{D_s}^2 h^{(1)}
+
p^\mu p^\nu h^{(2)}
\nonumber \\[8pt]
&
+
(p-\omega)^\mu (p-\omega)^\nu h^{(3)}
+
\left\{p^\mu (p-\omega)^\nu+(p-\omega)^\mu p^\nu\right\} h^{(4)}
\nonumber \\[8pt]
&
+
i\epsilon^{\mu\nu\alpha\beta} p_\alpha (p-\omega)_\beta h^{(5)}\;,
\label{eq:hff}
\end{flalign}
which, in our convention, are real and dimensionless. In the rest-frame of the $D_s$ meson the dependence of the form-factors upon the scalars $p\cdot \omega$ and $\omega^2$ can be traded for the dependence upon the variables $(\omega_0,\vec \omega^2)$. Therefore, by omitting the dependence upon $p^2=m_{D_s}^2$, we have  
\begin{flalign}
h^{(i)}\equiv h^{(i)}(\omega_0,\vec \omega^2)\;,
\qquad
i=1,\cdots,5\;.
\end{flalign}

In order to express the form-factors in terms of the different components of the hadronic tensor, i.e.\ to invert the system of \cref{eq:hff}, we consider the two unit vectors $\vec{\hat n}_r$   that are orthonormalized and orthogonal to $\vec{\hat \omega} = \vec \omega/\vert \vec \omega\vert$, i.e.
\begin{flalign}
\vec{\hat n}_r\cdot \vec{\hat n}_s=\delta_{rs}\;,
\qquad
\vec{\hat n}_r\cdot \vec{\hat \omega}=0\;,
\qquad
r,s=1,2\;,
\end{flalign}
and introduce the following quantities\footnote{In this paper we use a slightly different notation w.r.t.\ Ref.~\cite{Gambino:2022dvu}. The leptonic tensor in \cref{eq:leptens} differs (is larger) by a factor $4$ with respect to the one that was given in Ref.~\cite[Eq.~(2.3)]{Gambino:2022dvu}. The hadronic tensor was denoted as $W_{\mu\nu}$ in Ref.~\cite{Gambino:2022dvu} and we have the correspondence $H_{\mu\nu}=2\pi W_{\mu\nu}$. The correspondence between the hadronic form factors of \cref{eq:hff} and those of Ref.~\cite{Gambino:2022dvu} is $h^{(1)}=-2\pi m_{D_s} W_1$, $h^{(2)}=2\pi m_{D_s} W_2$, $h^{(3)}=2\pi m_{D_s} W_4$, $h^{(4)}=2\pi m_{D_s} W_5$,
$h^{(5)}=-2\pi m_{D_s} W_3$. Finally, the relations between the $\mathcal{Y}^{(i)}$ quantities defined here and the quantities denoted as $Y^{(i)}$ in Ref.~\cite[Eq.~(2.9)]{Gambino:2022dvu} are $\mathcal{Y}^{(1)}=Y^{(1)}/2$, $\mathcal{Y}^{(2)}=Y^{(2)}$, $\mathcal{Y}^{(3)}=Y^{(3)}$, $\mathcal{Y}^{(4)}=Y^{(4)}/2$, and $\mathcal{Y}^{(5)}=Y^{(5)}$.}
\begin{flalign}
&
\mathcal{Y}^{(1)}=-\frac{m_{D_s}}{2}\sum_{r=1}^2\sum_{i,j=1}^3 \hat n^i_r \hat n^j_r H^{ij}(p,\omega)\;,
\nonumber \\[8pt]
&
\mathcal{Y}^{(2)}=m_{D_s}H^{00}(p,\omega)\;,
\nonumber \\[8pt]
&
\mathcal{Y}^{(3)}=m_{D_s}\sum_{i,j=1}^3 \hat \omega^i \hat \omega^j H^{ij}(p,\omega)\;,
\nonumber \\[8pt]
&
\mathcal{Y}^{(4)}=-m_{D_s}\sum_{i=1}^3 \hat \omega^i H^{0i}(p,\omega)\;,
\nonumber \\[8pt]
&
\mathcal{Y}^{(5)}=-\frac{im_{D_s}}{2}\sum_{i,j,k=1}^3 \epsilon^{ijk} \hat \omega^k H^{ij}(p,\omega)\;,
\label{eq:Ydefs}
\end{flalign}
where $\epsilon^{ijk}$ is the totally antisymmetric three-index Levi--Civita symbol, with $\epsilon^{123}=1$. We then have
\begin{flalign}
&
h^{(1)}=\mathcal{Y}^{(1)}\;,
\nonumber \\[12pt]
&
h^{(2)}=
\frac{(1-\omega_0)^2-\vec \omega^2}{\vec \omega^2}\mathcal{Y}^{(1)}+\mathcal{Y}^{(2)}
\nonumber \\[4pt]
&
\phantom{h^{(2)}=
\frac{(1-\omega_0)^2}{\vec \omega^2}}
+\frac{(1-\omega_0)^2}{\vec \omega^2}\mathcal{Y}^{(3)}-\frac{2(1-\omega_0)}{\vert \vec \omega\vert} \mathcal{Y}^{(4)}\;,
\nonumber \\[12pt]
&
h^{(3)}=\frac{\mathcal{Y}^{(1)}+\mathcal{Y}^{(3)}}{\vec \omega^2}\;,
\nonumber \\[12pt]
&
h^{(4)}=-\frac{1-\omega_0}{\vec \omega^2}(\mathcal{Y}^{(1)}+\mathcal{Y}^{(3)})+\frac{\mathcal{Y}^{(4)}}{\vert \vec \omega\vert}\;,
\nonumber \\[12pt]
&
h^{(5)}=\frac{\mathcal{Y}^{(5)}}{\vert \vec \omega\vert}\;.
\label{eq:hsfromys}
\end{flalign}
In the previous two sets of equations, as already done in the case of the form-factors,  we have used the compact notation
\begin{flalign}
\mathcal{Y}^{(i)}\equiv \mathcal{Y}^{(i)}(\omega_0,\vec \omega^2)\;,
\qquad
i=1,\cdots,5\;.
\end{flalign}

By relying on the form-factors decomposition of \cref{eq:hff}, and by working out the phase-space kinematical constraints in the rest-frame of the $D_s$ meson, \cref{eq:gammfgstart} can be rewritten as
\begin{flalign}
&
\Gamma_{\bar f g}=
\int_0^{(\vert\vec \omega\vert_{\bar F G}^\mathrm{max})^2} d\vec \omega^2
\int_{\omega_{\bar FG}^\mathrm{min}}^{\omega^\mathrm{max}} d\omega_0
\int_{e_\ell^\mathrm{min}}^{e_\ell^\mathrm{max}} de_\ell
\frac{d\Gamma_{\bar f g}}{d\vec \omega^2 d\omega_0 de_\ell}\;,
\label{eq:gammfgtripleint}
\end{flalign}
where the triple-differential decay rate is given by
\begin{flalign}
&
\frac{d\Gamma}{d\vec \omega^2 d\omega_0 de_\ell}
=
\frac{m_{D_s}^5 G_F^2 S_\mathrm{EW}}{32\pi^4}\Bigg(
\nonumber \\[8pt]
&
-
2\left\{(1-\omega_0)^2-\vec \omega^2 \right\} h^{(1)}
+
\left\{
\vec \omega^2-(1-\omega_0-2e_\ell)^2
\right\} h^{(2)}
\nonumber \\[8pt]
&
+
2\left[(1-\omega_0)^2-\vec \omega^2 \right]\left[2e_\ell-(1-\omega_0)\right] h^{(5)}
\Bigg)\;.
\label{eq:tripleGhs}
\end{flalign}
By using Eqs.~(\ref{eq:hsfromys}) this quantity can also be expressed in terms of the independent components of the hadronic tensor, i.e.\ in terms of the distributions $\mathcal{Y}^{(i)}(\omega_0,\vec \omega^2)$.

The integration limits to be used in \cref{eq:gammfgtripleint} are given by the following expressions
\begin{flalign}
&
e_\ell^\mathrm{min}=\frac{1-\omega_0-\vert \vec \omega\vert}{2}\;,
\qquad
e_\ell^\mathrm{max}=\frac{1-\omega_0+\vert \vec \omega\vert}{2}\;,
\nonumber \\[8pt]
&
\omega_{\bar FG}^\mathrm{min}=\sqrt{r_{\bar FG}^2+\vec \omega^2}\;,
\qquad
\omega^\mathrm{max}=1-\sqrt{\vec \omega^2}\;,
\nonumber \\[8pt]
&
\vert\vec \omega\vert_{\bar FG}^\mathrm{max}= \left(\frac{1-r_{\bar FG}^2}{2}\right)\;.
\label{eq:limits1}
\end{flalign}
An important r\^ole in deriving \cref{eq:limits1} is played by the exclusive process in which the $D_s$ meson decays into the lightest possible hadronic state in each channel, that we call $P_{\bar F G}$. In all channels the lightest state is the QCD-stable pseudoscalar meson corresponding to the isolated single-particle eigenvalue of the Hamiltonian $\mathbb{H}$ with the given flavor. In the case of $\Gamma_{\bar c s}$ the lightest state $P_{\bar s s}$ is the neutral pion. In the case of $\Gamma_{\bar c d}$ the lightest state $P_{\bar s d}$ is a neutral kaon. In the case of $\Gamma_{\bar u s}$ the lightest state $P_{\bar u c}$ is a neutral $D$ meson and, since $m_{D_s}<m_{D}+m_\pi$, the ``inclusive'' channel $D_s\mapsto X_{\bar u c} \ell \bar \nu_\ell$ is in fact identical to the exclusive channel $D_s\mapsto D \ell \bar \nu_\ell$. 

The parameter $r_{\bar FG}$ appearing in \cref{eq:limits1} is
\begin{flalign}
r_{\bar FG} = \frac{m_{P_{\bar FG}}}{m_{D_s}}\,,
\label{eq:rdef}
\end{flalign}
i.e.\ the mass $m_{P_{\bar FG}}$ of the lightest state $P_{\bar F G}$ in units of $m_{D_s}$ and, therefore,
\begin{flalign}
r_{\bar ss} = \frac{m_{\pi}}{m_{D_s}}\,,
\qquad
r_{\bar sd} = \frac{m_{K}}{m_{D_s}}\,,
\qquad
r_{\bar uc} = \frac{m_{D}}{m_{D_s}}\,.
\end{flalign}

An important remark is now in order. In order to fully take into account the exclusive processes $D_s\mapsto P_{\bar F G} \ell \bar \nu_\ell$ in the calculation of $\Gamma_{\bar f g}$ the integration limits $\omega_{\bar FG}^\mathrm{min}$ and
$\vert\vec \omega\vert_{\bar FG}^\mathrm{max}$, which are in fact the energy and the maximum allowed spatial momentum of $P_{\bar F G}$ in units of $m_{D_s}$, have to be understood as
\begin{flalign}
\omega_{\bar FG}^\mathrm{min}
\mapsto \sqrt{r_{\bar FG}^2+\vec \omega^2} -\epsilon\;,
\quad
\vert\vec \omega\vert_{\bar FG}^\mathrm{max}\mapsto \left(\frac{1-r_{\bar FG}^2}{2}\right) + \epsilon\;,
\label{eq:interpretation}
\end{flalign}
with $\epsilon$ a small positive number.
Indeed, as we are going to explain at the end of the section, the contribution of the exclusive processes $D_s\mapsto P_{\bar F G} \ell \bar \nu_\ell$ to the differential decay rate, being associated with the isolated single-particle eigenvalue of the Hamiltonian in the given flavor channel, can be separated from the multi-particle contributions according to
\begin{flalign}
\frac{d\Gamma_{\bar f g}}{d\vec \omega^2 d\omega_0 de_\ell}=
\delta(\omega_0-\omega_{\bar FG}^\mathrm{min})
\frac{d\Gamma_{\bar f g}^\mathrm{excl}}{d\vec \omega^2 de_\ell}
+
\frac{d\Gamma_{\bar f g}^\mathrm{cont}}{d\vec \omega^2 d\omega_0 de_\ell}\;.
\label{eq:exclrate1}
\end{flalign}
By relying on the interpretation of the integration limits given in \cref{eq:interpretation}, one has
\begin{flalign}
\int_{\omega_{\bar FG}^\mathrm{min}-\epsilon}^{\omega^\mathrm{max}} d\omega_0\
\delta(\omega_0-\omega_{\bar FG}^\mathrm{min}) = 1\,,
\end{flalign}
which means that the exclusive contribution has been fully included. Notice that the shift of the limit $\vert\vec \omega\vert_{\bar FG}^\mathrm{max}$ is also necessary because at the end-point corner of the phase-space where $\vert \vec \omega\vert= \vert\vec \omega\vert_{\bar FG}^\mathrm{max}$ one has $\omega^\mathrm{max}=\omega_{\bar FG}^\mathrm{min}$.
As far as the parameter $\epsilon$ is concerned, from the theoretical perspective it has to be read as $0^+$, i.e.\ an infinitesimal shift that sets the prescription to calculate the integrals of the distributions $d\Gamma/d\vec \omega^2 d\omega_0$. From the phenomenological perspective $\epsilon$ can be identified with the energy-momentum resolution of the experimental apparatus.

We now provide the explicit expression of $d\Gamma_{\bar f g}^\mathrm{excl}/d\vec \omega^2 de_\ell$. In each flavor channel, the hadronic tensor $H^{\mu\nu}_{\bar fg}(p,\omega)$ can be written as
\begin{flalign}
H_{\bar fg}^{\mu\nu}(p,\omega)
=
\delta(\omega_0-\omega_{\bar FG}^\mathrm{\min})\,
\rho_{\bar fg}^{\mu\nu}(p,\Omega_{\bar FG})
+
\bar H_{\bar fg}^{\mu\nu}(p,\omega)\;.
\label{eq:Hsep}
\end{flalign}
In the previous expression we called $\bar H_{\bar fg}^{\mu\nu}(p,\omega)$ the contribution coming from the continuum spectrum and we have $\bar H_{\bar fg}^{\mu\nu}(p,\omega)=0$ for $\omega_0<\omega_{\bar FG}^\mathrm{\min}+\Delta$ where $\Delta=O(m_\pi)$ is the energy gap in the given flavor channel. Then we have introduced $\Omega_{\bar FG}=m_{D_s}(\omega_{\bar FG}^\mathrm{min},\vec \omega)$, i.e.\ the on-shell four-momentum of the state $P_{\bar F G}$ ($\Omega_{\bar FG}^2=m_{P_{\bar F G}}^2$), and the single-particle exclusive contribution
\begin{flalign}
&
\rho_{\bar fg}^{\mu\nu}(p,\Omega_{\bar FG}) =
\nonumber \\[8pt]
&
\frac{\pi}{2m_{D_s}^3 \omega_{\bar FG}^\mathrm{\min}}
\bra{D_s} (J^\mu_{\bar fg})^\dagger(0)\ket{P_{\bar F G}}
\bra{P_{\bar F G}} J^\nu_{\bar fg}(0)\ket{D_s}\;.
\end{flalign}
By using the standard decomposition 
\begin{flalign}
\bra{P_{\bar F G}} J^\mu_{\bar fg}(0)\ket{D_s}
=
(\Omega_{\bar FG}+p)^\mu f_{\bar fg}^+
+
(\Omega_{\bar FG}-p)^\mu f_{\bar fg}^-\;,
\end{flalign}
where the form factors $f_{\bar fg}^\pm$ depend on the masses of the $P_{\bar F G}$ and $D_s$ mesons and on $q^2=(\Omega_{\bar FG}-p)^2$ and therefore on $\vec \omega^2$ through $\omega_{\bar FG}^\mathrm{min}$, we have
\begin{flalign}
&\frac{d\Gamma_{\bar f g}^\mathrm{excl}}{d\vec \omega^2 de_\ell}
=
\nonumber \\[8pt]
&
\frac{m_{D_s}^5}{16\pi^3 \omega_{\bar FG}^\mathrm{min}}
\left[\vec \omega^2 -(1-\omega_{\bar FG}^\mathrm{min}-2e_\ell)^2 \right] \left[f_{\bar fg}^+(\vec \omega^2)\right]^2\;,
\label{eq:gammaexcl1}
\end{flalign}
where $d\Gamma_{\bar f g}^\mathrm{excl}/d\vec \omega^2 de_\ell$ is the differential decay rate of the exclusive process $D_s\mapsto P_{\bar F G} \ell \bar \nu_\ell$ introduced in \cref{eq:exclrate1}. 

%%%%%%%%%%%%%%%%%%%%%%%%%%%%%%%%%%%%%%%%%%%%%%%%%%%%%%%%%%%%%%%%%%%%%%%%%%%%%%%%%%%%%%%%%%%%%%%%
\section{
\label{sec:contrate}
The total decay rate 
}
%%%%%%%%%%%%%%%%%%%%%%%%%%%%%%%%%%%%%%%%%%%%%%%%%%%%%%%%%%%%%%%%%%%%%%%%%%%%%%%%%%%%%%%%%%%%%%%%
In order to compute the total rate $\Gamma$, the integrals appearing in \cref{eq:gammfgtripleint} have to be performed.
Given \cref{eq:tripleGhs}, and by using the fact that the hadronic form factors $h^{(i)}(\omega_0,\vec \omega^2)$ do not depend upon $e_\ell$, the lepton energy integral can be performed analytically and one finds
\begin{flalign}
&
\frac{1}{\bar \Gamma}
\frac{d \Gamma}{d \omega^0 d \vec \omega^2}
=
\nonumber \\[8pt]
&
\vert \vec \omega\vert^3\, Z^{(0)}
+
\vert \vec \omega\vert^2 (\omega^\mathrm{max}-\omega_0)\, Z^{(1)}
+
\vert \vec \omega\vert (\omega^\mathrm{max}-\omega_0)^2\, Z^{(2)},
\label{eq:doubleGhs}
\end{flalign}
where 
\begin{flalign}
\bar \Gamma= \frac{m_{D_s}^5 G_F^2 S_\mathrm{EW}}{48\pi^4}\;.
\end{flalign}
and where we have introduced the following three linear combinations of the five independent hadronic spectral densities $\mathcal{Y}^{(i)}(\omega_0,\vec \omega^2)$,
\begin{flalign}
&
Z^{(0)}=\mathcal{Y}^{(2)}+\mathcal{Y}^{(3)}-2\mathcal{Y}^{(4)}\;,
\nonumber \\[8pt]
&
Z^{(1)}=2\left(\mathcal{Y}^{(3)}-2\mathcal{Y}^{(1)}-\mathcal{Y}^{(4)}\right)\;,
\nonumber \\[8pt]
&
Z^{(2)}=\mathcal{Y}^{(3)}-2\mathcal{Y}^{(1)}\;.
\label{eq:Zgamma}
\end{flalign}
From the previous expressions it is evident that the parity-breaking form factor $h^{(5)}=\mathcal{Y}^{(5)}/\vert \vec \omega \vert$ does not contribute to the total rate.

To compute the $\omega_0$ integral in \cref{eq:gammfgtripleint} we first need to derive a mathematical representation of the decay rate that is suitable for a lattice evaluation. To this end, we start by introducing the kernels
\begin{flalign}
\Theta_\sigma^{(p)}(x) = x^p\, \Theta_\sigma(x)\;, 
\label{eq:defthetap}
\end{flalign}
where $p=0,1,2,\cdots$, is a non-negative integer and $\Theta_\sigma(x)$ is any Schwartz\footnote{That is, infinitely differentiable and vanishing, together with all of its derivatives, faster than any power of $x$ in the limit $x\mapsto -\infty$.} representation of the Heaviside step-function $\theta(x)$, which depends smoothly upon the smearing parameter $\sigma$ and which is such that
\begin{flalign}\label{eq:heaviside}
\lim_{\sigma\mapsto 0} \Theta_\sigma(x)=\theta(x)\;.
\end{flalign}
In this work we considered two different representations $\Theta_\sigma(x)$ which are explicitly given in \cref{eq:sigmoid,eq:erf}.

The introduction of this mathematical device allows to trade the $\omega_0$ phase-space integral, to be performed in the compact interval $[\omega^\mathrm{min}-\epsilon,\omega^\mathrm{max}]$ (see \cref{eq:gammfgtripleint,eq:limits1}), for convolutions of the distributions $Z^{(p)}(\omega_0,\vec \omega^2)$ with smooth smearing kernels,
\begin{flalign}
&
\frac{1}{\bar \Gamma}
\frac{d \Gamma^{(p)}(\sigma)}{d \vec \omega^2}
=
\nonumber \\[8pt]
&
\vert \vec \omega \vert^{3-p}\,
\int_{\omega^\mathrm{min}-\epsilon}^\infty d\omega_0\,
\Theta_\sigma^{(p)}(\omega^\mathrm{max}-\omega_0)\, Z^{(p)}(\omega_0,\vec \omega^2),
\label{eq:dGZint}
\end{flalign}
and with a limiting procedure,
\begin{flalign}
\Gamma
=
\sum_{p=0}^2
\int_0^{(\vert\vec \omega\vert^\mathrm{max}+\epsilon)^2} d\vec \omega^2\,
\lim_{\sigma\mapsto 0}
\frac{d\Gamma^{(p)}(\sigma)}{d \vec \omega^2}\;.
\label{eq:gammalimit}
\end{flalign}

We now rely on the Stone--Weierstrass theorem and observe that, for any positive value of the length scale $a$, the kernels $\Theta_\sigma^{(p)}(\omega^\mathrm{max}-\omega_0)$ can exactly be represented according to
\begin{flalign}
\Theta_\sigma^{(p)}(\omega^\mathrm{max}-\omega_0)
=
\lim_{N\mapsto \infty}\sum_{n=1}^{N} g^{(p)}_n(N)\, e^{-\omega_0 (a m_{D_s}) n} \;.
\label{eq:stoneZ}
\end{flalign}
The coefficients $g^{(p)}_n(N)$ appearing in the previous formula have to be read as the coordinates of the kernels $\Theta_\sigma^{(p)}(\omega^\mathrm{max}-\omega_0)$ on the discrete set of basis-functions $\exp[-\omega_0 (a m_{D_s}) n]$. The functional basis has been chosen in order to establish a direct connection between $d \Gamma^{(p)}(\sigma)/d \vec \omega^2$ and the primary data of a lattice simulation, i.e.\ Euclidean correlators at discrete time separations. Indeed, while it is not possible to compute the $Z^{(p)}(\omega_0,\vec \omega^2)$ distributions directly on the lattice, it is instead possible (see \cref{sec:correlators}) to compute the following  Euclidean correlators
\begin{flalign}
\hat Z^{(p)}(t,\vec \omega^2)
=
\int_{\omega^\mathrm{min}-\epsilon}^\infty d\omega_0\, e^{-\omega_0 (m_{D_s}t)}\, Z^{(p)}(\omega_0,\vec \omega^2)
\label{eq:Zcorr}
\end{flalign}
at the discrete Euclidean times $t=an$, where $a$ is the lattice spacing\footnote{See Ref.~\cite{Patella:2024cto} for the generalization of this strategy to the case in which the length scale $a$, called $\tau$ in that paper, is kept constant in physical units.}. By using \cref{eq:stoneZ} the connection can now easily be established,
\begin{flalign}\label{eq:lim_N_gamma}
\frac{1}{\bar \Gamma}
\frac{d \Gamma^{(p)}(\sigma)}{d \vec \omega^2}
=\vert \vec \omega \vert^{3-p}\,
\lim_{N\mapsto \infty}\sum_{n=1}^{N} g^{(p)}_n(N)\,
\hat Z^{(p)}(a n,\vec \omega^2)\; .
\end{flalign}

In order to determine the coefficients $g^{(p)}_n(N)$, and to study numerically the $N\mapsto \infty$ limit at fixed $\sigma>0$ and the associated systematic errors, we use the HLT algorithm of Ref.~\cite{Hansen:2019idp}, see \cref{sec:hlt}. In order to perform the necessary $\sigma\mapsto 0$ extrapolations we rely on the asymptotic formulae derived and discussed in \cref{sec:sigmato0}. Details concerning the numerical evaluation of the $\vec \omega^2$ integral will be provided in \cref{sec:conclusions}.

%%%%%%%%%%%%%%%%%%%%%%%%%%%%%%%%%%%%%%%%%%%%%%%%%%%%%%%%%%%%%%%%%%%%%%%%%%%%%%%%%%%%%%%%%%%%%%%%
\section{
\label{sec:contmoments}
The lepton-energy moments 
}
%%%%%%%%%%%%%%%%%%%%%%%%%%%%%%%%%%%%%%%%%%%%%%%%%%%%%%%%%%%%%%%%%%%%%%%%%%%%%%%%%%%%%%%%%%%%%%%%
The lepton-energy moments are defined as the integrals of the differential decay rate multiplied by a power of the lepton energy ($m_{D_s} e_\ell$) and normalized by the total rate, i.e.
\begin{flalign}
&
M_n=
\nonumber \\[8pt]
&
\int_0^{(\vert\vec \omega\vert^\mathrm{max}+\epsilon)^2} d\vec \omega^2
\int_{\omega^\mathrm{min}-\epsilon}^{\omega^\mathrm{max}} d\omega_0
\int_{e_\ell^\mathrm{min}}^{e_\ell^\mathrm{max}} de_\ell
\frac{dM_n}{d\vec \omega^2 d\omega_0 de_\ell}\;,
\end{flalign}
where
\begin{flalign}
\frac{dM_n}{d\vec \omega^2 d\omega_0 de_\ell} 
=\frac{\left(m_{D_s} e_\ell\right)^n}{\Gamma}
\frac{d\Gamma}{d\vec \omega^2 d\omega_0 de_\ell}\;.
\label{eq:moments}
\end{flalign}
In this work we have computed the first two moments, $M_1$ and $M_2$.
To do that, as already done in the case of the total rate, we performed the $e_\ell$ integrals analytically and then represented $M_1$ and $M_2$ in terms of the smearing kernels $\Theta_\sigma^{(p)}(\omega^\mathrm{max}-\omega_0)$. 

Concerning the first moment, we have
\begin{flalign}
M_1
=
\sum_{p=0}^3
\int_0^{(\vert\vec \omega\vert^\mathrm{max}+\epsilon)^2} d\vec \omega^2\,
\lim_{\sigma\mapsto 0}
\frac{dM_1^{(p)}(\sigma)}{d \vec \omega^2}\;,
\end{flalign}
where
\begin{flalign}
&
\frac{1}{\bar M_1}
\frac{d M_1^{(p)}(\sigma)}{d \vec \omega^2}
=
\nonumber \\[8pt]
&
\vert \vec \omega \vert^{4-p}\,
\int_{\omega_0^\mathrm{min}-\epsilon}^\infty d\omega_0\,
\Theta_\sigma^{(p)}(\omega^\mathrm{max}-\omega_0)\, Z^{(p)}_1(\omega_0,\vec \omega^2),
\label{eq:dM1Zint}
\end{flalign}
with the normalization given by
\begin{flalign}
\bar M_1= \frac{1}{\Gamma}\, \frac{m_{D_s}^6\, G_F^2\, S_\mathrm{EW}}{96\pi^4}\;,
\end{flalign}
and where we have introduced the following four linear combinations
\begin{flalign}
&
Z_1^{(0)}= \mathcal{Y}^{(2)}+\mathcal{Y}^{(3)}-2\mathcal{Y}^{(4)}\;,
\nonumber \\[8pt]
&
Z_1^{(1)}= -4 \mathcal{Y}^{(1)} + \mathcal{Y}^{(2)} + 3 \mathcal{Y}^{(3)} - 4 \mathcal{Y}^{(4)} + 2 \mathcal{Y}^{(5)}\;,
\nonumber \\[8pt]
&
Z_1^{(2)}= -6 \mathcal{Y}^{(1)} + 3 \mathcal{Y}^{(3)} - 2 \mathcal{Y}^{(4)} + \mathcal{Y}^{(5)}\;,
\nonumber \\[8pt]
&
Z_1^{(3)}=-2 \mathcal{Y}^{(1)} + \mathcal{Y}^{(3)}\;,
\label{eq:ZM1}
\end{flalign}
of the five independent hadronic spectral densities $\mathcal{Y}^{(i)}(\omega_0,\vec \omega^2)$. 

Concerning the second moment, we have
\begin{flalign}
M_2
=
\sum_{p=0}^4
\int_0^{(\vert\vec \omega\vert^\mathrm{max}+\epsilon)^2} d\vec \omega^2\,
\lim_{\sigma\mapsto 0}
\frac{dM_2^{(p)}(\sigma)}{d \vec \omega^2}\;,
\end{flalign}
with
\begin{flalign}
&
\frac{1}{\bar M_2}
\frac{d M_2^{(p)}(\sigma)}{d \vec \omega^2}
=
\nonumber \\[8pt]
&
\vert \vec \omega \vert^{5-p}\,
\int_{\omega_0^\mathrm{min}-\epsilon}^\infty d\omega_0\,
\Theta_\sigma^{(p)}(\omega^\mathrm{max}-\omega_0)\, Z^{(p)}_2(\omega_0,\vec \omega^2),
\label{eq:dM2Zint}
\end{flalign}
the normalization given by
\begin{flalign}
\bar M_2= \frac{1}{\Gamma}\, \frac{m_{D_s}^7\, G_F^2\, S_\mathrm{EW}}{960\pi^4}\;,
\end{flalign}
and the relevant hadronic spectral densities, which in this case are five, given by
\begin{flalign}
&
Z_2^{(0)}= 6 (\mathcal{Y}^{(2)} + \mathcal{Y}^{(3)} - 2 \mathcal{Y}^{(4)})\;,
\nonumber \\[8pt]
&
Z_2^{(1)}= 2 (-14 \mathcal{Y}^{(1)} + 5 \mathcal{Y}^{(2)} + 11 \mathcal{Y}^{(3)} - 16 \mathcal{Y}^{(4)} + 10 \mathcal{Y}^{(5)})\;,
\nonumber \\[8pt]
&
Z_2^{(2)}= -54 \mathcal{Y}^{(1)} + 5 \mathcal{Y}^{(2)} + 31 \mathcal{Y}^{(3)} - 30 \mathcal{Y}^{(4)} + 30 \mathcal{Y}^{(5)}\;,
\nonumber \\[8pt]
&
Z_2^{(3)}=-10(4 \mathcal{Y}^{(1)} - 2 \mathcal{Y}^{(3)} + \mathcal{Y}^{(4)} - \mathcal{Y}^{(5)})\;,
\nonumber \\[8pt]
&
Z_2^{(4)}=5 (-2 \mathcal{Y}^{(1)} + \mathcal{Y}^{(3)})\;.
\label{eq:ZM2}
\end{flalign}
Note that, in contrast to the total rate $\Gamma$, the first two lepton-energy moments do probe the parity-breaking form factor $h^{(5)}=\mathcal{Y}^{(5)}/\vert \vec \omega \vert$.

The connection between the differential moments $d M_l^{(p)}(\sigma)/d \vec \omega^2$, at fixed smearing parameter $\sigma$ and for $l=1,2$, and the Euclidean correlators that we have computed on the lattice is obtained by using, once again, the representation given in \cref{eq:stoneZ} of the smearing kernels $\Theta_\sigma^{(p)}(\omega^\mathrm{max}-\omega_0)$. We have
\begin{flalign}\label{eq:lim_N_moments}
\frac{1}{\bar M_l}
\frac{d M_l^{(p)}(\sigma)}{d \vec \omega^2}
=\vert \vec \omega \vert^{3+l-p}\,
\lim_{N\mapsto \infty}\sum_{n=1}^{N} g^{(p)}_n(N)\,
\hat Z^{(p)}_l(a n,\vec \omega^2)\,,
\end{flalign}
where the lattice correlators 
\begin{flalign}
\hat Z^{(p)}_l(t,\vec \omega^2)
=
\int_{\omega^\mathrm{min}-\epsilon}^\infty d\omega_0\, e^{-\omega_0 (m_{D_s}t)}\, Z^{(p)}_l(\omega_0,\vec \omega^2)
\label{eq:Zlcorr}
\end{flalign}
are the Laplace transforms of the spectral densities defined in \cref{eq:ZM1} for $l=1$ and in \cref{eq:ZM2} for $l=2$. These, as well as the ones of \cref{eq:Zcorr} entering the calculation of $\Gamma$, can be easily computed as linear combinations of the five independent amputated correlators
\begin{flalign}
\mathcal{\hat Y}^{(i)}(t,\vec \omega^2)
=
\int_{\omega^\mathrm{min}-\epsilon}^\infty d\omega_0\, e^{-\omega_0 (m_{D_s}t)}\, \mathcal{Y}^{(i)}(\omega_0,\vec \omega^2)\;.
\label{eq:Ys}
\end{flalign}
The procedure that we used to extract these quantities from lattice correlators is discussed in \cref{sec:correlators}. Before doing that, however, we derive in the next section the asymptotic formulae that we use to study numerically the $\sigma\mapsto 0$ extrapolations. These formulae will also motivate our choice of organizing the calculation in terms of the spectral densities $Z^{(p)}(\omega_0,\vec \omega^2)$ and $Z^{(p)}_l(\omega_0,\vec \omega^2)$ and not in terms of the $\mathcal{Y}^{(i)}(\omega_0,\vec \omega^2)$.

%%%%%%%%%%%%%%%%%%%%%%%%%%%%%%%%%%%%%%%%%%%%%%%%%%%%%%%%%%%%%%%%%%%%%%%%%%%%%%%%%%%%%%%%%%%%%%%%
\section{
\label{sec:sigmato0}
The \texorpdfstring{$\sigma\mapsto 0$}{s->0} asymptotic behaviour
}
%%%%%%%%%%%%%%%%%%%%%%%%%%%%%%%%%%%%%%%%%%%%%%%%%%%%%%%%%%%%%%%%%%%%%%%%%%%%%%%%%%%%%%%%%%%%%%%%
In the previous two sections, in order to compute the total rate and the lepton-energy moments on the lattice, we traded the compact $\omega_0$ phase-space integral for convolutions of the $Z^{(p)}(\omega_0,\vec \omega^2)$ and $Z^{(p)}_l(\omega_0,\vec \omega^2)$ distributions with the smooth kernels $\Theta_\sigma^{(p)}(\omega^\mathrm{max}-\omega_0)$ and with the $\sigma\mapsto 0$ limiting procedure. In order to understand how to perform numerically the required $\sigma\mapsto 0$ extrapolations we now study the asymptotic behavior for small values of $\sigma$ of the generic expression
\begin{flalign}
G^{(p)}(\sigma)
=
\int_{\omega^\mathrm{min}-\varepsilon}^\infty d\omega_0\,
\Theta^{(p)}_\sigma(\omega^\mathrm{max}-\omega_0)\, Z(\omega_0)
\label{eq:Oasympt1}
\end{flalign}
in which $G^{(p)}(\sigma)$ can be either $d\Gamma^{(p)}(\sigma)/d \vec \omega^2$ or $dM_l^{(p)}(\sigma)/d \vec \omega^2$ and, correspondingly, $Z(\omega_0)$ can be either $Z^{(p)}(\omega_0,\vec \omega^2)$ or $Z^{(p)}_l(\omega_0,\vec \omega^2)$ (see \cref{eq:dGZint,eq:dM1Zint,eq:dM2Zint}). 

As we are now going to explain, the behavior of $G^{(p)}(\sigma)$ for small values of $\sigma$ is governed by the behavior of the distribution $Z(\omega_0)$ for $\omega_0$ in a neighborhood of $\omega^\mathrm{max}$. A rigorous mathematical analysis of the possible singularities of the hadronic tensor (the distributions $Z^{(p)}(\omega_0,\vec \omega^2)$ and $Z^{(p)}_l(\omega_0,\vec \omega^2)$ are indeed linear combinations of $H^{\mu\nu}(p,\omega)$) goes far beyond the scope of this paper. Here, we study the $\sigma\mapsto 0$ limit of $G^{(p)}(\sigma)$ by starting from \cref{eq:Hsep} and by relying upon (well motivated) physics assumptions on the contributions to $Z(\omega_0)$ coming from the continuum part of the spectrum (multi-hadrons states) of $\mathbb{H}$. Indeed, the separation of the hadronic tensor $H^{\mu\nu}(p,\omega)$ into $\rho_{\bar fg}^{\mu\nu}(p,\Omega_{\bar FG})$ and $\bar H^{\mu\nu}(p,\omega)$ given in \cref{eq:Hsep} generates a corresponding separation for the distributions $\mathcal{Y}^{(i)}(\omega_0,\vec \omega^2)$ and, therefore, also for $Z^{(p)}(\omega_0,\vec \omega^2)$ and $Z^{(p)}_l(\omega_0,\vec \omega^2)$. This allows us to write
\begin{flalign}
Z(\omega_0) = \delta(\omega_0-\omega^\mathrm{\min}) Z_\mathrm{excl}
+\bar Z(\omega_0)
\label{eq:Zdecomposition}
\end{flalign}
and, correspondingly, to split the observable $G^{(p)}(\sigma)$ according to
\begin{flalign}
G^{(p)}(\sigma)
=
G^{(p)}_\mathrm{excl}(\sigma)
+
\bar G^{(p)}(\sigma)\;,
\end{flalign}
where
\begin{flalign}
&
G^{(p)}_\mathrm{excl}(\sigma)
=
\Theta^{(p)}_\sigma(\omega^\mathrm{max}-\omega^\mathrm{min})\, Z_\mathrm{excl}\;,
\nonumber \\[8pt]
&
\bar G^{(p)}(\sigma)=
\int_{\omega^\mathrm{min}}^\infty d\omega_0\,
\Theta^{(p)}_\sigma(\omega^\mathrm{max}-\omega_0)\, \bar Z(\omega_0)\;.
\label{eq:Oasympt2}
\end{flalign}
Our physics motivated\footnote{Assuming, as commonly done on the experimental side, that a differential decay rate can be measured at any energy is equivalent, on the theoretical side, to assume that the associated spectral density is a regular function in that energy range. From the physical perspective, it is very reasonable to assume that $\bar Z(\omega_0)$ is a continuous function with a countable (zero measure) set of $\theta$-function singularities located in correspondence of the thresholds of the allowed multi-particle states. In addition to these threshold singularities one has also to consider possible contributions to $\bar Z(\omega_0)$ coming from very narrow resonances. These are not true singularities but, in practice, might have the same effect of $\delta$-function singularities. There is no physical/kinematical reason to expect that one of these possible ($\delta$-) $\theta$-function (quasi) singularities can have an impact on our analysis.} working assumption concerns $\bar Z(\omega_0)$, which we shall consider analytical in a neighborhood of $\omega^\mathrm{max}$. 

Both the kernels $\Theta_\sigma(x)$ considered in this work (see \cref{eq:sigmoid,eq:erf}) satisfy the following properties
\begin{flalign}
&
\Theta_\sigma(x) = \Theta_1\left(\frac{x}{\sigma}\right)\,,
\qquad
\Theta_1(x)+\Theta_1(-x)=1\;,
\nonumber \\[8pt]
&
x^p\partial^q\left[\Theta_1(x)-1\right] 
\quad \stackrel{x\mapsto \infty}{\longrightarrow} \quad O(e^{-x})\;,
\label{eq:Thetaprops}
\end{flalign}
where $p$ and $q$ are generic non-negative integers.
Given our interpretation of the phase-space integration limits (see \cref{sec:contratemoments}), a direct consequence of these properties is that the single-particle contribution $G^{(p)}_\mathrm{excl}(\sigma)$ approaches its asymptotic limit $G^{(p)}_\mathrm{excl}(0)=(\omega^\mathrm{max}-\omega^\mathrm{min})^p Z_\mathrm{excl}$ with corrections that vanish faster than any power of $\sigma$.

The multi-particle contribution $\bar G^{(p)}(\sigma)$ requires a more careful analysis, that we start by considering the difference
\begin{flalign}
&
\Delta \bar G^{(p)}(\sigma)
=
\bar G^{(p)}(\sigma) -\bar G^{(p)}(0) \;,
\end{flalign}
between the observables $\bar G^{(p)}(\sigma)$ at $\sigma>0$ and the asymptotic result $\bar G^{(p)}(0)$. This can be rewritten as
\begin{flalign}
&
\bar G^{(p)}(0)
=
\int_{\omega^\mathrm{min}}^\infty d\omega_0\,
(\omega^\mathrm{max}-\omega_0)^p
\theta(\omega^\mathrm{max}-\omega_0)
\bar Z(\omega_0)\;,
\end{flalign}
so that, by using \cref{eq:Oasympt1}, the first of the properties listed in \cref{eq:Thetaprops} and by making the change of variables $x=(\omega^\mathrm{max}-\omega_0)/\sigma$, we have
\begin{flalign}
&
\Delta \bar G^{(p)}(\sigma)
=
\nonumber \\[8pt]
&
\sigma^{p+1}
\int_{-\infty}^{\frac{\omega^\mathrm{max}-\omega^\mathrm{min}}{\sigma}} dx\,
x^p\left[
\Theta_1\left(x\right)
-
\theta(x)
\right] \bar Z(\omega^\mathrm{max}-\sigma x)\;.
\end{flalign}
By relying again on \cref{eq:Thetaprops} we now  split the integral for $x<0$ and $x>0$ and extend the upper limit of integration up to corrections that vanish faster than any power of $\sigma$,
\begin{flalign}
\Delta \bar G^{(p)}(\sigma)
&=
\sigma^{p+1}
\int_0^\infty dx\,
x^p\left[
\Theta_1\left(x\right)
-
1
\right] \bar Z(\omega^\mathrm{max}-\sigma x)
\nonumber \\[8pt]
&
+
\sigma^{p+1}
\int_{-\infty}^0 dx\,
x^p\, 
\Theta_1\left(x\right)
\bar Z(\omega^\mathrm{max}-\sigma x)
\nonumber \\[8pt]
&
+O\left( e^{-\frac{\omega^\mathrm{max}-\omega^\mathrm{min}}{\sigma}}\right)
\;.
\end{flalign}
Finally, by changing variable $x\mapsto -x$ in the second integral of the previous expression, by relying on the (assumed) analyticity of $\bar Z(\omega)$ around $\omega=\omega^\mathrm{max}$ we arrive at
\begin{flalign}
&
\Delta \bar G^{(p)}(\sigma)
\nonumber \\[8pt]
&
=
\sum_{n=0}^\infty \sigma^{p+n+1}\, \left\{1 + (-1)^{p+n+1} \right\}\,
\bar Z^{(n)}(\omega^\mathrm{max}) I(p,n)
\nonumber \\[8pt]
&
+O\left( e^{-\frac{\omega^\mathrm{max}-\omega^\mathrm{min}}{\sigma}}\right)
\;,
\label{eq:ressigmaasymopt}
\end{flalign}
where
\begin{flalign}
\bar Z^{(n)}(\omega^\mathrm{max}) \equiv
\left.
\frac{d^n\bar Z(\omega)}{d\omega^n}
\right\vert_{\omega=\omega^\mathrm{max}}\;,
\end{flalign}
and we have introduced the finite numerical ``shape-integrals'' of the kernel
\begin{flalign}
I(p,n)
=
\frac{(-1)^n}{n!}
\int_0^\infty dx\,
x^{p+n}\left[
\Theta_1\left(x\right)
-
1
\right]\; .
\label{eq:numerical_shape_integrals}
\end{flalign}

\Cref{eq:ressigmaasymopt} is crucially important for the non-perturbative lattice calculation of $d\Gamma^{(p)}/d \vec \omega^2$ and $dM_l^{(p)}/d \vec \omega^2$ since it prescribes the functional forms to be used in order to extrapolate the results obtained at $\sigma>0$. Only even powers of $\sigma$ arise in the asymptotic expansions of $d\Gamma^{(p)}(\sigma)/d \vec \omega^2$ and $dM_l^{(p)}(\sigma)/d \vec \omega^2$ and, in particular, in the case of the rate we have
\begin{flalign}
\frac{d\Gamma^{(0,1)}(\sigma)}{d \vec \omega^2}
&=
\frac{d\Gamma^{(0,1)}}{d \vec \omega^2}
+O(\sigma^2)\;,
\nonumber \\[8pt]
\frac{d\Gamma^{(2)}(\sigma)}{d \vec \omega^2}
&=
\frac{d\Gamma^{(2)}}{d \vec \omega^2}
+O(\sigma^4)\;.
\label{eq:asymptG}
\end{flalign}
Similarly, in the case of the moments we have
\begin{flalign}
\frac{dM_l^{(0,1)}(\sigma)}{d \vec \omega^2}
&=
\frac{dM_l^{(0,1)}}{d \vec \omega^2}
+O(\sigma^2)\;,
\nonumber \\[8pt]
\frac{dM_l^{(2,3)}(\sigma)}{d \vec \omega^2}
&=
\frac{dM_l^{(2,3)}}{d \vec \omega^2}
+O(\sigma^4)\;,
\nonumber \\[8pt]
\frac{dM_2^{(4)}(\sigma)}{d \vec \omega^2}
&=
\frac{dM_2^{(4)}}{d \vec \omega^2}
+O(\sigma^6)\;.
\label{eq:asymptM}
\end{flalign}

The previous two sets of asymptotic relations explain our choice of organizing the calculation in terms of the kernels $\Theta^{(p)}_\sigma(\omega^\mathrm{max}-\omega_0)$ and, therefore, in terms of the distributions $Z^{(p)}(\omega_0,\vec \omega^2)$ and $Z^{(p)}_l(\omega_0,\vec \omega^2)$. Indeed, while it remains true that in taking the $\sigma\mapsto 0$ limits of our physical observables the leading corrections are $O(\sigma^2)$, the contributions $d\Gamma^{(p)}(\sigma)/d \vec \omega^2$ and $dM_l^{(p)}(\sigma)/d \vec \omega^2$ for $p>1$ can be computed more precisely by exploiting their faster rate of convergence toward the $\sigma=0$ physical point.

%%%%%%%%%%%%%%%%%%%%%%%%%%%%%%%%%%%%%%%%%%%%%%%%%%%%%%%%%%%%%%%%%%%%%%%%%%%%%%%%%%%%%%%%%%%%%%%%
\section{
\label{sec:correlators}
Lattice correlators
}
%%%%%%%%%%%%%%%%%%%%%%%%%%%%%%%%%%%%%%%%%%%%%%%%%%%%%%%%%%%%%%%%%%%%%%%%%%%%%%%%%%%%%%%%%%%%%%%%

%
\begin{table*}[t]
\centering
\begin{tabular}{lccccccc}
ensemble & $L/a$ & $a~[\rm fm]$ & $L~[\rm fm]$ &  $am_{ud}$ & $am_{s}$ & $am_{c}$ & $am_\mathrm{cr}$  \\
\hline \\[-2pt]
B48    & $48$ & ~ $0.07948(11)$ ~  & ~ 3.82 ~ &~ 0.0006669(28)  ~ & ~0.018267(53)~   &  ~0.23134(52) ~ & ~ -0.4138934(46)	~ \\
B64    & $64$ & ~ $0.07948(11)$ ~  & ~ 5.09 ~ &~ 0.0006669(28)  ~ & ~0.018267(53)~   &  ~0.23134(52) ~ & ~ -0.4138934(46)	~ \\
B96    & $96$    & ~ $0.07948(11)$ ~  & ~ 7.63 ~ & ~ 0.0006669(28)  ~ & ~0.018267(53)~   &  ~0.23134(52) 	~ & ~ -0.4138934(46)	~ \\[4pt]
% \hline
C80    & $80$    & ~ $0.06819(14)$ ~   & 5.46 &~ 0.0005864(34) ~
   &  ~0.016053(67)~  &  ~0.19849(64)	~  & ~ -0.3964534(41)  ~ \\[4pt]
D96    & $96$    & ~ $0.056850(90)$ ~  & 5.46 &~ 0.0004934(24) ~ 
   &  ~0.013559(39)~  &  ~0.16474(44)	~  & ~ -0.3761252(39) ~  \\[4pt]
% \hline
E112    & $112$    & ~ $0.04892(11)$ ~  & 5.48 &~ 0.0004306(23) ~  &  ~0.011787(55)~  &   ~0.14154(54)	~ & ~  -0.3613136(75) ~  \\[8pt]
\hline
\end{tabular}

%\begin{tabular}{lccccccccc}
 %   ensemble & $\qquad$ & $L/a$ & $a~[\rm fm]$ & $L~[\rm fm]$ & $\qquad$ & $am_{ud}$ & $am_{s}$ & $am_{c}$ & $am_\mathrm{cr}$  \\[2pt]
  %  \hline
   % \\[-2pt]
    %B48    && $48$  & $0.07948(11)$   & $\:\; 3.82 \:\;$ && $\:\; 0.00072 \:\;$ & $\:\; 0.016920000 \:\;$ &  $\:\; 0.236800 \:\;$ & $\:\; -0.4138934(46)   \:\;$ \\
    %B64    && $64$  & $0.07948(11)$   & $\:\; 5.09 \:\;$ && $\:\; 0.00072 \:\;$ & $\:\; 0.016920000 \:\;$ &  $\:\; 0.236800 \:\;$ & $\:\; -0.4138934(46)   \:\;$ \\
    %B96    && $96$  & $0.07948(11)$   & $\:\; 7.63 \:\;$ && $\:\; 0.00072 \:\;$ & $\:\; 0.016920000 \:\;$ &  $\:\; 0.236800 \:\;$ & $\:\; -0.4138934(46)   \:\;$ \\[4pt]
  %  C80    && $80$  & $0.06819(14) $   & $\:\; 5.46 \:\;$ && $\:\; 0.00060 \:\;$ & $\:\; 0.015290000 \:\;$ &  $\:\; 0.201900 \:\;$ & $\:\; -0.3964534(41)   \:\;$ \\
    %%C112   & $112$ & $\:\; 0.068 \:\;$   & $\:\; 7.64 \:\;$ & $\:\; 0.00060 \:\;$ & $\:\; 0.015290000 \:\;$ &  $\:\; 0.201900 \:\;$ & $\:\; -0.3964534(41)   \:\;$ \\[4pt]
   % D96    && $96$  & $0.056850(90)$   & $\:\; 5.46 \:\;$ && $\:\; 0.00054 \:\;$ & $\:\; 0.013260000 \:\;$ &  $\:\; 0.166600 \:\;$ & $\:\; -0.3761252(39)   \:\;$ \\[4pt]
    %E112   && $112$ & $0.04892(11) $   & $\:\; 5.48 \:\;$ && $\:\; 0.00044 \:\;$ & $\:\; 0.011825301 \:\;$ &  $\:\; 0.142554 \:\;$ & $\:\; -0.3613136(75)   \:\;$ \\[4pt]
    %\hline
%\end{tabular}
\caption{
ETMC gauge ensembles used in this work. We give the values of the lattice spacing $a$, of the spatial lattice extent $L$, of the simulated bare light ($m_u=m_d=m_{ud}$), strange ($m_s$) and charm ($m_c$) quark masses and of the critical mass $m_\mathrm{cr}$. The temporal extent of the lattice is always $T=2L$. Details concerning the determination of the lattice spacing and of the quark masses can be found in~\cite{ExtendedTwistedMass:2024nyi}.}
\label{tab:iso_EDI_FLAG}
\end{table*}

The lattice correlators needed to extract the decay rate and the lepton-energy moments have been computed on the physical-point gauge ensembles, listed in~\cref{tab:iso_EDI_FLAG}, that have been generated~\cite{Alexandrou:2018egz,ExtendedTwistedMass:2020tvp, ExtendedTwistedMass:2021qui,Finkenrath:2022eon} by the ETMC with $n_f = 2 + 1 + 1$ flavors of Wilson-Clover Twisted Mass (TM) sea quarks~\cite{Frezzotti:2000nk,Frezzotti:2003xj}. The bare parameters of the simulations have been tuned to match our scheme of choice for defining isoQCD, the so-called Edinburgh/FLAG consensus~\cite{FlavourLatticeAveragingGroupFLAG:2024oxs}, and therefore to match the inputs $m_\pi=135.0$~MeV, $m_K=494.6$~MeV, $m_{D_s}=1967$~MeV and $f_\pi=130.5$~MeV.

We adopted the mixed-action lattice setup introduced in~\cite{Frezzotti:2004wz} and described in full detail in the appendixes of Ref.~\cite{ExtendedTwistedMassCollaborationETMC:2024xdf}. In this setup the action of the valence quarks is discretized in the so-called Osterwalder--Seiler (OS) regularization,
\begin{flalign}
&
S_\mathrm{OS}
=
\nonumber \\[8pt]
&
a^4\sum_x \bar{\psi}_f\left\{
\gamma_\mu \bar{\nabla}_\mu[U]-ir_f\gamma_5 \left(
W^\mathrm{cl}[U]+m_\mathrm{cr} 
\right) +m_f
\right\}\psi_f \;,
\label{eq:OS_valence_quark_action}
\end{flalign}
where $f$ is the flavor index, the sum runs over the lattice points, $m_f$ is the bare quark mass, $m_\mathrm{cr}$ is the critical-mass counter-term and we refer to Refs.~\cite{Alexandrou:2018egz,ExtendedTwistedMass:2020tvp, ExtendedTwistedMass:2021qui,Finkenrath:2022eon} for the explicit definition of the covariant derivatives $\bar{\nabla}_\mu[U]$ and of the Wilson-Clover term $W^\mathrm{cl}[U]$, both depending upon the gauge links $U_\mu(x)$. Valence and sea quarks have been simulated with the same value of $m_\mathrm{cr}$, tuned to restore chiral symmetry, and the bare masses $m_f$ of the valence quarks have been tuned so that the corresponding renormalized masses match those of the sea quarks.
For each physical flavor $f$ we have two valence OS quark fields with opposite values of the Wilson parameters, $r_f=\pm 1$. The unitarity of the theory and the physical number of dynamical quarks is recovered in the continuum limit (see Ref.~\cite{ExtendedTwistedMassCollaborationETMC:2024xdf} for more details). 
We exploited this flexibility to optimize the numerical signal-to-noise ratios of the lattice correlators and, as explained in more details below, adopted the OS (TM) regularization of the weak currents to compute the quark connected (disconnected) Wick contractions.

To interpolate the $D_s$ meson we use the following pseudoscalar operator
\begin{flalign}
P(t,\vec{x}) = \sum_{\vec{y}}\bar \psi_c(t,\vec{x}) G^{N_\mathrm{sm}}_{t}(\vec{x},\vec{y}) \gamma_5 \psi_s(t,\vec{y}) \;, 
\end{flalign}
with $r_c=-r_s$. In the previous expression $G_{t}
(\vec{x},\vec{y})$ is the Gaussian smearing operator
\begin{flalign}
G_{t}(\vec{x},\vec{y}) = \frac{1}{1+ 6\kappa}\left( \delta_{\vec{x},\vec{y}} + \kappa H_{t}(\vec{x},\vec{y})    \right)~,
\end{flalign}
with
\begin{flalign}
H_{t}(\vec{x}, \vec{y}) = \sum_{\mu=1}^{3}\left( \mathcal{U}_{\mu}(t,\vec{x})\delta_{\vec{x}+\hat{\mu},\vec{y}} + \mathcal{U}^{\dagger}_{\mu}(t,\vec{x}-\hat{\mu})\delta_{\vec{x}-\hat{\mu},\vec{y}}    \right)~,
\end{flalign}
and we have indicated with $\mathcal{U}_{\mu}(x)$ the APE-smeared links,  defined as in Ref.~\cite{APE:1987ehd}. For this calculation, we employed the values $\kappa=0.5$ and fixed the number of smearing steps $N_\mathrm{sm}$ to obtain a smearing radius $a\sqrt{\frac{{N_\mathrm{sm}\kappa}}{{1+6\kappa}}}=0.18$~fm.

The two-point correlator
\begin{flalign}\label{eq:twopoint}
C(t)=\sum_{\vec x} T\bra{0} P(t,\vec x)\, P^\dagger(0) \ket{0}
\end{flalign}
is used to amputate the four-points functions from which we extract the correlators $\mathcal{\hat Y}^{(i)}(t,\vec \omega^2)$. To this end, from the asymptotic behavior for $0\ll t \ll T$ of $C(t)$,
\begin{flalign}
C(t)=\frac{R_P}{2 m_{D_s}}\, e^{-m_{D_s}t} + \cdots\;,
\label{eq:Csingleexp}
\end{flalign}
where the dots represent exponentially suppressed contributions, we extract the mass of the $D_s$ meson at finite lattice spacing and the residue $R_P$. 

The four-point correlators from which we extract the amputated correlators $\mathcal{\hat Y}^{(i)}(t,\vec \omega^2)$ are given by
\begin{flalign}
&
C_{\mu\nu}(t_\mathrm{snk},t,t_\mathrm{src},\vec \omega^2)
=
a^9\sum_{\vec x_\mathrm{snk},\vec x_\mathrm{src},\vec x} 
e^{im_{D_s} \vec \omega\cdot \vec x}\ \times
\nonumber \\[8pt]
&
\phantom{C_{\mu\nu}(t_\mathrm{snk},t,t_\mathrm{src})=
}
T\bra{0}
P(x_\mathrm{snk}) J_\mu^\dagger(x) J_\nu(0) P^\dagger(x_\mathrm{src})\ket{0}\;,
\end{flalign}
where $x=(t,\vec x)$, $x_\mathrm{snk}=(t_\mathrm{snk},\vec x_\mathrm{snk})$ and  $x_\mathrm{src}=(t_\mathrm{src},\vec x_\mathrm{src})$, and $J_\mu(x)$ is the  lattice discretized version of the weak current (see below). 

\begin{table*}[htbp]
    \centering
    \begin{tabular}{lccccccccc}
    ensemble  &$\qquad$& $N_\text{cnfg}$ & $N_\text{hit}$ & $\qquad$ &$(t_\text{snk}-t_\text{src})/a$   & $t_\text{src}/a$ &$\qquad$& $\theta_{\bar s d}^\text{max}$   & $\qquad\theta_{\bar u c}^\text{max}\qquad$\\[2pt]
        \hline
        \\[-2pt]
    B48           && $400$ & $5$ && $\quad 48 \quad$ & $\quad -12 \quad$ && $5.754410 $ & -          \\
    B64           && $300$ & $3$ && $\quad 56 \quad$ & $\quad -12 \quad$ && $7.672547$ & $0.684726$ \\
    B96           && $300$ & $2$ && $\quad 56 \quad$ & $\quad -12 \quad$ && $11.50882$ & -          \\
    [4pt]
    C80           && $600$ & $2$ && $\quad 65 \quad$ & $\quad -14 \quad$ && $8.221447$ & -          \\
    [4pt]
    D96           && $300$ & $2$ && $\quad 78 \quad$ & $\quad -16 \quad$ && $8.232777$ & $0.734723$ \\
    [4pt]
    E112          && $300$ & $2$ && $\quad 91 \quad$ & $\quad -19 \quad$ && $8.253267$ & -          \\
    [4pt]
    \hline
\end{tabular}
\caption{
Simulation parameters used in the calculation of the correlators $C(t)$ and $C_{\mu\nu}(t_\mathrm{snk},t,t_\mathrm{src},\vec \omega^2)$. $N_\text{cnfg}$ is the number of gauge configurations per ensemble and $N_\text{hit}$ is the number of stochastic sources used per configuration. $t_\text{snk}$ and $t_\text{src}$ are used as described in details in~\cref{fig:Ys}. The parameters $\theta_{\bar F G}^\text{max}$ are used for the insertion of the spatial momenta by using flavour-twisted boundary conditions (see explanation in the text).
\label{tab:supplementaryinfoconfigs}}
\end{table*}

The correlators $C(t)$ and $C_{\mu\nu}(t_\mathrm{snk},t,t_\mathrm{src},\vec \omega^2)$ have been computed on all the different gauge ensembles by using the values of the parameters listed in~\cref{tab:supplementaryinfoconfigs}. The parameters $\theta^\mathrm{max}_{\bar FG}$ have been used to compute the correlators $C^{\mu\nu}_{\bar f g}(t_\mathrm{snk},t,t_\mathrm{src},\vec \omega^2)$ with spatial momenta $m_{D_s}\vec \omega=(0,0,2\pi\theta/L)$ along the third spatial direction by using flavour-twisted boundary conditions~\cite{deDivitiis:2004kq} for the quark field $\psi_g$, i.e.\ $\psi_g(x+\hat 3 L) = \exp(2\pi \theta i)\psi_g(x)$. The values of $\theta^\mathrm{max}_{\bar sd}$ correspond, on the different ensembles, to the value of $\vert \vec \omega \vert^\mathrm{max}_{\bar s d}$ appearing in~\cref{tab:momenta}.

\begin{figure}[!t]
\includegraphics[width=\columnwidth]{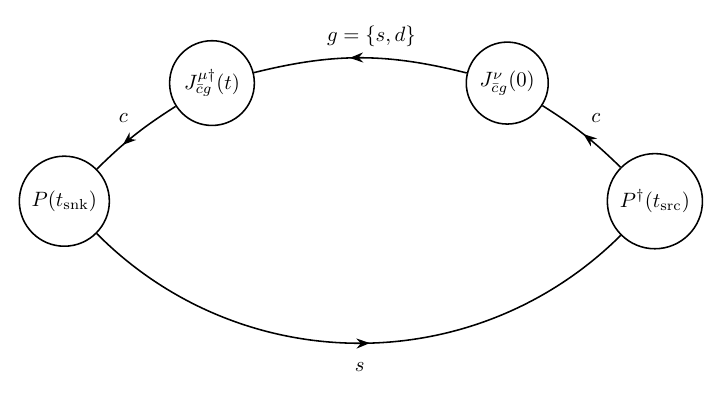}
\caption{Quark-connected Wick contraction contributing to $C^{\mu\nu}_{\bar c g}$ in the $\bar c s$ and $\bar c d$ channels. }
\label{fig:contdown}
\end{figure}
\begin{figure}[!t]
\includegraphics[width=\columnwidth]{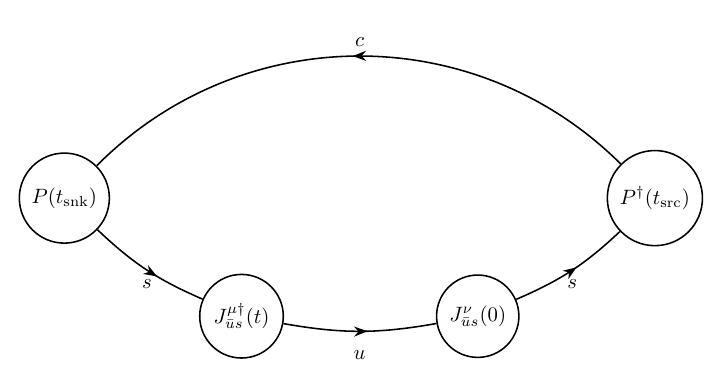}
\caption{Quark-connected Wick contraction contributing to $C^{\mu\nu}_{\bar u s}$. 
}
\label{fig:contup}
\end{figure}
By integrating out the quark fields, the correlator $C^{\mu\nu}_{\bar f g}$ gets decomposed into the gauge-invariant contributions corresponding to the different fermionic Wick contractions. The contractions corresponding to the quark-connected diagram shown in \cref{fig:contdown} contribute to both the dominant $\bar cs$ channel and to the Cabibbo suppressed $\bar cd$ channel.
The quark-connected contraction shown in \cref{fig:contup} contributes to the Cabibbo suppressed $\bar u s$ channel.
\begin{figure}[!t]
\includegraphics[width=\columnwidth]{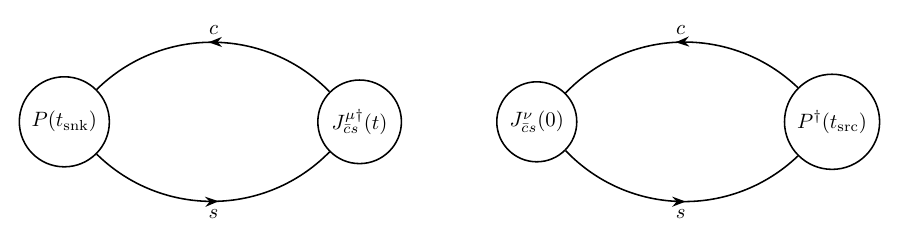}
\caption{The weak-annihilation contribution to $C^{\mu\nu}_{\bar c s}$.}
\label{fig:contwa}
\end{figure}
\begin{figure}[!t]
\includegraphics[width=\columnwidth]{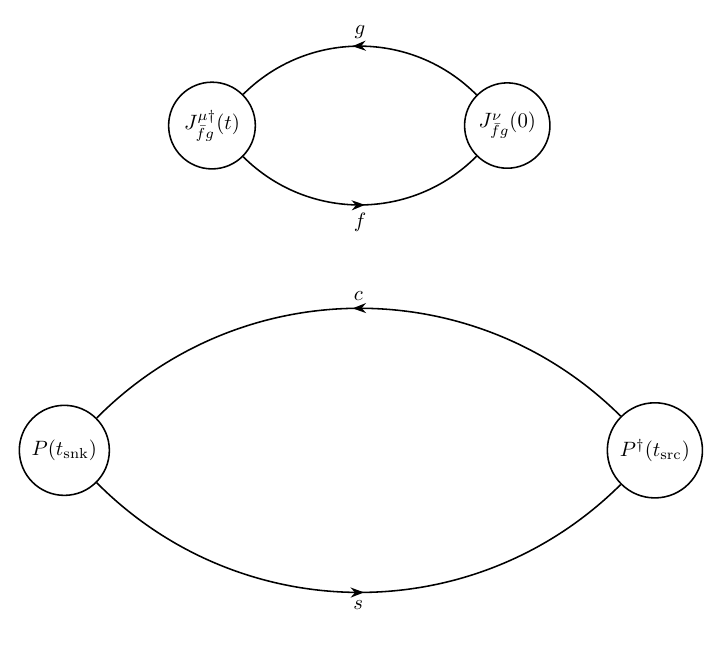}
\caption{The quark-disconnected current-current contribution to $C^{\mu\nu}_{\bar f g}$. }
\label{fig:contextra}
\end{figure}
The quark-disconnected contraction shown in \cref{fig:contwa} contributes only to the dominant $\bar c s$ channel. In the following, as commonly done in the phenomenological literature on the subject, we shall call this contribution the ``weak-annihilation'' diagram.   

\begin{table*}
    \begin{tabular}{lcc}
    ensemble $\qquad\qquad$ & $Z_V$ & $Z_A$   \\[2pt]
    \hline
    \\[-2pt]
    B48         & $\qquad  0.706354(54) \qquad$ & $\qquad  0.74296(19) \qquad$\\
    B64         & $\qquad  0.706354(54) \qquad$ & $\qquad  0.74296(19) \qquad$\\
    B96         & $\qquad  0.706406(52) \qquad$ & $\qquad  0.74261(19) \qquad$\\
    [4pt]
    C80         & $\qquad  0.725440(33) \qquad$ & $\qquad  0.75814(13) \qquad$\\
    [4pt]
    D96         & $\qquad  0.744132(31) \qquad$ & $\qquad  0.77367(10) \qquad$\\
    [4pt]
    E112        & $\qquad  0.758238(18) \qquad$ & $\qquad  0.78548(9)  \qquad$\\
    [4pt]
    \hline
    
\end{tabular}
\caption{
The values of the renormalization constants $Z_A$ and $Z_V$ used in this work. The three B ensembles share the same bare parameters and differ only in volume. This justifies the use of the same renormalization constants. Nevertheless, to verify the robustness of these determinations and the theoretical expectation that finite volume effects  on these short-distance quantities are negligible, we have recalculated them on the B96 ensemble.
\label{tab:configsrenormailzationconstants}}
\end{table*}
In our mixed-action setup the quark-connected contractions of \cref{fig:contdown,fig:contup} have been computed by employing the so-called OS discretization of the weak current, i.e.\
\begin{flalign}
J^{\mathrm{OS},\mu}_{\bar fg}(x)=\bar \psi_f(x) \gamma^\mu(Z_V-Z_A\gamma_5) \psi_g(x)\,,
\qquad r_f=r_g\,,
\end{flalign}
while the weak-annihilation diagram of \cref{fig:contwa} has been computed by employing the so-called TM discretization of the current, i.e.\ 
\begin{flalign}
J^{\mathrm{TM},\mu}_{\bar fg}(x)=\bar \psi_f(x) \gamma^\mu(Z_A-Z_V\gamma_5) \psi_g(x)\,,
\qquad r_f=-r_g\,.
\end{flalign}
The values of $Z_V$ and $Z_A$ used in this calculation are given in \cref{tab:configsrenormailzationconstants}. 
These have been 
determined in~\cite{ExtendedTwistedMass:2024nyi} following the Ward-identity method explained in Appendix B of Ref.~\cite{ExtendedTwistedMass:2022jpw}.

The analysis procedure that we used to estimate the statistical errors on our lattice data, to take into account systematic uncertainties and the correlations between the different results is explained in~\cref{sec:errors}. Concerning the renormalization constants, their uncertainties are at the $0.03$\% level and, therefore, totally negligible w.r.t.\ the statistical errors of our results for the differential decay rate and lepton-energy moments which (see \cref{sec:conclusions}) are at the few percent level. Nevertheless, we have taken into account these errors as explained in details in~\cref{sec:errors}. Concerning the uncertainties on the lattice spacings and on the bare quark masses, these are at the few permille level (see~\cref{tab:iso_EDI_FLAG}), i.e.\ of the same order of magnitude of the isospin breaking corrections that we are neglecting in this calculation. Albeit, for this reason, these uncertainties can consistently be neglected, we wanted to have a quantitative evidence of the fact that the  systematic errors induced by the uncertainties on the bare parameters are totally negligible w.r.t.\ the statistical errors of our physics results. To this end, we performed a dedicated study on the B64 ensemble by using the techniques described in the appendices of Ref.~\cite{ExtendedTwistedMass:2024nyi}
to vary the values of the bare parameters within their uncertainties. The study confirmed the expectations and, therefore, we don't keep track of the negligible fine-tuning and renormalization constants systematic uncertainties in our final error budgets.

\begin{figure}[t!]
\includegraphics[width=\columnwidth]{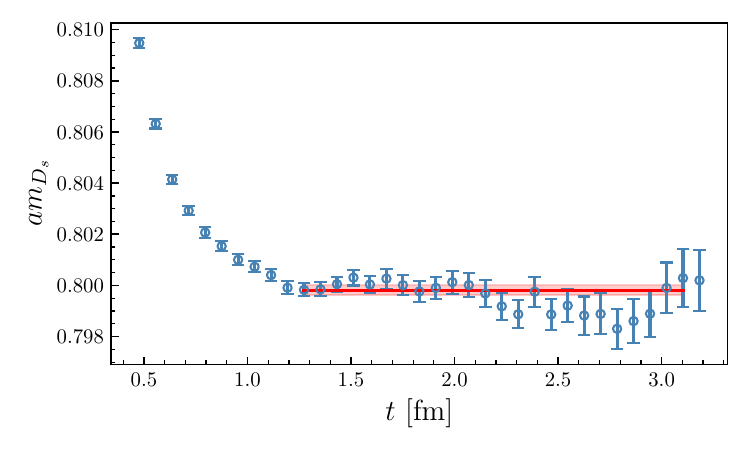}
\includegraphics[width=\columnwidth]{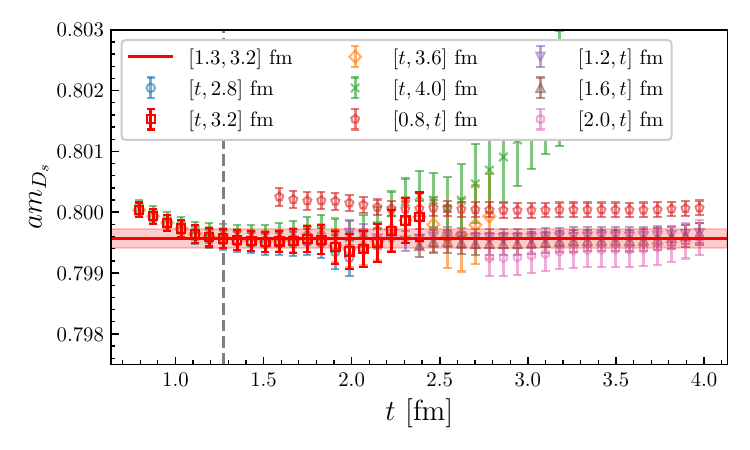}
\caption{\emph{Top-panel}: the blue points show the effective mass of the correlator $C(t)$ (see \cref{eq:twopoint}) on the ensemble B64. The red band shows our estimate of $am_{D_s}$ on this ensemble. \emph{Bottom-panel}: the plot shows the windowing procedure that we used to choose the range of the (correlated) plateau-fit shown in the top-panel. The red band is that of the fit already shown in the top-panel. The different datasets show the results of the plateau-fits performed by fixing the lower (upper) extreme of the fit interval and by varying the upper (lower) extreme.}
\label{fig:meffDs}
\end{figure}

The quark-disconnected contraction shown in \cref{fig:contextra} deserves some comments. In principle, this contraction contributes to the correlator $C^{\mu\nu}_{\bar f g}$ in all channels and, therefore, it should be computed. On the other hand, by interpreting  this diagram in the partially-quenched setup in which the quark fields of the current have the same mass of the physical quarks but different flavor, one has that the states propagating between the two currents have flavor $\bar c s \bar f g$. Given our previous knowledge of the QCD spectrum, a prerequisite to any decay rate or scattering amplitude calculation, this implies that these are states with energy $m_{D_s} \omega_0 > m_{D_s}$. Therefore, although the current-current contraction gives a contribution to the correlator $C^{\mu\nu}_{\bar f g}$ it doesn't contribute to the hadronic tensor $H^{\mu\nu}(p,\omega)$ for $\omega_0\le 1$. By relying on this argument we neglected the current-current contraction in our calculation of the decay rate\footnote{Strictly speaking, since the presence of ghosts prevents a straightforward interpretation of partially-quenched theories within the canonical formalism, this argument is not entirely rigorous. On the other hand, the argument is strongly supported by a very large amount of numerical evidence (e.g.\ any quenched calculation of hadronic quantities or any $n_f<4$ ($n_f<5$) calculation of $D_{(s)}$ ($B_{(s)}$) mesons observables has been performed by relying on the canonical interpretation of (partially) quenched correlators) and therefore we consider it fully satisfactory in practice.}.

The asymptotic behavior of the four-points correlator $C^{\mu\nu}(t_\mathrm{snk},t,t_\mathrm{src},\vec \omega^2)$ in the limits $T/2\gg t_\mathrm{snk} \gg t >0 \gg t_\mathrm{src} \gg -T/2$ is given by
\begin{flalign}
&
C^{\mu\nu}(t_\mathrm{snk},t,t_\mathrm{src},\vec \omega^2)
=
\frac{R_P}{4\pi m_{D_s}} e^{-m_{D_s}(t_\mathrm{snk}-t-t_\mathrm{src})}\, \times
\nonumber \\[8pt]
&
\phantom{C_{\mu\nu}(t_\mathrm{snk},t,t}
\int_{\omega^\mathrm{min}-\epsilon}^\infty d\omega_0\, e^{-\omega_0 (m_{D_s}t)}\,
H^{\mu\nu}(p,\omega) +\cdots\;,
\label{eq:Cmunuasympt}
\end{flalign}
where $H^{\mu\nu}(p,\omega)$ is the hadronic tensor and the dots represent again exponentially suppressed terms. From the previous relation, by using the values of $R_P$ and $m_{D_s}$ extracted from $C(t)$ (see \cref{eq:Csingleexp}) and by projecting the different components of $C_{\mu\nu}(t_\mathrm{snk},t,t_\mathrm{src})$ as done in \cref{eq:Ydefs} to define the five independent spectral densities $\mathcal{Y}^{(i)}(\omega_0,\vec \omega^2)$, we have extracted the correlators $\mathcal{\hat Y}^{(i)}(t,\vec \omega^2)$ (see \cref{eq:Ys}), e.g.
\begin{flalign}\label{eq:hat_Y}
&
\mathcal{\hat Y}^{(2)}(t,\vec \omega^2)
=
\nonumber \\[8pt]
&
\lim_{t_\mathrm{snk}\mapsto \infty}\lim_{t_\mathrm{src}\mapsto -\infty}\lim_{T\mapsto \infty}
\frac{4\pi\, m_{D_s}^2\, C^{00}(t_\mathrm{snk},t,t_\mathrm{src},\vec \omega^2)}
{R_P\, e^{-m_{D_s}(t_\mathrm{snk}-t-t_\mathrm{src})}}\;.
\end{flalign}
Then, by performing the linear combinations of the $\mathcal{\hat Y}^{(i)}(t,\vec \omega^2)$ correlators corresponding to \cref{eq:Zgamma,eq:ZM1,eq:ZM2}, we obtain the correlators $\hat Z^{(p)}(t,\vec \omega^2)$ and $\hat Z^{(p)}_l(t,\vec \omega^2)$.

In \cref{fig:meffDs} we show the extraction of the mass $m_{D_s}$ on the ensemble B64 from the correlator $C(t)$. The blue points correspond to the so-called effective mass of the correlator while the red band corresponds to the constant fit of the effective mass in the plateau-region and, therefore, to our estimate of $am_{D_s}$. Similar plots can be shown for all of the ensembles listed in \cref{tab:iso_EDI_FLAG}.

\begin{figure}[t!]
\includegraphics[width=\columnwidth]{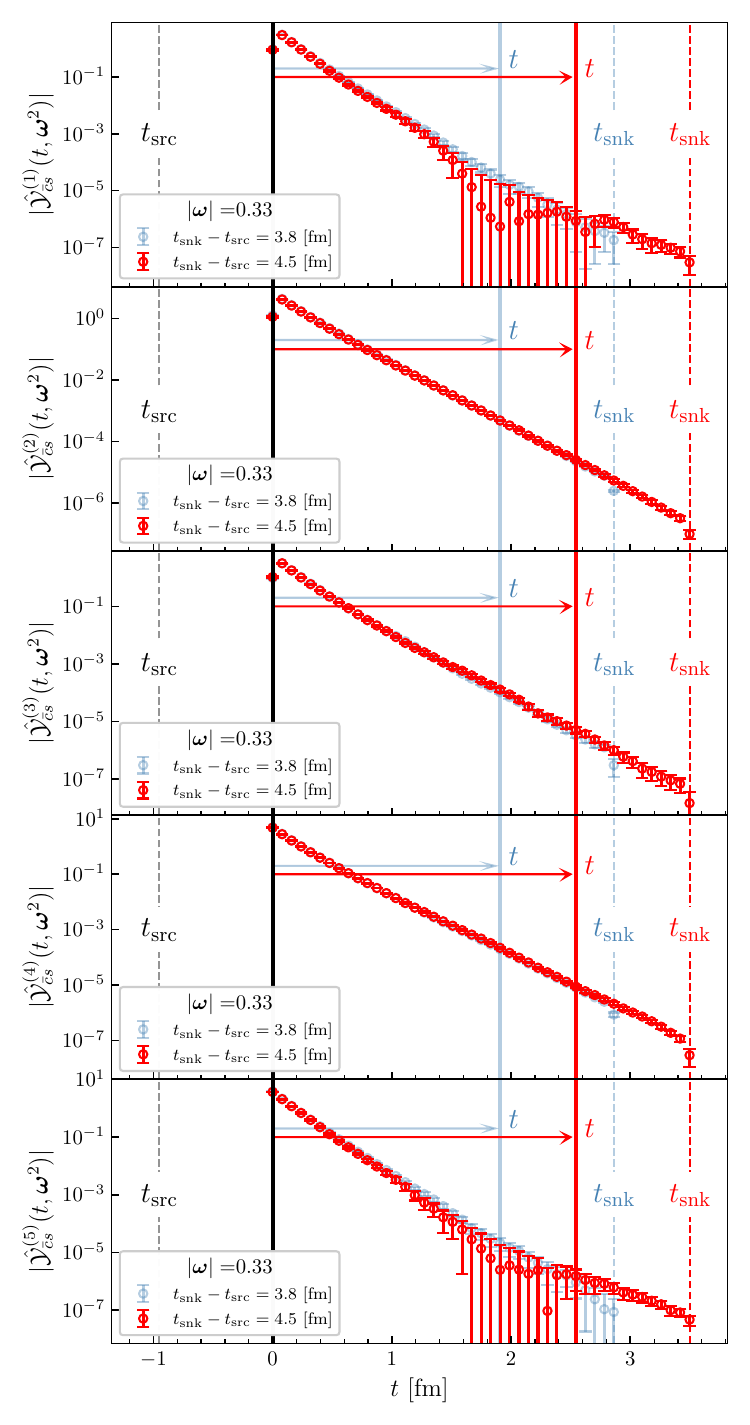}
\caption{From top to bottom we show the correlators defined in \cref{eq:Ys} for $i=1,\dots,5$. The data have been obtained from the B64 ensemble and correspond to the dominant contribution $\bar cs$ at spatial momentum $(m_{D_s}\vec \omega)^2=0.43$~GeV$^2$, or equivalently $|\vec \omega|=0.33$. 
The red points correspond to the separation $t_\mathrm{snk}-t_\mathrm{src}=56a\simeq 4.5$~fm while the light-blue points to $t_\mathrm{snk}-t_\mathrm{src}=48a\simeq 3.9$~fm. The solid vertical lines mark the points corresponding to the condition $t_\mathrm{snk}-t=0-t_\mathrm{src}=12a\simeq 1$~fm, i.e.\ to the values of $t$ ($t=32a$ red dataset and $t=24a$ blue dataset) such that the two separations between each interpolating operator and the closer current are equal. The vertical dashed and solid black lines correspond, respectively, to $t_\mathrm{src}$ and to the position of the current that we kept fixed.}
\label{fig:Ys}
\end{figure}
From the analysis of $C(t)$ on the different ensembles we extracted the information needed to compute the correlators $C^{\mu\nu}(t_\mathrm{snk},t,t_\mathrm{src},\vec \omega^2)$ in the interesting region of the parameter space, i.e.\ for values of $t_\mathrm{src}$ and $t_\mathrm{snk}$ such that the systematic errors associated with the asymptotic limits $T\mapsto \infty$, $t_\mathrm{src}\mapsto -\infty$ and $t_\mathrm{snk}\mapsto \infty$ can be kept under control. An example of this analysis is shown in \cref{fig:Ys}. The figure shows the five amputated correlators $\mathcal{\hat Y}^{(i)}_{\bar c s}(t,\vec \omega^2)$ extracted on the B64 ensemble from the quark-connected contraction of the correlator $C_{\bar c s}^{\mu\nu}(t_\mathrm{snk},t,t_\mathrm{src},\vec \omega^2)$ (see \cref{fig:contdown}) for two different values of the separation $t_\mathrm{snk}-t_\mathrm{src}$ between the interpolating operators and for $(m_{D_s}\vec{\omega})^2=0.43$~GeV$^2$. In both cases we set $t_\mathrm{src}=-12a\simeq -1$~fm while we set $t_\mathrm{snk}=36a\simeq 2.9$~fm in the case of the light-blue points and $t_\mathrm{snk}=44a\simeq 3.5$~fm in the case of the red points. The solid vertical lines mark the points corresponding to the condition $t_\mathrm{snk}-t=0-t_\mathrm{src}=12a$, i.e.\ the values of $t$ ($t=32a$ red dataset and $t=24a$ light-blue dataset) such that the two separations between each interpolating operator and the closer current are equal. As can be seen, the light-blue and red datasets are fully compatible within the statistical errors up to values of $t$ such that $t_\mathrm{snk}-t=a$. 
The separation $0-t_\mathrm{src}$ between the interpolating operator of the initial state and the first weak current has been fixed at $\simeq 1$~fm, a distance of the same order of the time separation where the plateau of the effective mass of the correlator $C(t)$ sets in (see \cref{fig:meffDs}). Then, by relying on the symmetries of our four-points correlator, we studied the dependence of our results upon $t_\mathrm{src}$ and $t_\mathrm{snk}$ by varying the distance $t_\mathrm{snk}-t$ between $P(x_\mathrm{snk})$ and the weak current inserted at time $t$. From this analysis, that we repeated for all considered values of the momenta $\vec \omega$ (see following sections) and also for the other flavor channels, we concluded that the systematic errors associated with the $t_\mathrm{src}\mapsto -\infty$ and $t_\mathrm{snk}\mapsto \infty$ limits are negligible with respect to the statistical errors of our correlators. Our estimates of the systematic errors associated with finite size effects, i.e.\ with the $T\mapsto \infty$ and $L\mapsto \infty$ limits, will be discussed in details in the following sections.

In order to extract the decay rate and the lepton-energy moments we used the data corresponding to the larger separation, i.e.\ to $t_\mathrm{snk}-t_\mathrm{src}\simeq 4.5$~fm, that we kept fixed in physical units on the different gauge ensembles. With this choice the systematics associated with the asymptotic limits can safely be neglected w.r.t.\ the statistical errors and, moreover, we can use larger values of $N$ to reconstruct the smearing kernels according to \cref{eq:stoneZ} and, hence, to study the systematics associated with the $N\mapsto \infty$ limits (see \cref{sec:hlt}).

%%%%%%%%%%%%%%%%%%%%%%%%%%%%%%%%%%%%%%%%%%%%%%%%%%%%%%%%%%%%%%%%%%%%%%%%%%%%%%%%%%%%%%%%%%%%%%%%
\section{
\label{sec:hlt}
The HLT algorithm and the \texorpdfstring{$N\mapsto \infty$}{N->inf} limit 
}
In this section we provide the details of the numerical implementation of the HLT algorithm~\cite{Hansen:2019idp} that we have used to extract the different contributions to the decay rate and to the lepton-energy moments according to \cref{eq:lim_N_gamma} and \cref{eq:lim_N_moments}. 
Here we focus the discussion on the decay rate. The case of the lepton-energy moments is totally analogous. 

We have considered two definitions of the smearing kernel $\Theta^{(p)}_\sigma(x)$ of \cref{eq:defthetap}. These have been obtained by starting from the following two regularizations of the Heaviside step-function,
\begin{flalign}
\Theta_\sigma(x)=\frac{1}{1+e^{-\frac{x}{\sigma}}} \;,
\label{eq:sigmoid}
\end{flalign}
and
\begin{flalign}
\Theta_\sigma(x)=\frac{1+\mathrm{erf}\left(\frac{x}{s\sigma}\right)}{2}\; ,
\label{eq:erf}
\end{flalign}
where the error-function is defined as
\begin{flalign}
\mathrm{erf}(x)=\frac{2}{\sqrt{\pi}}\int_0^x dt\, e^{-t^2}. \,
\end{flalign}
In the following we call ``sigmoid kernel'' and ``error-function kernel'' the smooth functions $\Theta^{(p)}_\sigma(x)$ obtained multiplying by $x^{p}$ respectively \cref{eq:sigmoid,eq:erf}. 
The two regularizations differ at $\sigma>0$ and become equivalent in the $\sigma\mapsto 0$ limit (see \cref{eq:heaviside}). Moreover, the properties of \cref{eq:Thetaprops} are satisfied in both cases and, by combining the numerical results corresponding to the two regularizations, we have a better control on the necessary $\sigma\mapsto 0$ extrapolations. To this end, we used the parameter $s>0$ appearing in \cref{eq:erf} to rescale the width of the error-function kernel w.r.t.\ that of the sigmoid kernel. Indeed, the shape of the smearing kernels is governed by the integrals of \cref{eq:numerical_shape_integrals} and we set $s=2.5$ in order to have
\begin{flalign}
I^\mathrm{sigmoid}(0,1) \simeq I^\mathrm{error-function}(0,1)\;,
\end{flalign}
see \cref{sec:cs_DGammaDq2} for more details.

The coefficients $g_n^{(p)}(N)$ appearing in \cref{eq:stoneZ}, that represent the smearing kernels on the basis functions $\exp(-\omega_0(am_{D_s }) n)$, are obtained by minimizing the linear combination
\begin{flalign}
W_\alpha^{(p)}[\bm g]=\frac{A_\alpha^{(p)}[\bm g]}{A_\alpha^{(p)}[\bm 0]}+\lambda B^{(p)}[\bm g]
\label{eq:W_functional}
\end{flalign}
of the so-called norm functional
\begin{flalign}
&
A_\alpha^{(p)}[\vec g] = \int_{\omega^\mathrm{th}}^{\infty} d\omega_0\, e^{\alpha (am_{D_s})\omega_0} \ \times
\nonumber \\[8pt]
&
\phantom{A_\alpha^{(p)}[\vec g]}
\left[ \Theta_\sigma^{(p)}(\omega^\mathrm{max}-\omega_0)
-\sum_{n=1}^{N}g_n e^{-\omega_0(am_{D_s})n}\right]^2
\label{eq:norm_functional}
\end{flalign}
and of the error functional
\begin{flalign}\label{eq:error_functional}
B^{(p)}[\vec g]= \sum_{n_1,n_2=1}^{N} g_{n_1} g_{n_2}\mathrm{Cov}^{(p)}(an_1,an_2),
\end{flalign}
where the matrix $\mathrm{Cov}^{(p)}$ is the statistical covariance of the correlator $\hat Z^{(p)}(an,\bm{w}^2;a)$ at finite lattice spacing. More precisely the coefficients $g^{(p)}_n(N)\equiv g^{(p)}_n(N;\vec \Sigma)$ are obtained by solving the linear system of equations 
\begin{flalign}
\left.\frac{\partial W_\alpha^{(p)}[\bm g]}{\partial \bm g} \right\vert_{\bm g =\bm g^{(p)}(N;\vec \Sigma)} =  0,
\label{eq:minimization}
\end{flalign}
and, therefore, at fixed $N$ and in presence of statistical errors, depend upon the HLT algorithmic parameters
\begin{flalign}
\vec \Sigma =\{\omega^\mathrm{th},\alpha,\lambda\}\;.
\end{flalign}

The parameter $\omega^\mathrm{th}$ appears in the definition of the norm functional, \cref{eq:norm_functional}, as the lower limit of the $\omega_0$ integral. In order to choose a value for $\omega^\mathrm{th}$ it is important to observe (see \cref{eq:Zcorr}) that the spectral density $Z^{(p)}(\omega_0,\vec \omega^2)$ vanishes for $\omega_0<\omega^\mathrm{min}$ and that, therefore, an error in the approximation of the smearing kernel $\Theta_\sigma^{(p)}(\omega^\mathrm{max}-\omega_0)$ for $\omega_0 < \omega^\mathrm{min}$ does not affect the physical result. By relying on this observation, for each flavor channel and for each contribution, we set $\omega^\mathrm{th}=0.9\ \omega_{\bar F G}^\mathrm{min}$.

We have considered a family of norm functionals, depending upon the parameter $\alpha$, by introducing in \cref{eq:norm_functional} the weight factor $\exp(\alpha am_{D_s} \omega_0)$. By considering sufficiently small values of $\sigma$, from the behavior of the kernels $\Theta_\sigma^{(p)}(\omega^\mathrm{max}-\omega_0)$ in the $\omega_0\mapsto \infty$ limit it follows that the integral of \cref{eq:norm_functional} is convergent for $\alpha<2$. For $0<\alpha <2$, the presence of the weight in the integrand forces the error in the approximation of the smearing kernel,
\begin{flalign}\label{{eq:kernel_approximation}}
\Theta_\sigma^{(p)}(\omega^\mathrm{max}-\omega_0;N,\vec \Sigma)=\sum_{n=1}^{N}g_{n}^{(p)}(N;\vec \Sigma)e^{-\omega_0(am_{D_s})n}\; ,
\end{flalign}
to decrease exponentially in the $\omega_0\mapsto \infty$ limit (see \cref{fig:kernel}). 

This feature is particularly important in order to keep under control the cutoff effects on our physical observables. Indeed, the decay rate and the lepton-energy moments are on-shell quantities that probe the QCD spectrum for energies smaller than $m_{D_s}$. Therefore, in principle, to keep under control cutoff effects, given our $O(a)$-improved lattice setup, it would be enough to have $(am_{D_s})^2\ll 1$ on the finer simulated lattices and, in fact, this condition is satisfied in our case (see \cref{tab:iso_EDI_FLAG}). On the other hand, given our representations of the decay rate and of the lepton-energy moments (see \cref{eq:dGZint,eq:dM1Zint}), it is important to avoid large errors in the approximation of the smearing kernels $\Theta_\sigma^{(p)}(\omega^\mathrm{max}-\omega_0)$ for $\omega_0\gg 1/(a m_{D_s})$ that could enhance the cutoff effects by interfering with the distortions of the lattice spectral densities $Z^{(p)}(\omega_0,\vec \omega^2; a)$ at energies of the order of the lattice cutoff. Actually, in our approach (see Ref.~\cite{Patella:2024cto} for a different possibility) the limits
\begin{flalign}
&
\frac{d \Gamma^{(p)}(\sigma)}{d \vec \omega^2}
=
\lim_{a\mapsto 0} \lim_{N\mapsto \infty}
 \lim_{\lambda\mapsto 0}
\frac{d \Gamma^{(p)}(\sigma;a,N,\vec \Sigma)}{d \vec \omega^2}\;,
\label{eq:aNlimits}
\end{flalign}
where
\begin{flalign}
&
\frac{d \Gamma^{(p)}(\sigma;a,N,\vec \Sigma)}{d \vec \omega^2}
=
\nonumber \\[8pt]
&
\qquad
\bar \Gamma\, \vert \vec \omega \vert^{3-p}\,
\sum_{n=1}^{N} g^{(p)}_n(N;\vec \Sigma)\,
\hat Z^{(p)}(a n,\vec \omega^2;a)\; ,
\label{eq:dgammaa}
\end{flalign}
have to be taken by first performing the $\lambda \mapsto 0$ and $N\mapsto \infty$ limits, that can safely be interchanged and that we perform jointly with the so-called \emph{stability analysis} procedure (see below), and \emph{then} by performing the continuum extrapolation. Notice that the dependence upon the parameter $\alpha$ disappears after performing the $N\mapsto \infty$ limit because, according to the Stone-Weierstrass theorem, the systematic error associated with the imperfect reconstruction of the smearing kernel at finite $N$ can be made arbitrarily small by increasing $N$ for any definition of the $L_2$-norm and therefore, in our language, for any definition of the functional $A_\alpha^{(p)}[\vec g]$. Unfortunately, this is not the case for the statistical error
\begin{flalign}
\Delta_\mathrm{stat}\left[ 
\frac{1}{\bar \Gamma}
\frac{d \Gamma^{(p)}(\sigma;a,N,\vec \Sigma)}{d \vec \omega^2}
\right]
=
\sqrt{
B^{(p)}[\vec g^{(p)}(N;\vec \Sigma)]} \;.
\end{flalign}
\begin{figure}[t!]
\centering
\includegraphics[width=\columnwidth]{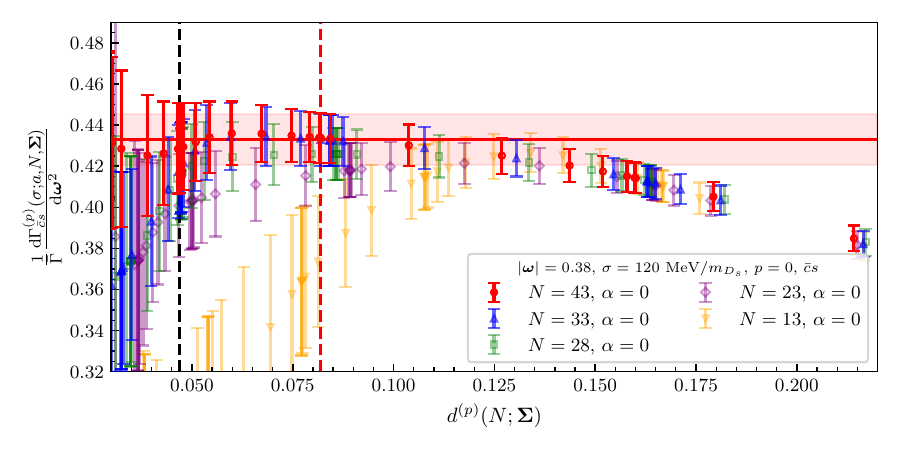}\\
\includegraphics[width=\columnwidth]{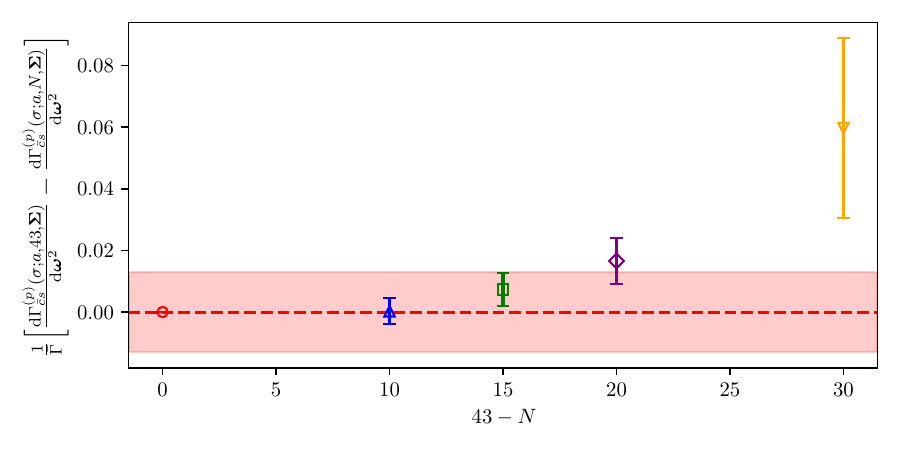}
\includegraphics[width=\columnwidth]{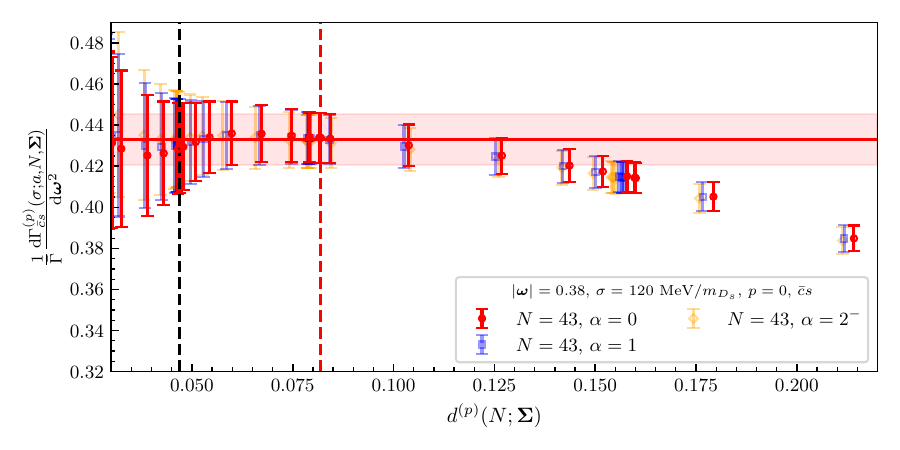}\\
\caption{ 
Stability analysis for the contribution $p=0$ to the total decay rate for the $\bar c s$  channel with smearing parameter 
$\sigma m_{D_s}=120$~MeV, spatial momentum $|\vec \omega|=0.38$, sigmoid kernel and D96 ensemble.  See the main text for the complete description and interpretation of the figure. \emph{Top-panel}: study of the limit $N\mapsto \infty$ by changing $\lambda$ at fixed $\alpha=0$. \emph{Middle-panel}: study of the dependence on $\alpha$, i.e.\ on the definition of the norm functional of \cref{eq:norm_functional}, by changing $\lambda$ at fixed $N=43$. \emph{Bottom-panel}: the plot shows the difference of the $N=43$ and $N\le 43$ results obtained at the value of $d^{(p)}(N;\vec \Sigma)$ corresponding to the red vertical dashed line in the top-panel. The errors on the points take into account the correlations between the different datasets.}
\label{fig:stability_analysis}
\end{figure}
\begin{figure}[t!]
\includegraphics[width=\columnwidth]{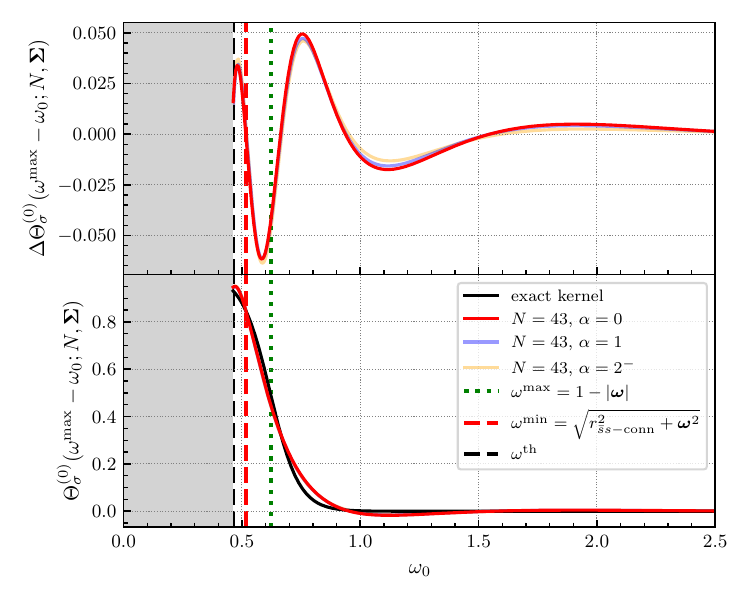}
\caption{
The plots of this figure have been obtained by using the same data of \cref{fig:stability_analysis}. The \emph{top-panel} shows the difference between the approximated and the exact kernels for $N=43$ and for the three norms $\alpha=0$, $\alpha=1$ and $\alpha=2^{-}$. The coefficients that define the approximations in the plot are associated with the points in correspondence of the red vertical dashed line in \cref{fig:stability_analysis} and, therefore, to different approximations at fixed $d^{(p)}(N;\vec \Sigma)\simeq0.08$. The  vertical dashed red and black lines correspond to the lightest state in the spectrum, $\omega^\mathrm{min}=\sqrt{r^2_{\bar s s-\mathrm{conn}}+\vec \omega^2}$, and to the parameter $\omega^\mathrm{th}$, respectively. The error in the approximation of the kernel in the gray area, $\omega_0<\omega^\mathrm{th}$, is irrelevant for the physical result. The vertical dotted green line corresponds to $\omega^\mathrm{max}$. The \emph{bottom-panel} shows the direct comparison between the approximated and exact kernels.}
\label{fig:kernel} 
\end{figure}

Within the HLT algorithm statistical errors are tamed by implementing the regularization mechanism originally proposed by Backus and Gilbert in Ref.~\cite{SMBackus}. This is done by introducing the so-called trade-off parameter $\lambda$ and by adding the term proportional to the error functional in \cref{eq:W_functional}. There are two facts that have to be considered in order to understand the r\^ole of the trade-off parameter within the HLT algorithm. The first is that the Backus-Gilbert regularization is \emph{statistically unbiased}: in the idealized situation in which the correlators $\hat Z^{(p)}(a n,\vec \omega^2;a)$ have no errors the functional $B^{(p)}[\vec g]$ is identically zero and, therefore, the same result is obtained for any value of $\lambda$. The second fact is that, for small values of the smearing parameter $\sigma$, the coefficients obtained by solving \cref{eq:minimization} with increasingly smaller values of $\lambda$ tend to become huge in magnitude and oscillating in sign~\cite{Hansen:2019idp}. Consequently, by using these coefficients in \cref{eq:dgammaa}, the statistical errors on the differential decay rate tend to be unacceptably large and, moreover, the estimates of the central values cannot be trusted in this regime because even tiny rounding errors on the lattice correlators $\hat Z^{(p)}(a n,\vec \omega^2;a)$ get enormously enhanced. The stability analysis, introduced in Ref.~\cite{Bulava:2021fre} (see also Refs.~\cite{ExtendedTwistedMassCollaborationETMC:2022sta,Evangelista:2023fmt,ExtendedTwistedMass:2024myu,Frezzotti:2023nun,Barone:2023tbl,Bonanno:2023ljc,Bonanno:2023thi,Bulava:2023brj,DelDebbio:2024lwm,Bennett:2024cqv,Blum:2024hyr,Almirante:2024lqn,Smecca:2025hfw}), is a procedure that allows to perform the $\lambda \mapsto 0$ and $N\mapsto \infty$ limits appearing in \cref{eq:aNlimits} by leveraging on these two facts. 

An example of stability analysis is shown in \cref{fig:stability_analysis}. The data correspond to the $d\Gamma^{(0)}_{\bar c s}(\sigma;a,N,\vec \Sigma)/d \vec \omega^2$ contribution to the decay rate evaluated on the D96 ensemble for $\sigma m_{D_s}=120$~MeV and $|\vec \omega|=0.38$. In the top and middle panels the differential decay rate is plotted as a function of the variable 
\begin{flalign}
d^{(p)}\left(N;\vec \Sigma\right)=\sqrt{\frac{A_0^{(p)}\left[\vec g^{(p)}(N;\vec \Sigma)\right]}{A_0^{(p)}[\vec 0]}}\;,
\end{flalign}
measuring the deviation of the reconstructed kernel $\Theta_\sigma^{(p)}(\omega^\mathrm{max}-\omega_0;N,\vec \Sigma)$ from the target one. By choosing increasingly smaller values of $\lambda$ one gets smaller values of $d^{(p)}\left(N;\vec \Sigma\right)$ and, therefore, smaller systematic errors on the differential decay rate. Conversely, by reducing $d^{(p)}\left(N;\vec \Sigma\right)$ the statistical errors rapidly increase. In the top-panel we show the data corresponding to $\alpha=0$ and to increasingly larger values of $N$. As can be seen, for sufficiently small values of  $d^{(p)}\left(N;\vec \Sigma\right)$ (on the left of the vertical dashed black line) and for $N>13$ the results for the differential decay rate become independent upon $N$ within the statistical errors. The differences between the $N=43$ result and the results at the other considered values of $N$, obtained at the value of $d^{(p)}\left(N;\vec \Sigma\right)$ corresponding to the vertical red dashed line in the top-panel, are also shown in the bottom-panel. The red band in this panel corresponds to the statistical error of the $N=43$ result and the errors on the points take into account the correlations of the different datasets. As can be seen, the $N=33$ result is compatible with the $N=43$ result at the level of the small error of the correlated difference. This means that by using $N=43$ on this ensemble, the systematic error associated with the $N\mapsto \infty$ limit is totally irrelevant w.r.t.\ the statistical errors of our results. This fact is also corroborated by the results shown in the middle-panel, that correspond to $N=43$ and to different values of the norm parameter $\alpha$. As can be seen, there is no significant dependence upon the choice of the norm parameter and this is another evidence that, within the statistical errors, the onset of the $N\mapsto \infty$ limit has been reached.  

In order to quote the central value, the statistical error and to estimate the residual systematic error on the differential decay rate we search for a plateau-region in which the results do not show any significant dependence upon $d^{(p)}\left(N;\vec \Sigma\right)$. The absence of such a plateau-region would prevent us from quoting a result but, in the case of our data, a plateau-region is clearly visible for all contributions, all flavor channels, all considered values of $\sigma$ and of $\vec \omega$. In the case shown in the top and middle panels of \cref{fig:stability_analysis}, we extracted our estimate of the physical differential decay rate, i.e.\ the $\lambda\mapsto 0$ and $N\mapsto \infty$ result, from the red dataset, corresponding to $\alpha=0$ and $N=43$, that clearly exhibits a plateau on the left of the vertical red line. For any point in the plateau-region the systematic error on the differential decay rate can safely be neglected w.r.t.\ the corresponding statistical error. 
Given the strong statistical correlations of the different points, we do not consider a constant fit of the plateau-region but, in order to quantify a possible residual systematic error, we select two points. The first point, whose coefficients are denoted by $\vec g^{(p)}_\star$, is selected at the beginning of the plateau-region (the red vertical line in \cref{fig:stability_analysis}). We then select a second point, whose coefficients are denoted by $\vec g^{(p)}_{\star\star}$, corresponding to the condition   
\begin{flalign}\label{eq:AoB}
 \frac{
 A_\alpha^{(p)}\left[\vec g^{(p)}_{\star\star}\right]
 }{
 B^{(p)}\left[\vec g^{(p)}_{\star\star}\right]
 }= 
 \frac{1}{10} 
  \frac{
 A_\alpha^{(p)}\left[\vec g^{(p)}_{\star}\right]
 }{
 B^{(p)}\left[\vec g^{(p)}_{\star}\right]
 }\;,
 \end{flalign}
and therefore to a (ten times) better reconstruction of the smearing kernel (the black vertical line in \cref{fig:stability_analysis}). From these two points we obtain a conservative estimate of the residual systematic error associated with our results as we are now going to explain.

Let us consider a given quantity $O$ for which we have different determinations $O_i$ that we expect to differ by an amount comparable to the systematic error. In order to obtain a data-driven estimate of this systematic error we consider the pull variables
\begin{flalign}
\mathcal{P}^{ij}_\mathrm{sys} = \frac{O_i-O_j}{\Delta_{ij}}\;,
\label{eq:pull}
\end{flalign}
where $\Delta_{ij}$ is a conservative estimate of the error of the difference $O_i-O_j$ (depending upon the observable we consider either the error of one of the terms or the sum in quadrature of the errors of the two terms (see \cref{eq:PHLT}, \cref{eq:PFSE}, \cref{eq:pull_a} and \cref{eq:PSIGMA})). We then estimate the systematic error by using the formula
\begin{flalign}
    &
    \Delta_\mathrm{sys}=
    \max_{ij}\left[
    \left|
    O_i-O_j
    \right|
    \mathrm{erf}\bigg(\frac{\mathcal{P}^{ij}_\mathrm{sys}}{\sqrt{2}}\bigg)
    \right]\; .
\label{eq:systematic_error}    
\end{flalign}
The error-function weights the difference with a (rough) estimate of the probability that the observed value is not due to a fluctuation. To ensure a reliable estimate of the systematic error, the observables $O_{i}$ must 
  have different sensitivities to the given systematic error. For example, in the case of finite-size effects (FSE) we considered the determinations of our observables obtained on significantly different physical volumes.

In the case of the HLT stability analysis we estimate both the statistical errors and the central values of our results from the results at the $\vec g^{(p)}_{\star}$ point,
\begin{flalign}
\Delta^{(p)}_\mathrm{stat}(\vec \omega,\sigma)
\equiv
\Delta_\mathrm{stat}\left[ 
\frac{1}{\bar \Gamma}
\frac{d\Gamma^{(p)}_\star(\sigma)}{d\vec \omega^2}
\right]\;,
\label{eq:defdeltagammastar}
\end{flalign}
and the systematic error by using the results at $\vec g^{(p)}_{\star}$ and $\vec g^{(p)}_{\star\star}$ in \cref{eq:systematic_error} with the pull variable
\begin{flalign}\label{eq:PHLT}
\mathcal{P}^{(p)}_\mathrm{HLT}(\vec \omega,\sigma)=
\frac{1}{\Delta^{(p)}_\mathrm{stat}(\vec \omega,\sigma)\, \bar \Gamma}
\left(
    \frac{d\Gamma^{(p)}_\star\big(\sigma\big)}{d\vec \omega^2}
    -
    \frac{d\Gamma^{(p)}_{\star\star}\big(\sigma\big)}{d\vec \omega^2}
\right)
\,.
\end{flalign}

In \cref{fig:kernel} we compare the exact kernel $\Theta_\sigma^{(p)}(\omega^\mathrm{max}-\omega_0)$ with the reconstructed ones at the $\vec g^{(p)}_\star$ point for the different considered values of $\alpha$.

\begin{figure}[t]
\includegraphics[width=\columnwidth]{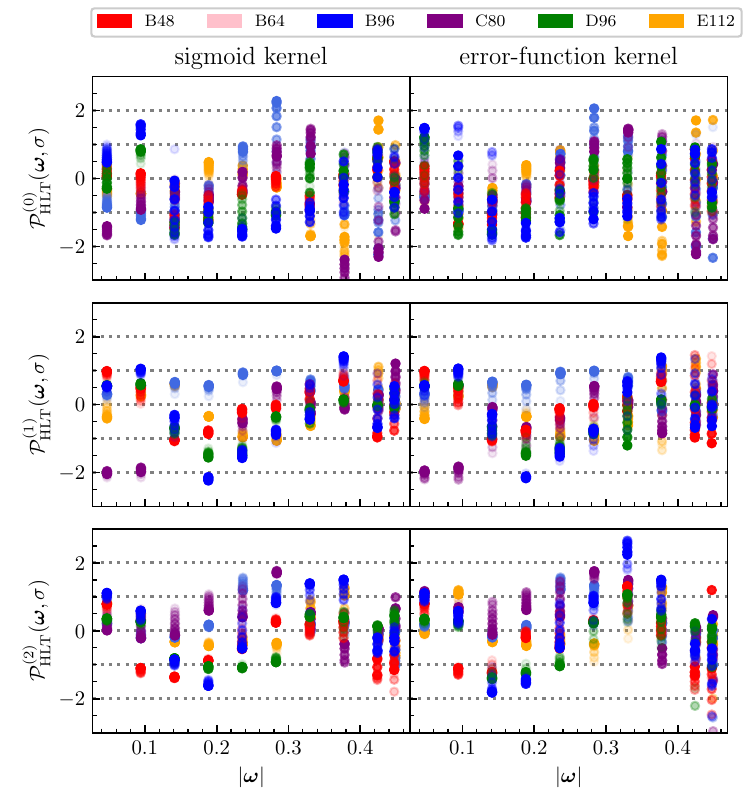}
\caption{Pull variable $\mathcal{P}_\mathrm{HLT}^{(p)}(\vec \omega,\sigma)$ for the quark-connected contributions  $d\Gamma^{(p)}_{\bar c s}(\sigma;a,L)/d\vec \omega^2$. Different colors correspond to different ensembles while different gradations of the same color correspond to different values of $\sigma$ (darker points correspond to smaller values of $\sigma$, see \cref{tab:sigmas}).}
\label{fig:cs_pull_HLT}
\end{figure}

In order to compute the decay rate and the lepton-energy moments for each flavor channel, for all of the considered values of $\vec \omega^2$ and of $\sigma$, on all of the lattice ensembles and for the two different definitions of the smearing kernel (sigmoid and error-function), we performed more than 24000 stability analyses. Aggregated information concerning these analyses, that are totally analogous to the one discussed in full details in this section, will be given in the following sections (see e.g.\ \cref{fig:cs_pull_HLT}). 

%%%%%%%%%%%%%%%%%%%%%%%%%%%%%%%%%%%%%%%%%%%%%%%%%%%%%%%%%%%%%%%%%%%%%%%%%%%%%%%%%%%%%%%%%%%%%%%%
\section{
\label{sec:cs_DGammaDq2}
Analysis of the $\Gamma_{\bar c s}$ contribution
}

\begin{table}[t]
\begin{tabular}{cc|c}
& $\qquad|\vec \omega|\qquad$ & $|\vec \omega|m_{D_s}$ [GeV] \\ [4pt] \hline
\\
  & 0.05  &  0.09 \\ [2pt]
  & 0.09  &  0.19 \\ [2pt]
  & 0.14  &  0.28 \\ [2pt]
  & 0.19  &  0.37 \\ [2pt]
  & 0.24  &  0.47 \\ [2pt]
  & 0.28  &  0.56 \\ [2pt]
  & 0.33  &  0.65 \\ [2pt]
  & 0.38  &  0.75 \\ [2pt]
  & 0.42  &  0.84 \\ [2pt]
  & 0.45  &  0.89 \\ [4pt]
  \hline
  \\
$\vert\vec \omega\vert_{\bar sd}^\mathrm{max}$  &   0.47      &   0.93     \\  
\end{tabular}
\caption{Values of the spatial momenta of the hadronic state used in the lattice calculation of the differential decay rate and of the lepton-energy moments.
\label{tab:momenta}}
\end{table}

\begin{table}[]
\begin{tabular}{cc|c}
& $\sigma$ & $\sigma m_{D_s}$[MeV]\\ [4pt] \hline
\\
& 0.005 &10   \\ [2pt]
& 0.010 &20   \\ [2pt]
& 0.020 &40   \\ [2pt]
& 0.031 &60   \\ [2pt]
& 0.041 &80   \\ [2pt]
& 0.051 &100   \\ [2pt]
& 0.061 &120   \\ [2pt]
& 0.071 &140   \\ [2pt]
& 0.081 &160   \\ [2pt]
& 0.102 &200                 
\end{tabular}
\caption{ The table shows the values of the smearing parameter $\sigma$ that we used for the two different smearing kernels. In the case of the error-function kernel only values of $\sigma \ge 0.020$ have been considered. }
\label{tab:sigmas}
\end{table}

In this section we present and discuss our results for the dominant $\Gamma_{\bar c s}$ contribution to the decay rate. We discuss separately the quark-connected contribution, extracted from the Wick contraction shown in \cref{fig:contdown}, and the weak-annihilation contribution, extracted from the diagram of \cref{fig:contwa}. 

%%%%%%%%%%%%%%%%%%%%%%%%%%%%%%%%%%%%%%%
\subsection{The quark-connected contribution}

In the numerical calculation it is convenient to separate the quark-connected contribution to $\Gamma_{\bar c s}$ from the weak-annihilation contribution. When this is done one has to take into account, though, that the lightest possible hadronic state $P_{\bar s s}$ appearing in the quark-connected contribution is not the neutral pion but the unphysical $\eta_{\bar s s}$ meson (which is lighter than a two-kaon state). Indeed, while in the case of the weak-annihilation contribution there are no strange propagators between the two weak currents (see \cref{fig:contwa}), and a single neutral pion can be generated from the sea, this cannot happen in the case of the quark-connected contribution (see \cref{fig:contdown}). We have extracted the mass of the  $\eta_{\bar s s}$ meson from the quark-connected contribution to the correlator  
\begin{flalign}\label{eq:twopointss}
C(t)=\sum_{\vec x} T\bra{0} \bar s\gamma_5s(t,\vec x)\, \bar s\gamma_5s(0) \ket{0}
\end{flalign}
obtaining $r_{\bar s s-\mathrm{conn}}\simeq 0.35$ and, consequently, $\vert\vec \omega\vert_{\bar ss-\mathrm{conn}}^\mathrm{max}\simeq 0.44$ (see \cref{eq:limits1}). By using this information, and the fact that $\vert\vec \omega\vert_{\bar sd}^\mathrm{max}>\vert\vec \omega\vert_{\bar ss-\mathrm{conn}}^\mathrm{max}$ (see next section), to be able to cover the full phase space we have then computed the quark-connected Wick contraction of the correlators $C^{\mu\nu}_{\bar c s}(t_\mathrm{snk},t,t_\mathrm{src},\vec \omega^2)$ for the 10 values of $\vert\vec \omega\vert$ given in \cref{tab:momenta}.

In order to provide information concerning the quality of the HLT stability analyses that we have performed to extract the quark-connected contribution to $\Gamma_{\bar c s}$, we show in \cref{fig:cs_pull_HLT} the pull variable $\mathcal{P}^{(p)}_\mathrm{HLT}(\vec \omega,\sigma)$, defined in \cref{eq:PHLT}, for the three different quark-connected contributions $d\Gamma^{(p)}_{\bar c s}(\sigma;a,L)/d\vec \omega^2$ (that at this stage depend upon the lattice spacing and the volume), for all  gauge ensembles, for all of the values of $\sigma$ and $\vec \omega$ that we considered, and for both smearing kernels. As can be seen, in all cases we have $|\mathcal{P}^{(p)}_\mathrm{HLT}(\vec \omega,\sigma)|<3$ and only in very few cases $|\mathcal{P}^{(p)}_\mathrm{HLT}(\vec \omega,\sigma)|>2$. This means that, at the level of two standard deviations, our results are in the statistically dominated regime.

\begin{figure}[t]
\includegraphics[width=\columnwidth]{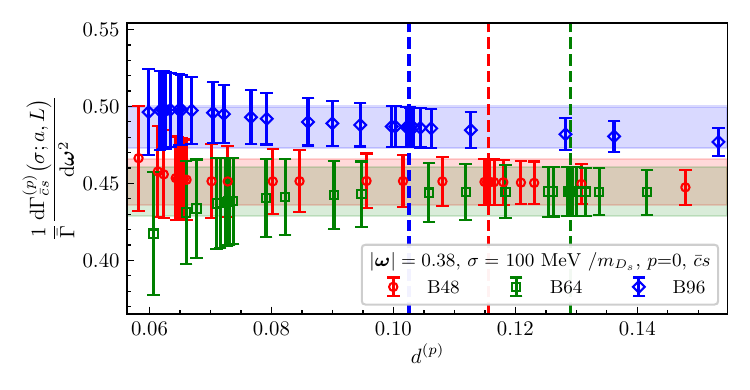}
\includegraphics[width=\columnwidth]{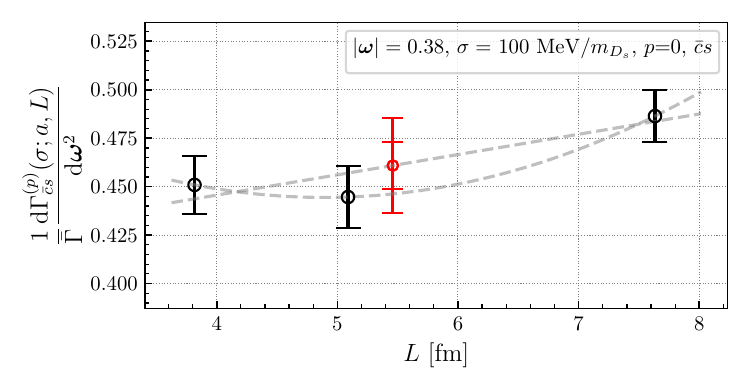}
\caption{
\emph{Top-panel}: stability analyses of the quark-connected contribution $d\Gamma^{(0)}_{\bar c s}(\sigma;a_B,L)/d\vec \omega^2$ on the B48, B64 and B96 ensembles that have the same lattice spacing ($a_B$) but different physical volumes. The data correspond to $|\vec \omega|=0.38$ and $\sigma m_{D_s}=100$~MeV. The dashed vertical lines correspond to the $\vec g^{(p)}_{\star}$ points. \emph{Bottom-panel}: interpolation of the results $d\Gamma^{(0)}_{\bar c s}(\sigma;a_B,L)/d\vec \omega^2$, extracted from the stability analyses shown in the top-panel, at the reference volume $L_\star\simeq 5.5$~fm. The red point is the result of the linear interpolation and the larger error bar takes into account our estimate of the FSE systematic error associated with the interpolation and with the $L\mapsto \infty$ limit.}
\label{fig:cs_FSE}
\end{figure}
\begin{figure}[t]
\includegraphics[width=\columnwidth]{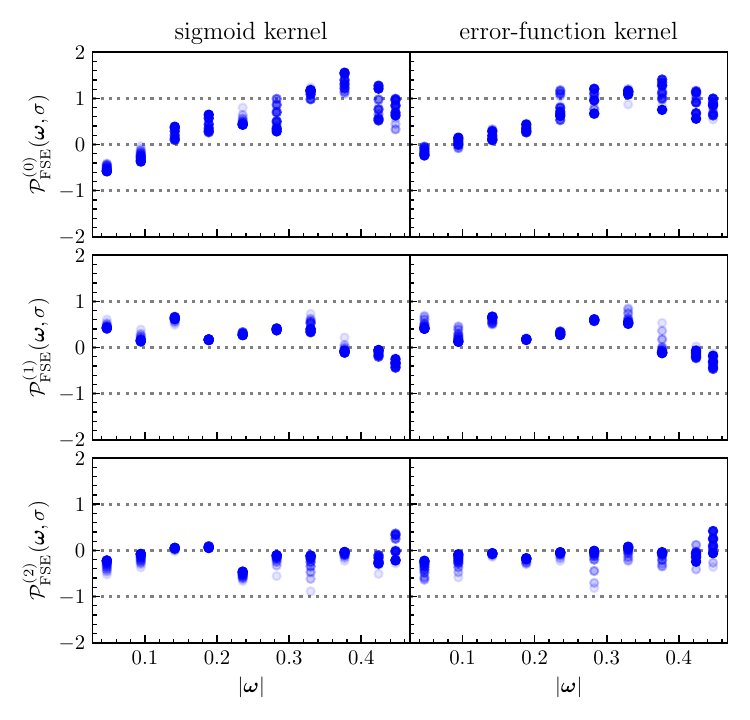}
\caption{Pull variable $\mathcal{P}^{(p)}_\mathrm{FSE}(\vec \omega,\sigma)$ of \cref{eq:PFSE} for the connected contribution $d\Gamma^{(p)}_{\bar c s}(\sigma)/d\vec \omega^2$ to the differential decay rate.}
\label{fig:cs_FSEpull}
\end{figure}
\begin{figure}[h!]
\includegraphics[width=\columnwidth]{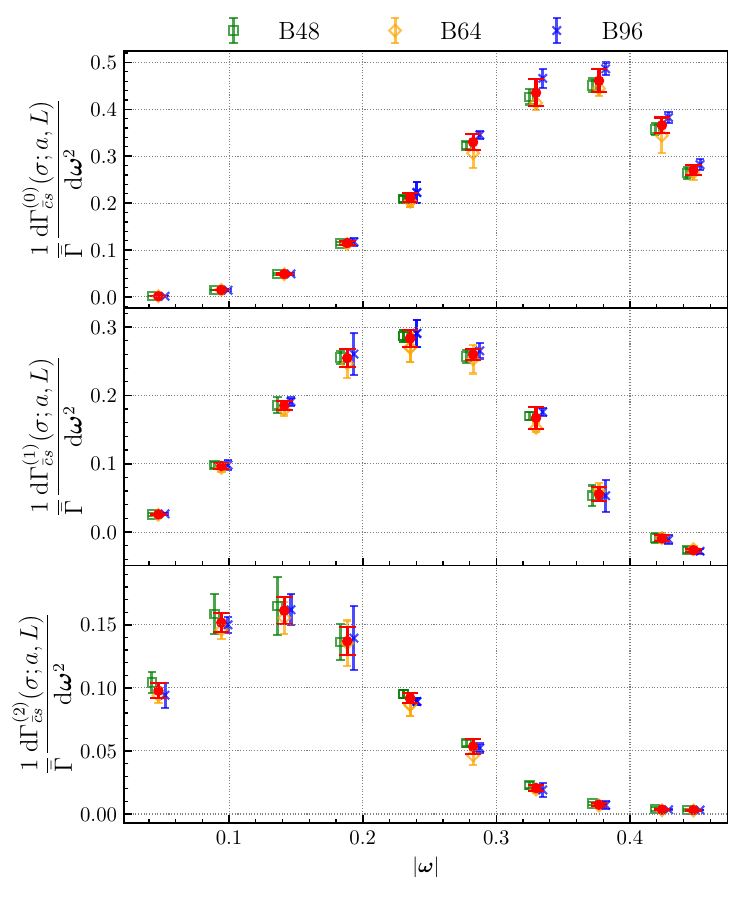}
\caption{
In the case shown in~\cref{fig:cs_FSE} the FSE systematic error is particularly important. The data of this figure also correspond to $\sigma m_{D_s}=100$~MeV but here we show, as functions of $\vert \vec \omega\vert$, the results for the three contributions to the differential decay rate obtained on the different volumes. The red points are the results of the interpolation and their errors take into account our estimate of the FSE systematic uncertainty.}
\label{fig:cs_FSE2}
\end{figure}
In order to estimate the FSE systematic errors $\Delta_L^{(p)}(\vec \omega,\sigma)$ we used the three ensembles B48, B64 and B96 at the coarsest simulated value of the lattice spacing (see \cref{tab:iso_EDI_FLAG}). While the ensembles C80, D96 and E112 (with lattice spacings $a_C$, $a_D$ and $a_E$) have been generated at the same reference physical volume $L_\star\simeq 5.5$~fm, the volume of the B48 ensemble is $L\simeq 3.8$~fm, that of the B64 ensemble is $L\simeq 5.1$~fm and that of the B96 ensemble is $L\simeq 7.6$~fm. In \cref{fig:cs_FSE} we illustrate the procedure that we use to quote our results at the coarsest value of the lattice spacing ($a_B\simeq 0.08$~fm) and to estimate the FSE systematic errors. The top-panel shows the stability analyses from which we extract the results on the B48 (red), on the B64 (green) and on the B96 (blue) ensembles. We then perform both linear and quadratic interpolations of these results. From the fits shown in the bottom-panel of \cref{fig:cs_FSE} we obtain $d\Gamma^{(p)}(\sigma;a_B,L_\star)/d\vec \omega^2$, by taking the central value from the linear fit and by adding in quadrature to the error of the linear interpolation a systematic error estimated from the spread between the linear and the quadratic interpolation, according to \cref{eq:systematic_error}. 
We stress that this systematic error, the first that in our calculation is associated with the volume dependence, is needed to properly quantify the total error on our $L_\star$ results at the coarsest value of the lattice spacing but not to estimate the systematic error associated with the $L\mapsto \infty$ extrapolation. To estimate the latter, the one that in the following we 
call FSE systematic error, we use again \cref{eq:systematic_error} with
\begin{flalign}\label{eq:PFSE}
\mathcal{P}^{(p)}_\mathrm{FSE}(\vec \omega,\sigma)=
\frac{
\frac{d\Gamma^{(p)}\big(\sigma;a_B,L_\star\big)}{d\vec \omega^2}
    -
    \frac{d\Gamma^{(p)}\big(\sigma;a_B,7.6~\mathrm{fm}\big)}{d\vec \omega^2}
}{\Delta^{(p)}_\mathrm{stat}(\vec \omega,\sigma;a_B,L_\star)\, \bar \Gamma}
\,,
\end{flalign}
where $d\Gamma^{(p)}\big(\sigma;a_B,7.6~\mathrm{fm}\big)/d\vec \omega^2$ is the B96 result. By relying upon the separation of ultraviolet and infrared physics in a local quantum field theory setup, we use this estimate of the FSE systematic error ($\Delta_L^{(p)}(\vec \omega,\sigma)$), obtained on the B ensembles, to correct our results after having performed the continuum extrapolations (see below). We show the values of the pull variable $\mathcal{P}^{(p)}_\mathrm{FSE}(\vec \omega,\sigma)$ for the quark-connected $\bar c s$ contribution to the decay rate in \cref{fig:cs_FSEpull}. As can be seen, in all cases we have $|\mathcal{P}^{(p)}_\mathrm{FSE}(\vec \omega,\sigma)|<2$ and in most of the cases $|\mathcal{P}^{(p)}_\mathrm{FSE}(\vec \omega,\sigma)|<1$. This means that the FSE systematic errors on our results are, on average, much smaller than the corresponding statistical errors. Indeed, the case shown in~\cref{fig:cs_FSE} corresponds to one of the points with larger values of $\mathcal{P}^{(0)}_\mathrm{FSE}(\vec \omega,\sigma)$ in the top-left panel of~\cref{fig:cs_FSEpull}. In~\cref{fig:cs_FSE2}, for the same value $\sigma$ considered in~\cref{fig:cs_FSE}, 
we show as functions of $\vert \vec \omega \vert$ the results for the three different contributions to the differential decay rate on the three different volumes. The red points in the figure correspond to our estimates of $d\Gamma^{(p)}(\sigma;a_B,L_\star)/d\vec \omega^2$ and their errors include our estimates of the $L\mapsto \infty$ FSE systematic error. As can be seen, the points corresponding to the B48, B64 and B96 results are always compatible with the corresponding red point within its error, also in the case shown in~\cref{fig:cs_FSE} in which, because of unavoidable statistical fluctuations, the B48 and B96 results differ at the level of 2 standard deviations.

\begin{figure}[t]
\includegraphics[width=\columnwidth]{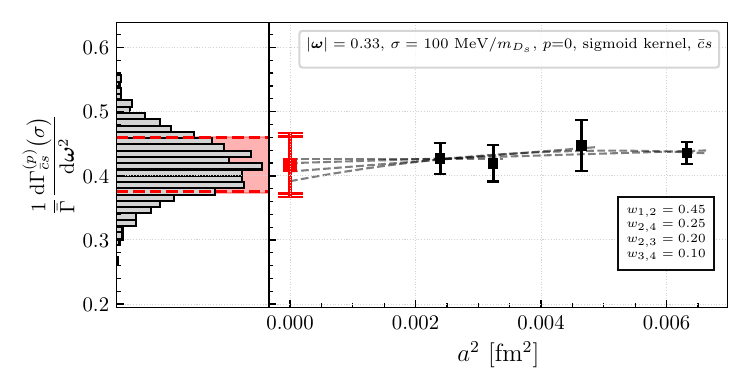}
\caption{
Continuum extrapolation of the quark-connected $d\Gamma^{(0)}_{\bar c s}(\sigma;L_\star)/d\vec \omega^2$ contribution to the decay rate. The data correspond to $|\vec \omega|=0.33$, to $\sigma m_{D_s}=100$~MeV and to the sigmoid smearing kernel. The four different dashed lines correspond to the different fits that we combine by using the Bayesian Model Average procedure discussed in the text. The histogram shows the distribution of the weighted bootstrap samples, the horizontal red dashed lines are the 16\% and 84\% percentiles while the red band is the statistical error.  The red point is the continuum result with the larger error bar taking into account our estimate of the systematic error associated with FSE (that instead is not added to the fitted points).}
\label{fig:cs_a}
\end{figure}
\begin{figure}[h!]
\includegraphics[width=\columnwidth]{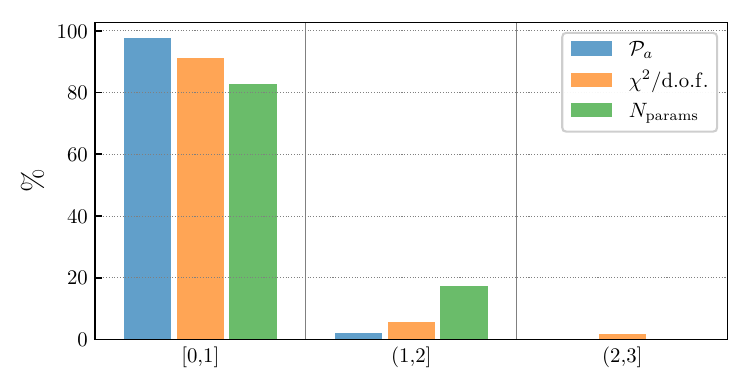}
\caption{
Aggregated information concerning the quality of all our continuum extrapolations of the quark-connected contribution $d\Gamma^{(p)}_{\bar c s}(\sigma;L_\star)/d\vec \omega^2$ (all values of $p$, $\sigma$, $\vec \omega$ and for the two smearing kernels). The blue histogram corresponds to the $\mathcal{P}_a$ variable defined in \cref{eq:pull_a}. The orange histogram corresponds to the reduced $\chi^2$ of the dominant continuum extrapolation fit. The green histogram corresponds to the number of free parameters ($N_\mathrm{params}$) of the dominant continuum extrapolation fit. }
\label{fig:cs_pull_a}
\end{figure}

In \cref{fig:cs_a} we show an example of the continuum extrapolation of our results $d\Gamma^{(p)}\big(\sigma;a,L_\star\big)/d\vec \omega^2$.  
We have four points (see \cref{tab:iso_EDI_FLAG}) that, at this stage of the analysis, are totally uncorrelated. Indeed, as already remarked in the previous paragraph, while we add to the results at the coarsest value of the lattice spacing the systematic uncertainty associated with the interpolation of the B48, B64 and B96 results, we do not add the FSE systematic error to any of the points and perform the continuum extrapolations at the fixed physical volume $L=L_\star$. We then add the FSE systematic error to the continuum extrapolated result and, in this way, take into account the fact that $L_\star<\infty$.

With these four points we perform four different extrapolations: a constant fit of the two finer points (corresponding to $a_E$ and $a_D$); a fit linear in $a^2$ of the three finer points (corresponding to $a_E$ and $a_D$ and $a_C$); a fit linear in $a^2$ and a fit quadratic in $a^2$ of all  points. The different fits are combined by employing the Bayesian Model Average~\cite{Akaike:1974vps} (see also Ref.~\cite{Evangelista:2023fmt}) that we are now going to explain.

Given $N$ different fits, the central value of the extrapolated result is given by
\begin{flalign}
    \bar{x} = \sum_{k=1}^{N} w_k x_k,
    \label{eq:AICaverage}
\end{flalign}
where $x_k$ are the extrapolated results of each separate fit. The weights $w_k$ are such that
\begin{flalign}
&
   w_k \propto \exp\big[-(\chi^2_k+2N^k_\mathrm{params}-N^k_\mathrm{points})/2\big]\;, 
   \nonumber \\[8pt]
&
   \sum_{k=1}^{N}w_k = 1 \;,
    \label{eq:AICaverage2}
\end{flalign}
where $\chi^2_k$, $N^k_\mathrm{params}$, $N^k_\mathrm{points}$ are the $\chi^2$-variable, the number of parameters and the number of points of the different fits. The total error is estimated by using
\begin{flalign}
    \Delta^2_\mathrm{tot} = \sum_{k=1}^{N}w_k\Delta_k^2 +  \sum_{k=1}^{N}w_k(x_k-\bar x)^2\;, 
    \label{eq:AICaverage3}
\end{flalign}
where the first sum is the weighted average of the square of the errors $\Delta_k$ on $x_k$ coming from the different fits. The second sum, the weighted sum of the square of the spread between each fit and the central value, provides an estimate for the systematic error. We employ the same procedure to extrapolate our results to the $\sigma \mapsto 0$ limit (see below).

Aggregated information concerning the quality of all our continuum extrapolations is provided in \cref{fig:cs_pull_a}. The figure shows three histograms, collecting the information on the values of three ``quality variables'' coming from the continuum extrapolations of all our results for the quark-connected contribution $d\Gamma^{(p)}_{\bar c s}(\sigma;L_\star)/d\vec \omega^2$, i.e.\ for each value of $p$, $\sigma$, $\vec \omega$ and for the two smearing kernels. The blue bars correspond to the pull variable
\begin{flalign}
\mathcal{P}_a = \frac{\left\vert \bar x -x(a_E)\right\vert}{\Delta_\mathrm{tot}} \;,    
\label{eq:pull_a}
\end{flalign}
where $\bar x$ again represents the result of the combined continuum extrapolation, $\Delta_\mathrm{tot}$ its error while $x(a_E)$ is the result at the finer value of the lattice spacing (that in our case is the one obtained on the E112 ensemble). \cref{fig:cs_pull_a} shows that $\mathcal{P}_a\le 1$ in more than 95\% of the cases and that we never observe $\mathcal{P}_a> 2$. This means that (almost) all our continuum extrapolated results are compatible with the points at the finest lattice spacing within one standard deviation. The orange bars correspond to the reduced $\chi^2$ of the dominant (larger weight) fit entering the weighted average of \cref{eq:AICaverage}. The Figure shows that in more than 90\% of the cases the dominant fit has $\chi^2/\mathrm{d.o.f}\le 1$. The green bars correspond to the $N_\mathrm{params}$ variable of the dominant fit. We have $N_\mathrm{params}=1$ in the case of the constant fit, $N_\mathrm{params}=2$ in the case of the linear fits and $N_\mathrm{params}=3$ for the quadratic fits. The figure shows that in more than 80\% of the cases the dominant fit is the constant one of the two finer points, i.e.\ the one providing the larger statistical error on the continuum extrapolated result. In summary, \cref{fig:cs_pull_a} provides evidence that our continuum extrapolations are rather flat, i.e.\ that we observe rather small cutoff effects within our estimates of the statistical and HLT systematic errors, and makes us very confident on the quality of our continuum extrapolations.

After having performed the continuum extrapolations the error of the continuum results (that already takes into account our estimates of the HLT and continuum-extrapolation systematic uncertainties) is added in quadrature to our estimates of the FSE systematic errors. This allows us to neglect the dependence upon the volume of our results and, therefore, we call them $d\Gamma^{(p)}(\sigma)/d\vec \omega^2$.

\begin{figure}[t!]
\includegraphics[width=\columnwidth]{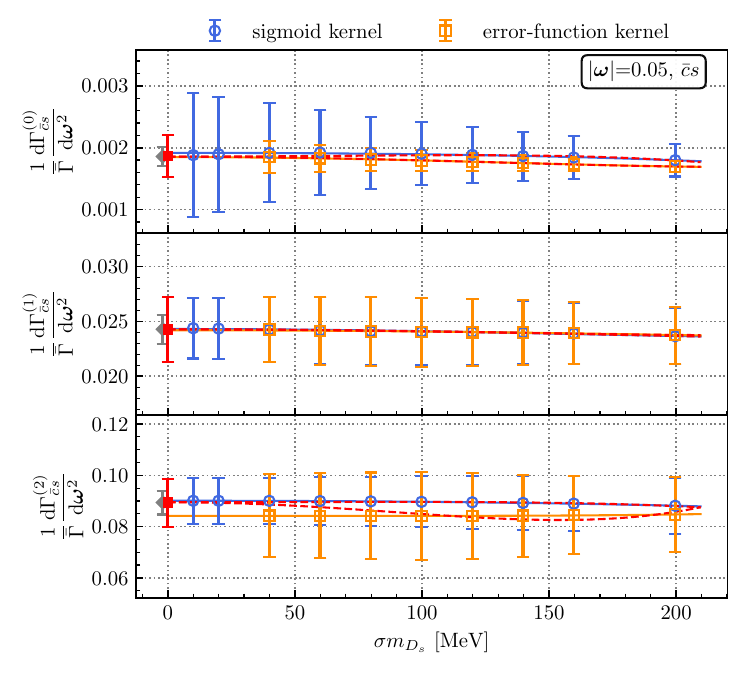}
\caption{$\sigma \mapsto 0$ extrapolation of the connected $d\Gamma^{(p)}_{\bar c s}/d\vec \omega^2$ contribution to the differential decay rate. The data correspond to $\vert \vec \omega\vert=0.05$. The blue and orange solid lines are the separate fits of the results obtained by using the sigmoid and the error-function smearing kernels, respectively. The red line is the combined fit of both datasets. The red point is the extrapolated result, the one that we use to quote our physical result, and its error includes our estimate of the systematic error associated with the extrapolation. In order to stress the fact that the results at the different values of $\sigma$ are strongly correlated, and that our error analysis procedure (see~\cref{sec:errors}) properly takes into account these correlations, we also show the gray points. These correspond to the extrapolations performed by treating the fitted points as uncorrelated. As it can be seen, the gray errors are much smaller than the errors that we quote on our physics results.}
\label{fig:cs_sigma_1}
\end{figure}
\begin{figure}[t!]
\includegraphics[width=\columnwidth]{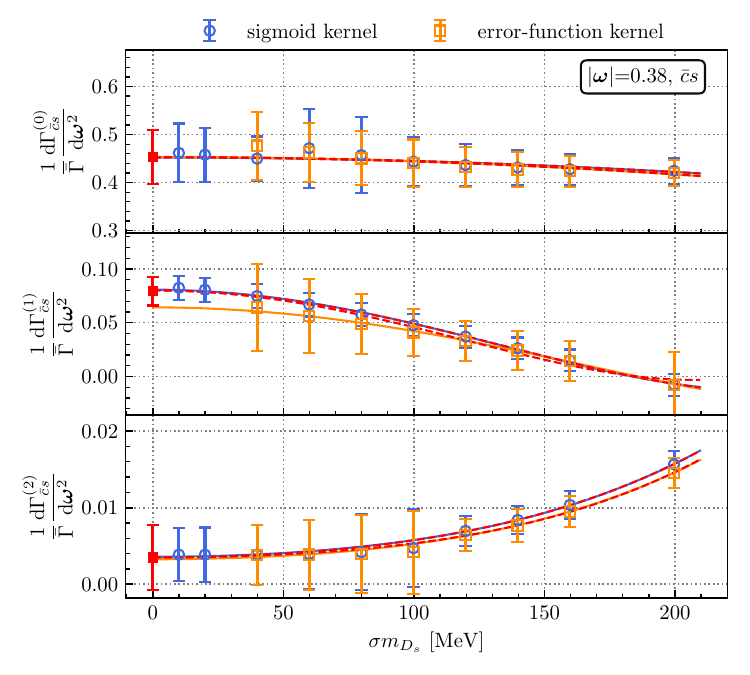}
\caption{Same as \cref{fig:cs_sigma_1} but for $\vert \vec \omega\vert=0.38$. The non-monotonous behaviour of the errors observed in some cases, e.g.\ in the bottom-panel, is due to the fact that the procedure that we use to estimate the HLT systematic errors (see~\cref{sec:hlt}) is data-driven and, therefore, sensitive to statistical fluctuations. In fact, the condition of \cref{eq:AoB} can generate in some cases quite large, and therefore very conservative, errors.}
\label{fig:cs_sigma_2}
\end{figure}
\begin{figure}[t!]
\includegraphics[width=\columnwidth]{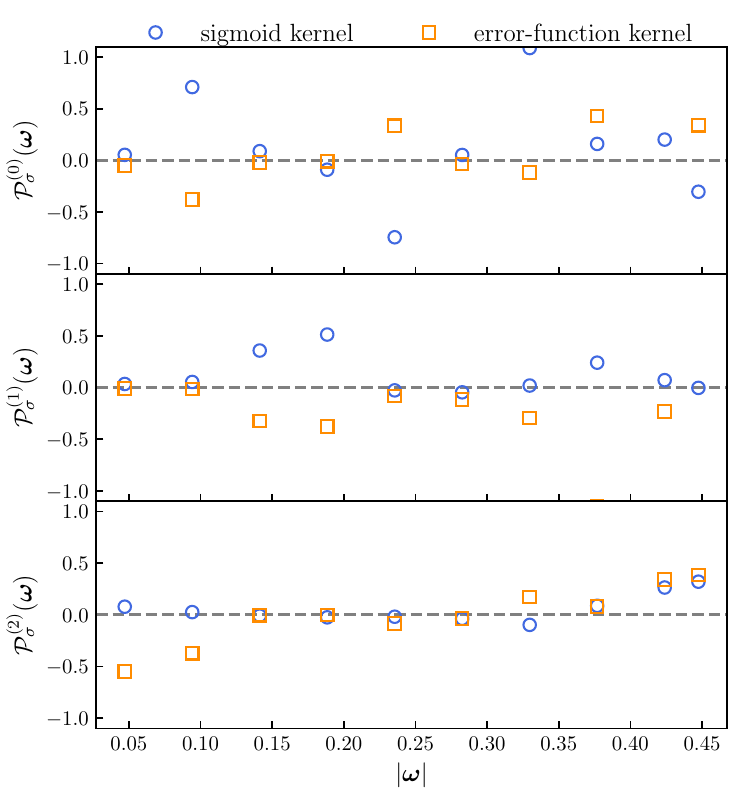}
\caption{Pull variable $\mathcal{P}^{(p)}_\sigma(\vec \omega)$ for the connected $d\Gamma^{(p)}_{\bar c s}/d\vec \omega^2$ contribution to the differential decay rate.}
\label{fig:cs_sigma_pull}
\end{figure}

The last step of the analysis consists in performing the necessary $\sigma\mapsto 0$ extrapolations. To this end, we use the asymptotic formulae  of \cref{eq:asymptG} and consider the following fitting functions 
\begin{flalign}\label{eq:sigma_ansatz3}
\frac{d\Gamma^{(0),\mathrm{I}}(\sigma)}{d \vec \omega^2}
&=
C_0^{(0),\mathrm{I}}+C_1^{(0),\mathrm{I}}\sigma^2+C_2^{(0),\mathrm{I}}\sigma^4\;,
\nonumber \\[8pt]
\frac{d\Gamma^{(1),\mathrm{I}}(\sigma)}{d \vec \omega^2}
&=
C_0^{(1),\mathrm{I}}+C_1^{(1),\mathrm{I}}\sigma^2+C_2^{(1),\mathrm{I}}\sigma^4\;,
\nonumber \\[8pt]
\frac{d\Gamma^{(2),\mathrm{I}}(\sigma)}{d \vec \omega^2}
&=
C_0^{(2),\mathrm{I}}+C_1^{(2),\mathrm{I}}\sigma^4+C_2^{(2),\mathrm{I}}\sigma^6,
\nonumber \\[8pt]
\end{flalign}
where I=$\{$sigmoid, error-function$\}$ is the label associated to the two different smearing kernels. For each quark-connected contribution $d\Gamma^{(p)}_{\bar c s}(\sigma)/d\vec \omega^2$, we perform three different fits: 
the first two correspond to separate polynomial extrapolations of the results obtained with the sigmoid and the error-function smearing kernels. The third fit is a combined extrapolation in which the coefficient of the constant term is the same for the two datasets, i.e.\  $C_0^{(p),\mathrm{sigmoid}}=C_0^{(p),\mathrm{error-function}}$. The three fits are then combined by using \cref{eq:AICaverage,eq:AICaverage2,eq:AICaverage3} to obtain our estimates of the connected contributions $d\Gamma^{(p)}_{\bar c s}/d\vec \omega^2$ to the physical differential decay rate. Examples of these extrapolations are shown in \cref{fig:cs_sigma_1,fig:cs_sigma_2}. The data in \cref{fig:cs_sigma_1} correspond to a point close to the lower-end of the phase-space integration interval $[0,\vert\vec \omega\vert_{\bar ss-\mathrm{conn}}^\mathrm{max}=0.44]$. As can be seen, in this kinematic configuration our results show a very mild dependence upon $\sigma$, almost negligible within the errors that, at this stage, include our estimates of the systematics associated with the HLT stability analysis, with FSE and with the continuum extrapolations. The data in \cref{fig:cs_sigma_2} correspond to a point close to the upper-end of the phase-space integration interval. In this case, while the dependence upon $\sigma$ is significant w.r.t.\ the errors, it is reassuringly consistent with the expected asymptotic behavior. We do not observe a significant difference between the results of the two smearing kernels and this makes us confident on the robustness of our extrapolated results. Actually, as explained in \cref{sec:hlt},
we matched the $O(\sigma^2)$ corrections associated with the two kernels by choosing the value $s=2.5$ for the shape parameter appearing in the definition of the error-function kernel given in \cref{eq:erf}. Therefore, the fact that at all the chosen values of $\sigma$ we do not observe significant differences between the two kernels means that $O(\sigma^4)$ corrections are rather small, and can be read as a reassuring evidence that our data can be extrapolated by relying on the expected theoretical asymptotic behaviour. 

In \cref{fig:cs_sigma_pull} we show, for each smearing kernel, the pull variable
\begin{flalign}\label{eq:PSIGMA}
\mathcal{P}^{(p)}_\sigma(\vec \omega)=
\frac{1}{\Delta^{(p)}(\vec \omega)\, \bar \Gamma}
\left(
\frac{d\Gamma^{(p)}}{d\vec \omega^2}
-
\frac{d\Gamma^{(p)}\big(\sigma^\mathrm{min}\big)}{d\vec \omega^2}
\right)
\,,
\end{flalign}
obtained by taking the ratio between the difference of the extrapolated point and of the result at the smallest considered value of $\sigma$ with the error $\Delta^{(p)}(\vec \omega)$ of the extrapolated point. As can be seen, almost all our data have $\vert \mathcal{P}^{(p)}_\sigma(\vec \omega)\vert<0.5$, and this corroborates our confidence on the robustness of our $\sigma\mapsto 0$ extrapolations.

\begin{figure}[t!]
    \includegraphics[width=\columnwidth]{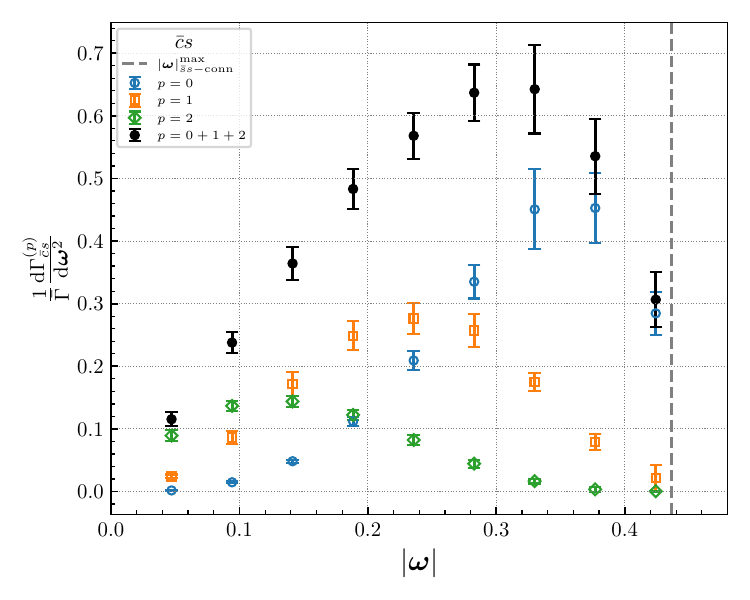}
    \caption{Quark-connected contribution $d\Gamma_{\bar c s}/d\vec \omega^2$ to the physical differential decay. The black points correspond to the sum of the three quark-connected contributions $d\Gamma^{(p)}_{\bar c s}/d\vec \omega^2$ that are also shown in different colors. The error-bars correspond to the total error, i.e.\ to the sum in quadrature of the statistical errors and of the HLT, FSE, $a\mapsto 0$ and $\sigma\mapsto 0$ systematic errors. The central values and the associated errors of the black points are listed in the column denoted by $\bar c s$ of~\cref{tab:numbers_gamma}.}
    \label{fig:cs_final}
\end{figure}
Our final results for the quark-connected contribution $d\Gamma_{\bar c s}/d\vec \omega^2$ to the physical differential decay rate are shown in \cref{fig:cs_final}.

\subsection{The weak-annihilation contribution}
\begin{figure}[t]
    \centering
    \includegraphics[width=\columnwidth]{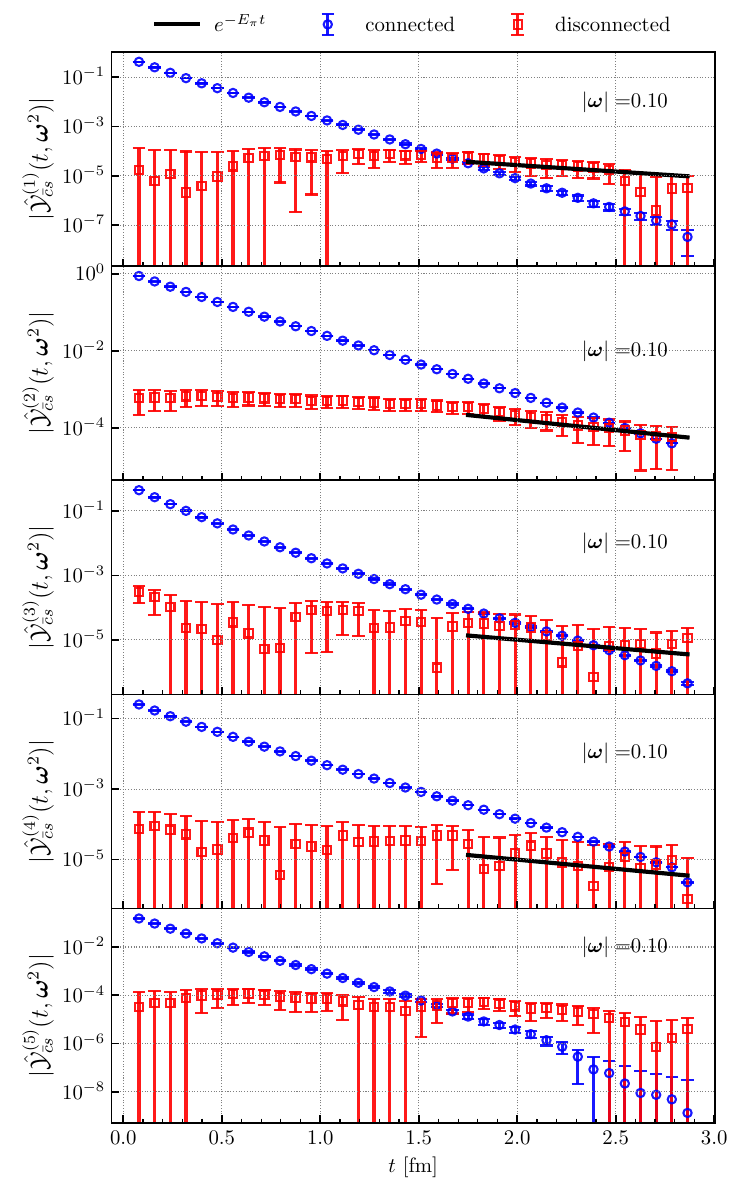}
    \caption{Comparison of the fermion connected (blue) and disconnected (red) amputated correlators $\mathcal{\hat Y}^{(i)}_{\bar c s}(t,\vec \omega^2)$. The data have been obtained on the B64 ensemble at $|\vec \omega|=0.10$. The slope of the black straight line is $-tE_\pi$, with $E_\pi^2=m_\pi^2+|\vec{\omega}|^2m_{D_s}^2$ while the intercept has been tuned in the different panels to match the value of one of the red points. From the nice agreement of the behavior of the weak-annihilation correlators at large times with the corresponding black lines we deduce that, as expected, the lightest hadronic state propagating in this channel is the neutral pion that, instead, does not appear in the fermion connected channel where we have the heavier $\eta_{\bar s s}$ meson. 
    The single-pion state is not expected to contribute to the parity-breaking correlator $\mathcal{\hat Y}^{(5)}_{\bar c s}(t,\vec \omega^2)$ which is shown in the bottom panel.}
    \label{fig:Ys_wann2}
\end{figure}
\begin{figure}[t]
    \centering
    \includegraphics[width=\columnwidth]{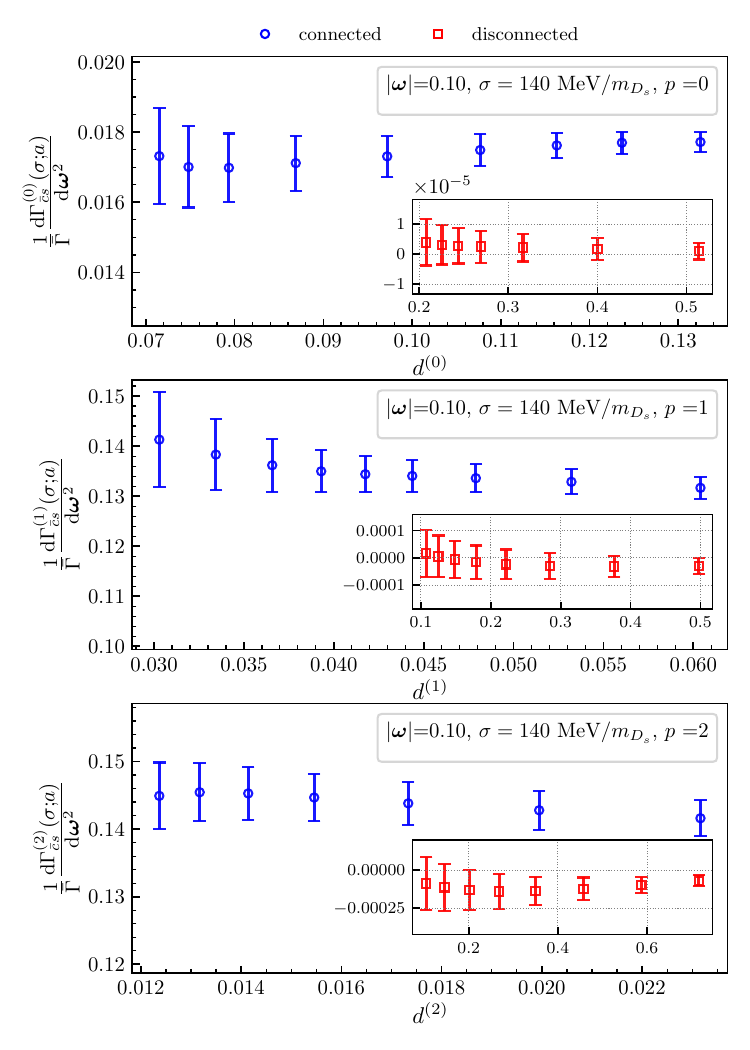}
    \caption{HLT stability analyses for the quark-connected (blue) and weak-annihilation (red) contributions to $d\Gamma^{(p)}_{\bar c s}(\sigma)/d\vec \omega^2$. The data have been obtained on the B64 ensemble for $|\vec \omega|=0.10$ and $\sigma m_{D_s}=140$~MeV. The plots are focused on the plateau-regions, where the statistical errors are dominant, and show that the weak-annihilation contribution is three orders of magnitude smaller than, and therefore totally negligible w.r.t.\ the errors of, the quark-connected contribution.}
    \label{fig:stability_wann2}
\end{figure}
\begin{figure}[t]
    \centering
    \includegraphics[width=\columnwidth]{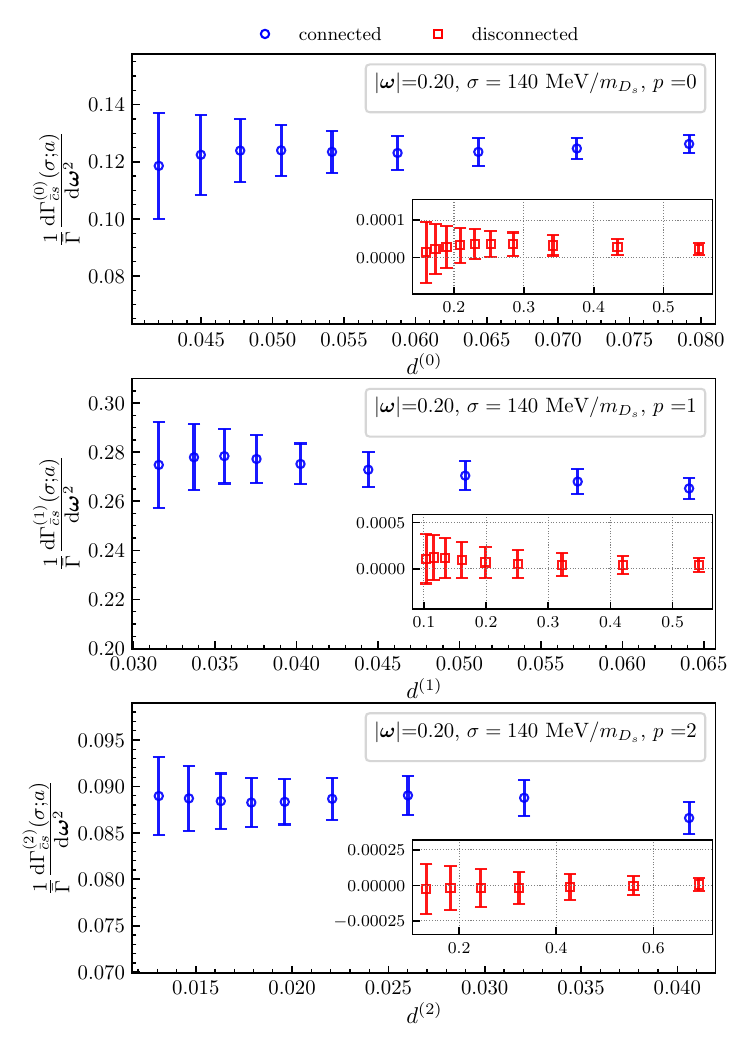}
    \caption{Same as \cref{fig:stability_wann2} but for $|\vec \omega|=0.20$.}
    \label{fig:stability_wann4}
\end{figure}

The lattice evaluation of the weak-annihilation contraction of \cref{fig:contwa} is much more challenging and computationally demanding than the quark-connected contractions of \cref{fig:contdown} that has been discussed in the previous subsection. Additionally, the weak-annihilation contribution is expected \cite{Bigi:1993bh,Ligeti:2010vd,Gambino:2010jz, Bigi:2009ym} to be $O(\Lambda_\mathrm{QCD}^3/m_c^3)$ suppressed w.r.t.\ the dominant contribution. For these reasons, we limited the calculation of the weak-annihilation contribution to a single gauge ensemble, the B64, and to two values of the momentum, corresponding to $|\vec \omega|=\{0.10,0.20\}$. As we are now going to show, although obtained on a restricted set of the parameter's space, our first-principles non-perturbative lattice results show that the weak-annihilation contribution is strongly suppressed w.r.t.\ the quark-connected one. In fact, within the errors that we quote on the dominant quark-connected $d\Gamma_{\bar c s}/d\vec \omega^2$ contribution, the weak-annihilation contribution can be safely neglected.

In \cref{fig:Ys_wann2}, the analogous of \cref{fig:Ys}, we show the five amputated correlators $\mathcal{\hat Y}^{(i)}_{\bar c s}(t,\vec \omega^2)$ extracted from both the quark-connected (blue) and weak-annihilation (red) contractions of the correlator $C_{\bar c s}^{\mu\nu}(t_\mathrm{snk},t,t_\mathrm{src},\vec \omega^2)$ at the same value of the momentum\footnote{The data presented in the previous subsection have been obtained by using twisted boundary conditions~\cite{deDivitiis:2004kq} in order to calculate the quark-connected correlators at the values of momenta listed in \cref{tab:momenta}. This is not possible in the case of the weak-annihilation correlators that have been evaluated at two values of the momentum allowed by periodic boundary conditions. In order to have a direct comparison of the fermion connected and disconnected contributions, we have generated the blue data in \cref{fig:Ys_wann2} at the same values of the momenta used in the calculation of the weak annihilation diagram.}, $|\vec \omega|=0.10$. As can be seen, although much more noisy than the quark-connected ones, the weak-annihilation correlators provide statistically significant physical information and are nicely consistent with the expected asymptotic behavior at large times, i.e.\ with the fact that the lightest hadronic state in this channel is the neutral pion (black solid line). A similar plot can be shown for the other considered value of the momentum, $|\vec \omega|=0.20$.

In \cref{fig:stability_wann2,fig:stability_wann4} we compare the HLT stability analyses of the quark-connected and weak-annihilation contributions to $d\Gamma^{(p)}_{\bar c s}(\sigma)/d\vec \omega^2$, for the two considered values of $\vec \omega$ and for $\sigma m_{D_s}=140$~MeV in the case of the sigmoid smearing kernel. As can be seen, at both the considered values of the momenta (that cover up to the middle of the phase-space integration interval of the quark-connected contribution, see \cref{fig:cs_final}), the weak-annihilation contribution is a factor $O(10^{-3})$ smaller than the connected one. Similar results can be shown for different values of the smearing parameter $\sigma$.

The results discussed in this subsection, obtained from a non-perturbative lattice evaluation of the weak-annihilation diagram, allow us to neglect the weak-annihilation contribution w.r.t.\  the errors that we have on the dominant quark-connected $d\Gamma^{(p)}_{\bar c s}/d\vec \omega^2$ contribution to the decay rate.

\section{
\label{sec:cd_DGammaDq2}
Analysis of the $\Gamma_{\bar c d}$ contribution
}
\begin{figure}[t]
\includegraphics[width=\columnwidth]{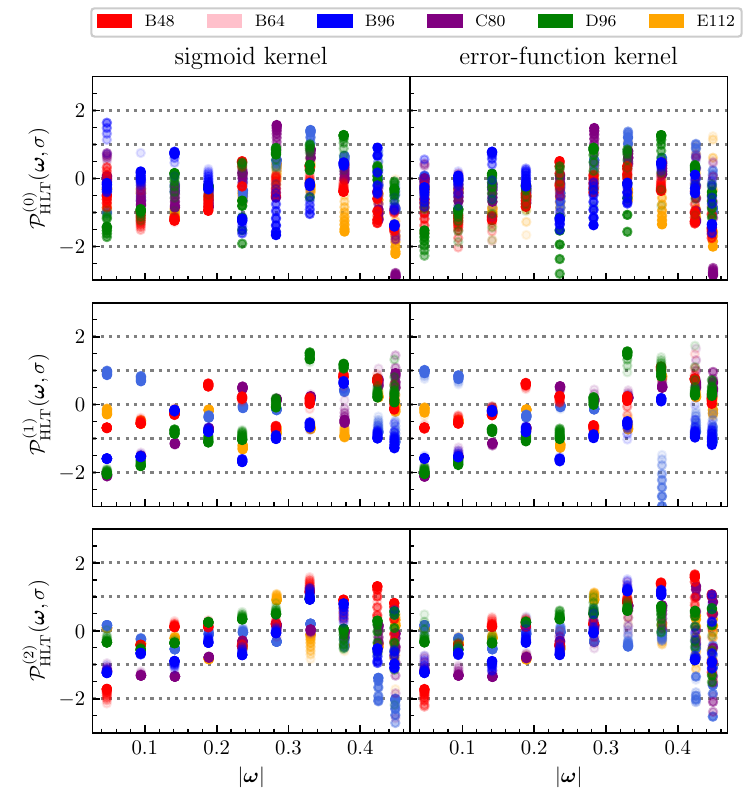}
\caption{Same as \cref{fig:cs_pull_HLT} for the $\Gamma_{\bar c d}$ contribution.}
\label{fig:cd_pull_HLT}
\end{figure}
\begin{figure}[t]
\includegraphics[width=\columnwidth]{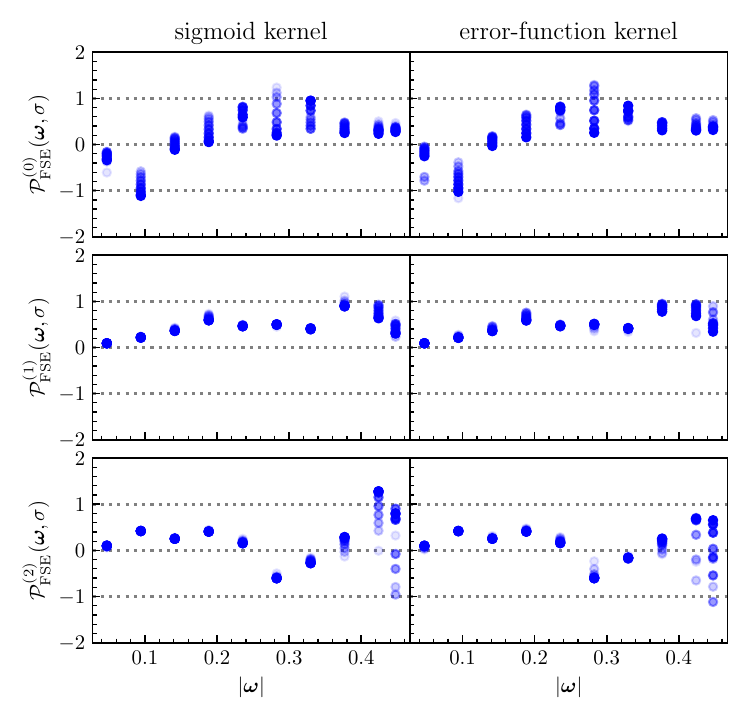}
\caption{Pull variable $\mathcal{P}^{(p)}_\mathrm{FSE}(\vec \omega,\sigma)$ for the contribution $d\Gamma^{(p)}_{\bar c d}(\sigma)/d\vec \omega^2$ to the differential decay rate.  See also \cref{fig:cs_FSE,fig:cs_FSEpull}.} 
\label{fig:cd_FSE}
\end{figure}
\begin{figure}[t]
\includegraphics[width=\columnwidth]{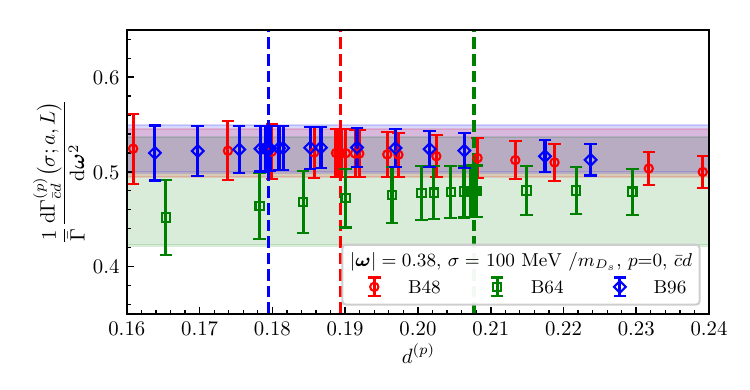}
\includegraphics[width=\columnwidth]{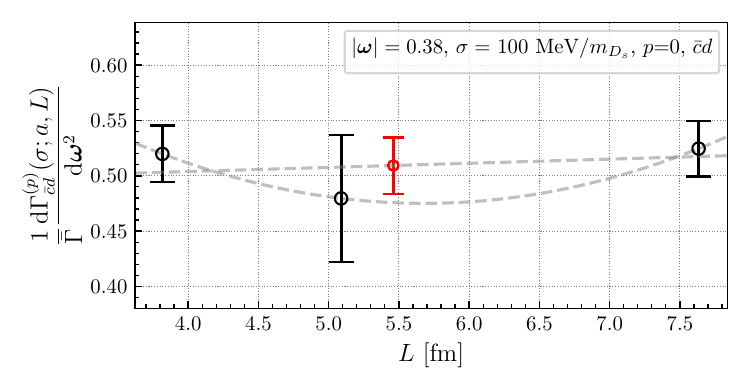}
\caption{\emph{Top-panel}: stability analyses of contribution $d\Gamma^{(0)}_{\bar c d}(\sigma;a_B,L)/d\vec \omega^2$ on the B48, B64 and B96 ensembles. The data correspond to $|\vec \omega|=0.38$ and $\sigma m_{D_s}=100$~MeV.  \emph{Bottom-panel}: interpolation of the results $d\Gamma^{(0)}_{\bar c d}(\sigma;a_B,L)/d\vec \omega^2$, extracted from the stability analyses shown in the top-panel, at the reference volume $L_\star\simeq 5.5$~fm.} 
\label{fig:cd_FSE2}
\end{figure}
\begin{figure}[h!]
\includegraphics[width=\columnwidth]{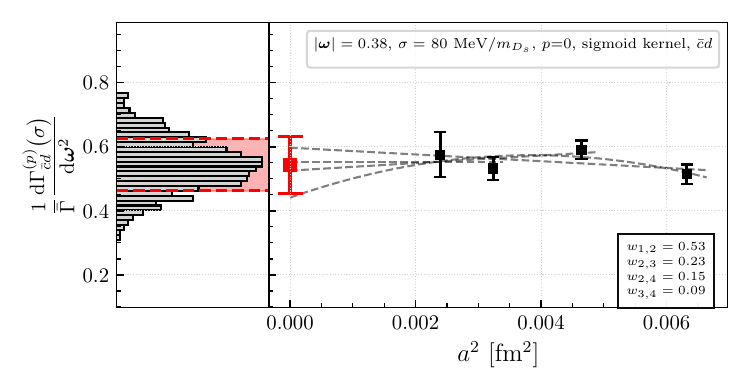}
\includegraphics[width=\columnwidth]{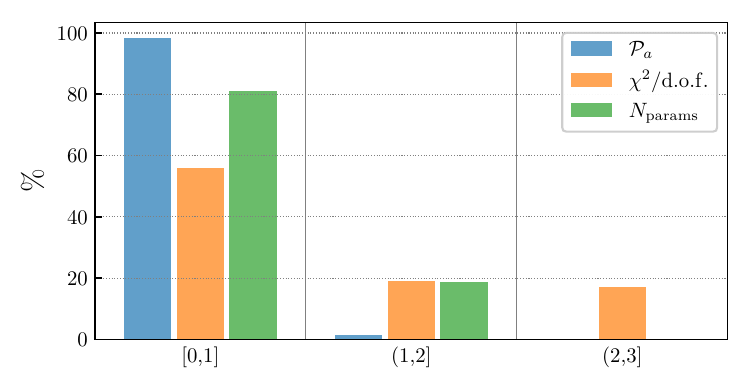}
\caption{Same as \cref{fig:cs_a,fig:cs_pull_a} for the $\Gamma_{\bar c d}$ contribution.}
\label{fig:cd_a}
\end{figure}
\begin{figure}[h!]
\includegraphics[width=\columnwidth]{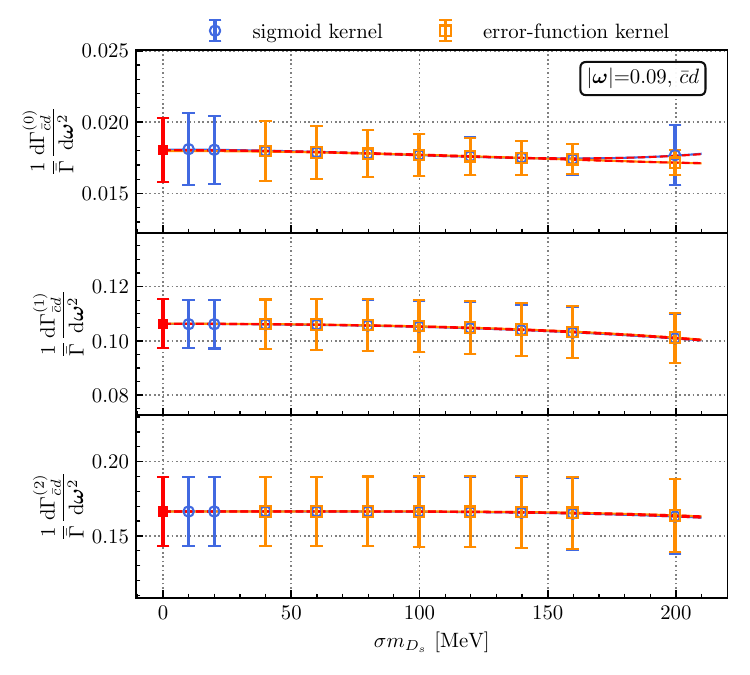}
\caption{$\sigma \mapsto 0$ extrapolation of the  $d\Gamma^{(p)}_{\bar c d}/d\vec \omega^2$ contribution to the differential decay rate for $|\vec \omega|=0.09$. See the analogous \cref{fig:cs_sigma_1}.}
\label{fig:cd_sigma_1}
\end{figure}
\begin{figure}[h!]
\includegraphics[width=\columnwidth]{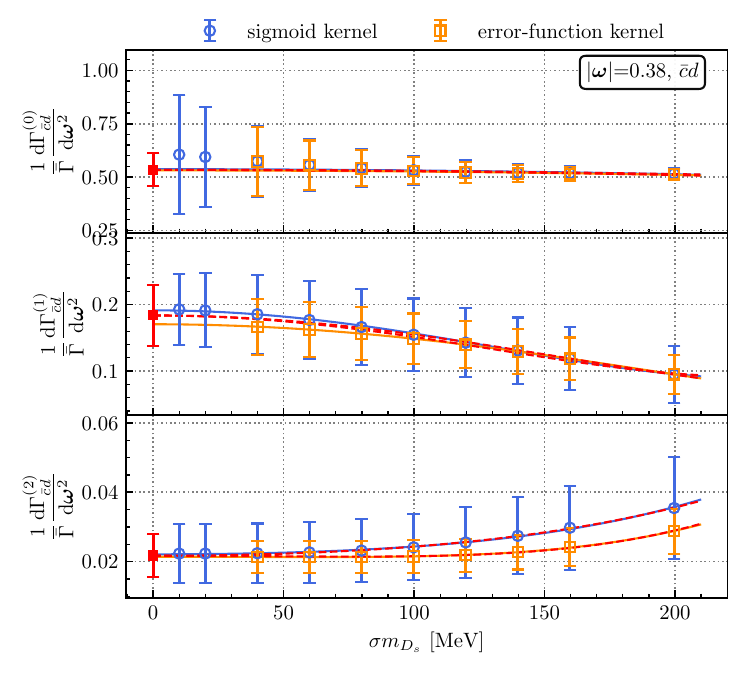}
\caption{Same as \cref{fig:cd_sigma_1} but for $\vert \vec \omega\vert=0.38$}
\label{fig:cd_sigma_2}
\end{figure}
\begin{figure}[h!]
\includegraphics[width=\columnwidth]{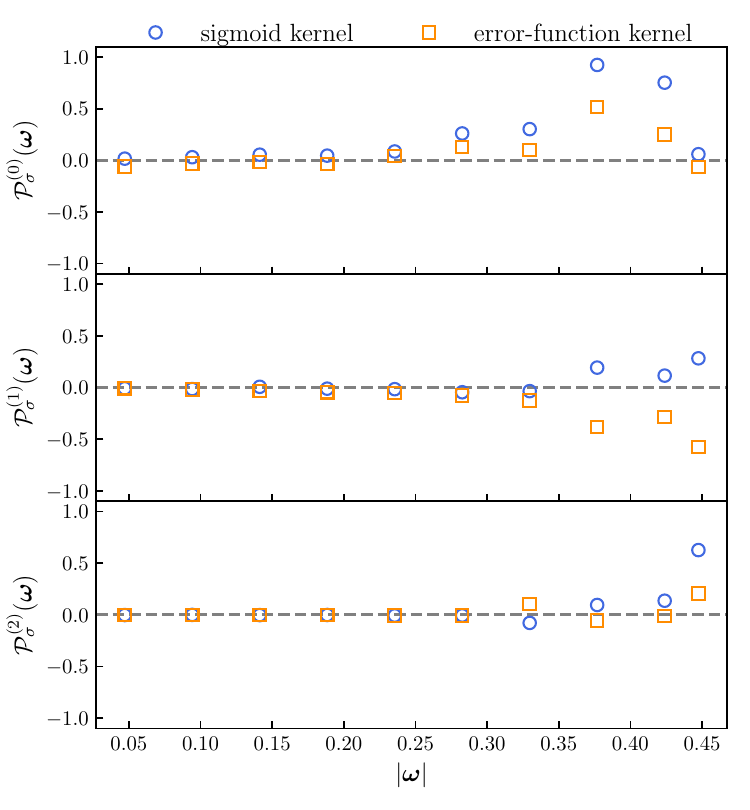}
\caption{Pull variable $\mathcal{P}^{(p)}_\sigma(\vec \omega)$ for the  $d\Gamma^{(p)}_{\bar c d}/d\vec \omega^2$ contribution to the differential decay rate.}
\label{fig:cd_sigma_pull}
\end{figure}
\begin{figure}[h!]
    \includegraphics[width=\columnwidth]{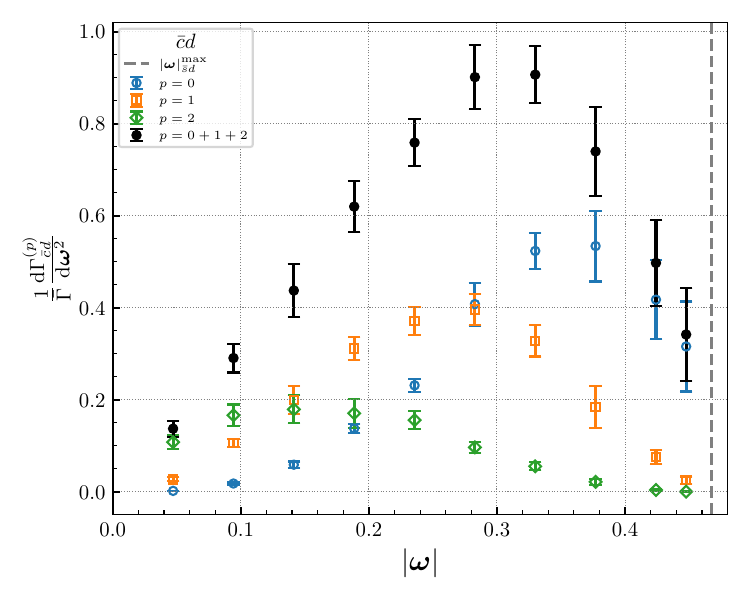}
    \caption{$d\Gamma_{\bar c d}/d\vec \omega^2$ contribution to the physical differential decay. See the analogous \cref{fig:cs_final}. The central values and the associated errors of the black points are listed in the column denoted by $\bar c d$ of~\cref{tab:numbers_gamma}.}
    \label{fig:cd_final}
\end{figure}
In this section we present our results for the $\Gamma_{\bar c d}$ contribution to the decay rate. These have been obtained by repeating all the steps of the analysis extensively discussed in \cref{sec:cs_DGammaDq2}. 

In this flavor channel we have only the quark-connected diagram and the lightest hadronic state $P_{\bar s d}$ is the neutral kaon, for which we have $r_{\bar s d}\simeq 0.26$ and, consequently, $|\vec \omega|^\mathrm{max}_{\bar s d}\simeq 0.47$. We have considered the same values of $|\vec \omega|$ and $\sigma$ that we used in the case of the $\bar c s$ channel which are given respectively in \cref{tab:momenta,tab:sigmas}.

The quality of the HLT stability analyses is illustrated in \cref{fig:cd_pull_HLT}, where the plot shows the pull variable $\mathcal{P}_\mathrm{HLT}^{(p)}(\vec \omega,\sigma)$ of \cref{eq:PHLT} for the three contributions $d\Gamma^{(p)}_{\bar c d}(\sigma;a,L)/d\vec \omega^2$, for all the values of $|\vec \omega|$ and $\sigma$, all the ensembles and for the two smearing kernels. The plot shows that $|\mathcal{P}_\mathrm{HLT}^{(p)}(\vec \omega,\sigma)|>2$ in very few cases and thus provides numerical evidence that also in this channel the statistical error is dominant over the HLT systematic error (defined in \cref{eq:systematic_error}).

In \cref{fig:cd_FSE} we show the pull variable $\mathcal{P}^{(p)}_\mathrm{FSE}(\vec \omega,\sigma)$, defined in \cref{eq:PFSE}, for the three contributions $d \Gamma^{(p)}_{\bar c d}(\sigma)/d \vec \omega^2$ for all the values of $|\vec \omega|$, $\sigma$ and the two smearing kernels.  
As can be seen, $|\mathcal{P}^{(p)}_\mathrm{FSE}(\vec \omega,\sigma)|<1.5$ in all cases, a reassuring quantitative evidence of the fact that also in this channel FSE are smaller than the statistical errors. 
In \cref{fig:cd_FSE2} we also show an example of the required stability analyses and of the estimation of the FSE.

\Cref{fig:cd_a} shows an example of continuum extrapolation for the contribution $d \Gamma^{(0)}_{\bar cd}(\sigma)/d\vec\omega^2$. The data correspond to $|\vec \omega|=0.38$, $\sigma m_{D_s}=80$~MeV and to the sigmoid smearing kernel. The figure also shows the  distributions of the ``quality variables'' $\mathcal{P}_a$, $\chi^2/\mathrm{d.o.f.}$ and $N_\mathrm{params}$, introduced in \cref{sec:cs_DGammaDq2}. The variable $\mathcal{P}_a$ is smaller than 1 in more than 95\% of the cases and never larger than 2, a quantitative evidence of the compatibility between the extrapolated points and the corresponding ones at the finest lattice spacing at the level of one standard deviation  in almost all the cases. The reduced $\chi^2$ of the dominant fit is smaller than 1, between 1 and 2, between 2 and 3 and larger than 3 in 50\%, 25\%, 15\% and 10\% of the cases, respectively. These numbers highlight a slight worsening of the quality of the continuum extrapolations compared to the quark-connected $\Gamma_{\bar cs}$ contribution  to the decay rate, see \cref{fig:cs_pull_a}. This trend can presumably be correlated with the fact that the amputated correlators $\hat{\mathcal{Y}}^{(p)}(t,\vec \omega^2)$ for the $\bar c d$ channel exhibit a larger noise-to-signal ratio compared to those corresponding to the quark-connected diagram of the $\bar cs$ channel since $M_{K_0}<M_{\eta_{\bar s s}}$.  
In fact, the slight worsening of the quality of the continuum extrapolations is due to slightly larger statistical fluctuations in the $\bar cd$ channel w.r.t.\ the quark-connected $\bar cs$ channel and, at the same time, the $\bar c d$ correlators are more noisy than the $\bar c s$ ones.
The dominant fits are constant and linear in 75\% and 25\% of the cases, respectively, as it is shown by the distribution of the variable $N_\mathrm{params}$.

The $\sigma \mapsto 0$ extrapolations have been performed as explained in \cref{sec:cs_DGammaDq2} for the connected $d\Gamma^{(p)}_{\bar c s}/d \vec\omega^2$ contributions to the decay rate.

Two examples are shown in \cref{fig:cd_sigma_1,fig:cd_sigma_2}. As can be seen, the behavior of $d \Gamma^{(p)}_{\bar c d}(\sigma)/d\vec \omega^2$ as $\sigma \mapsto 0$ is accurately reproduced by the theoretical small-$\sigma$ expansion worked out in \cref{sec:sigmato0}.
In \cref{fig:cd_sigma_pull} we show the pull variable $\mathcal{P}_\sigma^{(p)}(\vec \omega)$ defined in \cref{eq:PSIGMA}, for the contribution $d \Gamma^{(p)}_{\bar c d}/d\vec \omega^2$. As can be seen, $\mathcal{P}_\sigma^{(p)}(\vec \omega)<1$ in all the cases and $\mathcal{P}_\sigma^{(p)}(\vec \omega)<0.5$ in most of the cases, a strong quantitative evidence of the robustness of our $\sigma\mapsto 0$ extrapolations.
The final result for $d\Gamma^{(p)}_{\bar c d}/d\vec\omega^2$ is shown in \cref{fig:cd_final}.

\section{
\label{sec:su_DGammaDq2}
Analysis of the $\Gamma_{\bar u s}$ and of the $\Gamma_{\bar c s}^\mathrm{excl}$ contributions
}
\begin{figure}[t!]
    \includegraphics[width=\columnwidth]{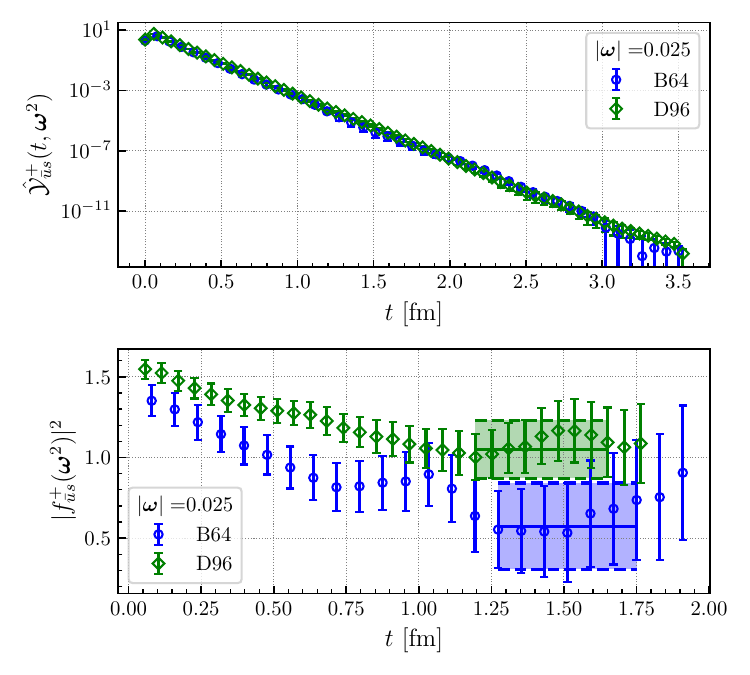}
    \caption{\emph{Top-panel}: $\hat{\mathcal{Y}}^+_{\bar u s}(t,\vec \omega^2)$ correlators at $|\vec\omega|=0.025$ computed on the B64 (blue) and D96 (green) ensembles. \emph{Bottom-panel}: $[f^+_{\bar u s}(\vec\omega^2)]^2$ extracted from two correlators in the top-panel by fitting to a constant the large-$t$ behavior of the ``effective residue'' $\omega^\mathrm{min}_{\bar us} e^{\omega^\mathrm{min}_{\bar us} t}\hat{\mathcal{Y}}^+_{\bar u s}(t,\vec \omega^2)/2\pi$ (see \cref{eq:gammaexcl3}).}
    \label{fig:su_panel}
\end{figure}
\begin{figure}[b!]
    \includegraphics[width=\columnwidth]{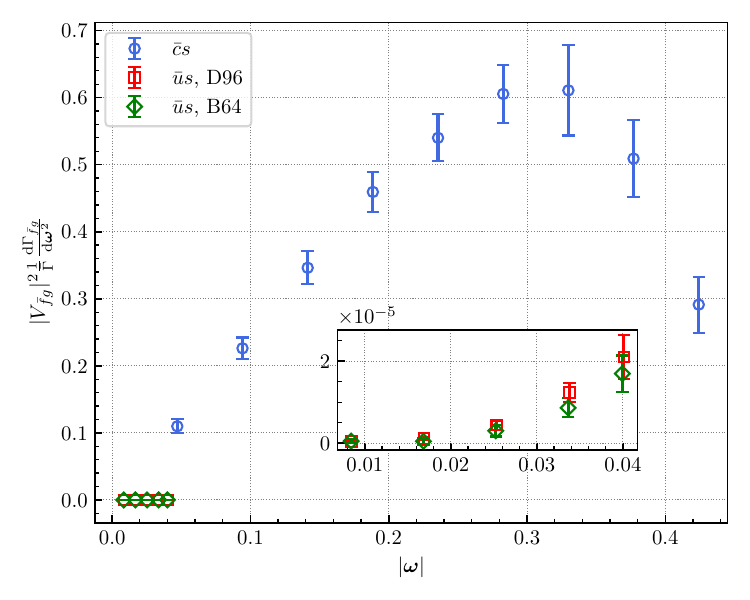}
    \caption{The blue points correspond to the dominant  $d\Gamma_{\bar c s}/d\vec \omega^2$ contribution to the differential decay rate and have been obtained by multiplying the data discussed in \cref{sec:cs_DGammaDq2} for the current best-estimate value of $\vert V_{cs} \vert^2$ taken from Ref.~\cite{ParticleDataGroup:2024cfk}. The green and red points correspond to the negligible $d\Gamma_{\bar u s}/d\vec \omega^2$ contribution and have been obtained by using our lattice determinations of the form-factor $f^+_{\bar us}(\vec \omega^2)$ and the current best-estimate value of $\vert V_{us} \vert^2$ also taken from Ref.~\cite{ParticleDataGroup:2024cfk}.}
    \label{fig:su_comparison}
\end{figure}

As discussed in \cref{sec:contratemoments}, the $\Gamma_{\bar u s}$ contribution is totally saturated from the exclusive process $D_s\mapsto D\ell \bar \nu$,
\begin{flalign}
\frac{d\Gamma_{\bar u s}}{d\vec \omega^2}    
\equiv
\frac{d\Gamma_{\bar u s}^\mathrm{excl}}{d\vec \omega^2} \;,
\end{flalign}
with the available phase space limited to the narrow interval $0\le \vert \vec \omega\vert \le \vert \vec \omega\vert^\mathrm{max}_{\bar u c}\simeq 0.05$. Moreover, $\Gamma_{\bar u s}$ is Cabibbo suppressed w.r.t.\ the dominant $\Gamma_{\bar c s}$ contribution. For these reasons, $\Gamma_{\bar u s}$ represents a negligible contribution to the total decay rate. Nevertheless, we have explicitly computed $d\Gamma_{\bar u s}^\mathrm{excl}/d\vec \omega^2$ on the B64 and D96 ensembles. 

In order to compute the exclusive contribution to the differential decay rate, we extracted the form factor $f^+$ appearing in \cref{eq:gammaexcl1} from the asymptotic behavior at large $t$ of the amputated correlator
\begin{flalign}
\mathcal{\hat Y}^+(t,\vec \omega^2)
&=
\mathcal{\hat Y}^{(2)}(t,\vec \omega^2)
+\frac{(1-\omega^\mathrm{min})^2}{\vec \omega^2} \mathcal{\hat Y}^{(3)}(t,\vec \omega^2)
\nonumber \\[8pt]
&-\frac{2(1-\omega^\mathrm{min})}{\vert \vec \omega\vert} \mathcal{\hat Y}^{(4)}(t,\vec \omega^2)\;,
\label{eq:gammaexcl2}
\end{flalign}
which is given by
\begin{flalign}
\mathcal{\hat Y}^+(t,\vec \omega^2)= 
\frac{2\pi}{\omega^\mathrm{min}} 
\left[f^+(\vec \omega^2)\right]^2\, e^{-\omega^\mathrm{min} t} +\cdots\;,
\label{eq:gammaexcl3}
\end{flalign}
where the dots represent exponentially suppressed contributions.

\Cref{fig:su_panel} shows the extraction of the form-factor $f^+_{\bar us}(\vec \omega^2)$ from the amputated correlator $\mathcal{\hat Y}^+_{\bar us}(t,\vec \omega^2)$ on both the B64 and D96 ensembles for $\vert \vec \omega \vert=0.025$. In \cref{fig:su_comparison} we provide quantitative evidence that $d\Gamma_{\bar u s}/d\vec \omega^2$ is in fact negligible w.r.t.\ the errors that we have on the dominant $d\Gamma_{\bar c s}/d\vec \omega^2$ contribution.

\begin{figure}
    \centering
    \includegraphics[width=\columnwidth]{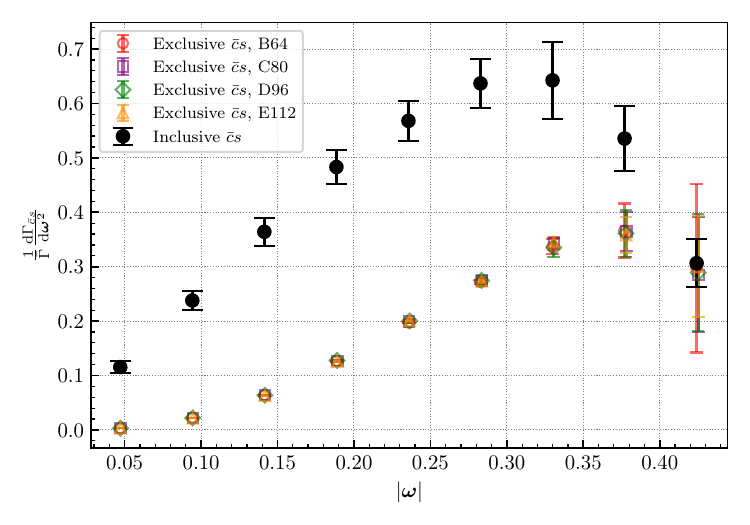}
    \caption{Comparison between the inclusive (black) and the exclusive differential decay rates for the different ensembles in the dominant $\bar cs$ flavor channel. }
    \label{fig:exclusive_channel}
\end{figure}
Before closing this section we show in \cref{fig:exclusive_channel} the comparison of the dominant contribution $d\Gamma_{\bar c s}/d\vec \omega^2$ to the inclusive differential decay rate  with the exclusive contribution in the same flavor channel, i.e.\  with $d\Gamma_{\bar c s}^\mathrm{excl}/d\vec \omega^2$. The exclusive results, that we show separately for the different ensembles, have been obtained by using the same analysis procedure that we used to compute $d\Gamma_{\bar u s}^\mathrm{excl}/d\vec \omega^2$, i.e.\ by extracting the form-factor $f^+_{\bar cs}(\vec \omega^2)$ from the amputated correlator $\mathcal{\hat Y}^+_{\bar cs}(t,\vec \omega^2)$. As expected (see \cref{sec:contratemoments}), the inclusive and exclusive contributions are fully compatible within errors at the end-point of the phase-space, i.e.\ at $\vec \omega=\vec \omega_{\bar ss-\mathrm{conn}}^\mathrm{max}$. This is a reassuring evidence concerning the robustness of the procedure that we used to estimate the systematic errors. Particularly important in this case is the systematic uncertainty associated with the $\sigma\mapsto 0$ extrapolations that become steeper when $\vec \omega$ gets closer to $\vec \omega_{\bar ss-\mathrm{conn}}^\mathrm{max}$ (see \cref{fig:cs_sigma_1,fig:cs_sigma_2}).
In the bulk of the phase space, i.e.\ for $|\vec \omega|< |\vec \omega|_{\bar ss-\mathrm{conn}}^\mathrm{max}$, the inclusive decay rate is substantially larger than the exclusive contribution. 
This is a strong evidence that the method that we have used in our lattice calculation allows to study from first-principles truly-inclusive processes, i.e.\ processes that cannot be approximated by considering a single exclusive channel, at a level of accuracy which is relevant for phenomenology.

%%%%%%%%%%%%%%%%%%%%%%%%%%%%%%%%%%%%%%%%%%%%%%%%%%%%%%%%%%%%%%%%%%%%%%%%%%%%%%%%%%%%%%%%%%%%%%%%
\section{
\label{sec:conclusions}
Summary and Outlooks}
\begin{figure}
    \centering
    \includegraphics[width=\columnwidth]{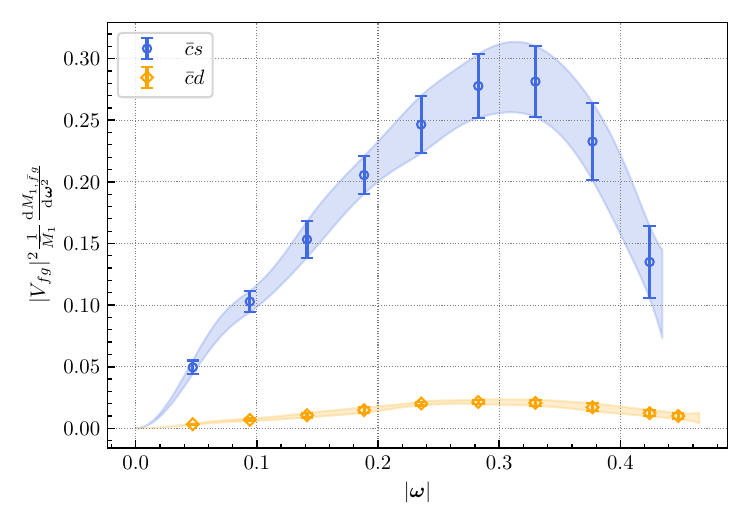}
    \caption{
    Differential first lepton-energy moment for the two dominant channels. In this plot we inserted the CKM factors $|V_{cs}|^2$ and $|V_{cd}|^2$ taken from Ref.~\cite{ParticleDataGroup:2024cfk} (PDG 2024). The filled bands represent the results of a cubic spline interpolation of the corresponding points.
    }
    \label{fig:final_differential_DMDq2}
\end{figure}
\begin{figure}
    \centering
    \includegraphics[width=\columnwidth]{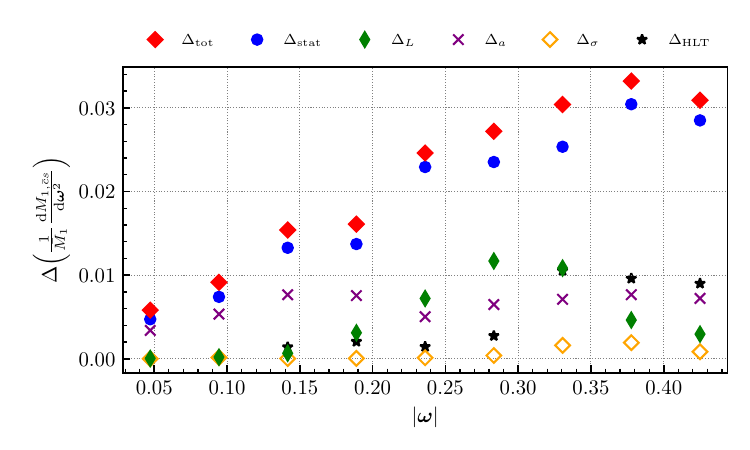}
    \includegraphics[width=\columnwidth]{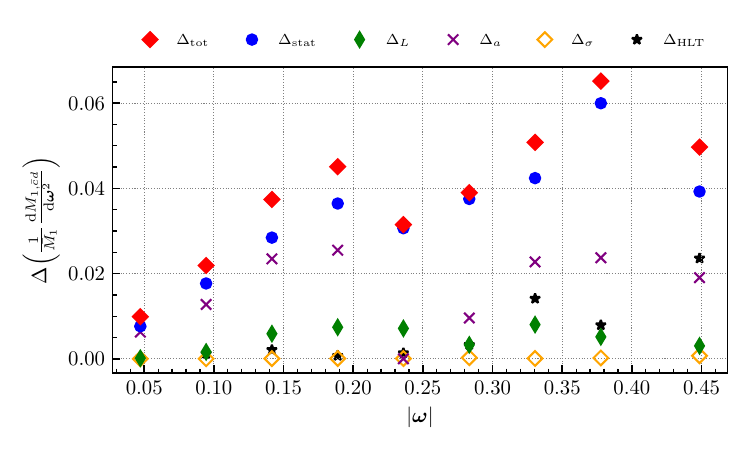}
    \caption{
    Error budgets of the differential first lepton-energy moment for the channels $\bar c s$ (top-panel) and $\bar c d$ (bottom-panel). The red points correspond to the total error $\Delta_\mathrm{tot}$, the blue points to the statistical error $\Delta_\mathrm{stat}$, the green points to the finite size systematic error $\Delta_L$, the purple points to the continuum extrapolation systematic error $\Delta_a$, the yellow points to the $\sigma\mapsto 0$ extrapolation systematic error $\Delta_\sigma$ and the black points to the HLT systematic error $\Delta_\mathrm{HLT}$. $\Delta_a$ and $\Delta_\sigma$ are given by the second term in \cref{eq:AICaverage3}.
    }
    \label{fig:DMDq2_error_budget}
\end{figure}

In this work we have computed from first-principles on the lattice the decay rate and the first two lepton-energy moments of the inclusive $D_s\mapsto X\ell \bar \nu_\ell$ semileptonic process. We have studied separately the different flavor channels that contribute to the total rate and investigated carefully all sources of systematic uncertainties. Our quantitative analysis has shown that, at the present level of accuracy, the $\Gamma_{\bar u s}$ contribution is negligible w.r.t.\ the dominant $\Gamma_{\bar c s}$ and the Cabibbo-suppressed $\Gamma_{\bar c d}$ contributions. 

Our final results for $dM_{1,\bar c s}/d \vec \omega^2$ and $dM_{1,\bar c d}/d \vec \omega^2$ are shown in \cref{fig:final_differential_DMDq2} while the associated error-budgets are shown in \cref{fig:DMDq2_error_budget}. The corresponding plots for the differential decay rate are shown in the companion paper~\cite{DeSantis:2025yfm} while those for the second moment are shown in \cref{sec:anamom}. The tables containing the numerical values of the results shown in these figures, as well as the associated covariance matrices, can be found in \cref{sec:tables}. As can be seen, our errors are statistically dominated and, therefore, the overall accuracy can certainly be improved. 

\begin{table}[t]
\begin{tabular}{lcc}
$\bar f g$ & $\quad\qquad\bar c s\quad\qquad$ & $\quad\qquad\bar c d\quad\qquad$ \\ [4pt] 
\hline
\\
 $10^{14}\times\Gamma_{\bar f g}$ [GeV]               & $ 8.53(46)(30)[55]$  & $ 12.60(79)(49)[93]$ \\ [8pt]
 $\Gamma M_{1,\bar f g}/\Gamma_{\bar cs}$ [GeV]      & $ 0.453(21)(11)[24]$ & $ 0.731(53)(30)[61]$   \\ [8pt]
 $\Gamma M_{2,\bar f g}/\Gamma_{\bar cs}$ [GeV$^2$]  & $ 0.223(9)(6)[11]$ & $ 0.416(37)(22)[43] $  \\ [2pt]
\end{tabular}
\caption{ Our final determinations of the decay rate and the first two lepton-energy moments for the two dominant channels. The CKM factors are not included in this table. The covariance matrix of these results is given in \cref{tab:covariance_results}.
\label{tab:results}}
\end{table}
In order to obtain our predictions for $M_{1,\bar fg}$ we performed a numerical integration of the differential moment $d M_{1,\bar fg}/d\vec{\omega}^2$. The same procedure has been used also in the case of the rate and of the second moment. More precisely, for each flavor channel we have interpolated $d M_{1,\bar fg}/d\vec{\omega}^2$ which we have computed for the discrete set of momenta listed in \cref{tab:momenta}. We used a cubic spline (see Ref.~\cite{CubicSpline}  and references therein for further information) by sampling the interval  $\big[0,|\vec\omega|^\mathrm{max}_{\bar F G}\big]$ uniformly with 200 points. The endpoints of the interval have been included in the sampled set. Notice that the points above the largest simulated momentum (respectively $0.42$ and $0.45$ for the $\bar c s$ and $\bar cd$ channels) up to $|\vec\omega|^\mathrm{max}_{\bar F G}$ have been extrapolated. We imposed the theoretical constraint that the differential decay rate has to vanish for $|\vec\omega|=0$. The filled bands in \cref{fig:final_differential_DMDq2} are the results of these interpolations.

The interpolated points have then been used to perform the numerical integration by applying both the trapezoid method and the Simpson's rule. The difference between the two results is totally negligible w.r.t.\ our statistical errors and, therefore, we do not quote below a systematic error associated with this step of the analysis. Finally, the integrated results are multiplied by the respective normalization factors\footnote{We have used $G_F= 1.1663788(6)\times 10^{-5}\,\mathrm{GeV}^{-2}$ from Ref.~\cite{ParticleDataGroup:2024cfk}} $\bar \Gamma$, $\bar M_1$ and $\bar M_2$. The  results with errors for the three observables and for the two channels are given in \cref{tab:results}. 

By using the current best estimates of the relevant CKM matrix elements from Ref.~\cite{ParticleDataGroup:2024cfk} (namely $|V_{cs}|=0.975(6)$ and $|V_{cd}|=0.221(4)$) and by combining the results of \cref{tab:results} we get
\begin{flalign}\label{eq:final_numbers}
&\Gamma=8.72(47)(31)[56]\times 10^{-14}\mathrm{GeV}\;,
\nonumber \\[8pt]
&
M_1= 0.456(19)(11)[22]\mathrm{GeV}\;,
\nonumber \\[8pt]
&
M_2= 0.227(9)(5)[10]\mathrm{GeV}^2\;.
\end{flalign}

Our first-principles theoretical results compare very well with the corresponding experimental results, obtained by the CLEO~\cite{CLEO:2009uah} and BES-III~\cite{BESIII:2021duu} collaborations,
\begin{flalign}
    &  \Gamma^\mathrm{CLEO} = 8.56(55)\times 10^{-14}\,\mathrm{GeV}\;,
    \nonumber\\[4pt]
    &  \Gamma^\mathrm{BES-III} =8.27(22)\times 10^{-14}\,\mathrm{GeV}\;,
    \nonumber\\[12pt]
    &
    M_1^\mathrm{CLEO} = 0.456(11) \,\mathrm{GeV}\;,
    \nonumber\\[4pt]
    &
    M_1^\mathrm{BES-III} = 0.439(9) \,\mathrm{GeV}\;,
    \nonumber\\[12pt]
    &
    M_2^\mathrm{CLEO} = 0.239(12) \,\mathrm{GeV}^2.
    \nonumber\\[4pt]
    &
    M_2^\mathrm{BES-III} = 0.222(5) \,\mathrm{GeV}^2\;.
    \label{eq:expresults}
\end{flalign}
The experimental results for the decay rate have been obtained by using the experimental branching-ratios as explained in \cref{sec:exp}. The experimental results for the lepton-energy moments have been obtained by repeating also in the case of the BES-III data the analysis performed in Ref.~\cite{Gambino:2010jz} in the case of the CLEO results.

The analysis of the phenomenological implications of our theoretical results is the subject of the companion paper~\cite{DeSantis:2025yfm}. The main goal of this work was to provide robust evidence concerning the fact that inclusive semileptonic decays of heavy mesons can nowadays be studied on the lattice at a phenomenologically relevant level of accuracy. Given the very careful analysis of all sources of systematic errors that we described in the previous sections, and given the very good agreement of our first-principles lattice results with the available experimental determinations, we can state with confidence that the goal has been reached.

As already stressed, the total error of our results is dominated by the statistical uncertainty and, therefore, it can be reduced (likely at the level of the accuracy of the BES-III measurements). We postpone this task to future work on the subject. Indeed, our results open a brilliant perspective for future lattice calculations of inclusive B mesons decays. We have already started a project in which we will compute the inclusive semileptonic decay rates of the $B_{(s)}$ mesons by extrapolating the results obtained at increasingly heavier quark masses. This will also give us the chance to reduce the errors on the $D_s$ inclusive observables computed in this work.

\section*{Acknowledgments}

The authors gratefully acknowledge the Gauss Centre for Supercomputing e.V. (www.gauss-centre.eu) for funding this project by providing computing time on the GCS Supercomputer JUWELS~\cite{JUWELS} at Jülich Supercomputing Centre (JSC) and on the GCS Supercomputers SuperMUC-NG at Leibniz Supercomputing Centre, and the granted access to the Marvin cluster hosted by the University of Bonn. The authors acknowledge the Texas Advanced Computing Center (TACC) at The University of Texas at Austin for providing HPC resources (Project ID PHY21001). The authors gratefully acknowledge PRACE for awarding access to HAWK at HLRS within the project with Id Acid 4886. We acknowledge the Swiss National Supercomputing Centre (CSCS) and the EuroHPC Joint Undertaking for awarding this project access to the LUMI supercomputer, owned by the EuroHPC Joint Undertaking, hosted by CSC (Finland) and the LUMI consortium through the Chronos programme under project IDs CH17-CSCS-CYP. We acknowledge EuroHPC Joint Undertaking for awarding the project ID EHPC-EXT-2023E02-052 access to MareNostrum5 hosted by at the Barcelona Supercomputing Center, Spain.

This work has been supported  by the MKW NRW under the funding code NW21-024-A as part of NRW-FAIR and by the Italian Ministry of University and Research (MUR) and the European Union (EU) – Next Generation EU, Mission 4, Component 1, PRIN 2022, CUP F53D23001480006 and CUP D53D23002830006.
We acknowledge support from the ENP, LQCD123, SFT, and SPIF Scientific Initiatives of the Italian Nuclear Physics Institute (INFN). F.S. is supported by ICSC-Centro Nazionale di Ricerca in High Performance Computing, Big Data and Quantum Computing, funded by European Union-NextGenerationEU  and by Italian Ministry of University and Research (MUR) projects FIS 0000155. A.S. is supported by STFC grant ST/X000648/1. We thank Paolo Garbarino for helpful discussions.

\appendix

\section{Error analysis
\label{sec:errors}}

In order to analyze our lattice data we use the bootstrap procedure explained in this appendix. This procedure allows to combine results obtained from different simulations and, at the same time, to properly take into account statistical correlations when combining results extracted from the same set of gauge configurations. Furthermore, the procedure also allows to easily incorporate and properly take into account systematic errors.

The starting point of our analysis is the calculation, on all the the gauge ensembles listed in~\cref{tab:iso_EDI_FLAG}, of the required two and four points correlators, our primary observables. Let's call $X$ a generic observable and let's introduce the ensemble index $G\in \{$B48, B64, B96, C80, D96, E112$\}$. At the beginning of the analysis a correlator, at fixed values of the time and spatial momenta variables, is represented as a double-index array $X^c_G$ where $c=1,\dots, N_\mathrm{cnfg}^G$ is the configuration index and $N_\mathrm{cnfg}^G$ is the number of the gauge configurations of the ensemble $G$ (see~\cref{tab:supplementaryinfoconfigs}).      

We then fix the same number $N_\mathrm{b}$ of samples for all the ensembles and bootstrap-resample the primary data $X^c_G$ in order to get the ensemble-dependent bootstrap samples $X^b_G$. The index $b=1,\dots,N_\mathrm{b}$ now runs on the bootstrap samples. By introducing the bootstrap average
\begin{flalign}
\left\langle X\, Y\, \cdots\right\rangle \equiv \frac{1}{N_\mathrm{b}} 
\sum_{b=1}^{N_\mathrm{b}} \left(X^b\, Y^b\, \cdots \right)\;, 
\end{flalign}
and by considering two different ensembles $G\neq G^\prime$, we have
$\left\langle X_G \, Y_{G^\prime} \right\rangle =0$ 
because, at this stage, there is no correlation between the samples $X_G^b$ and $Y_{G^\prime}^b$. Indeed these have been generated by starting from independent, and therefore uncorrelated, Monte Carlo simulations. 

Given a bootstrap sample $f^b$, that can either be an observable computed on a single ensemble ($f^b=X_G^b$) or the combination (e.g.\ a fit or a linear combination) of results coming from different ensembles ($f^b=f(X_{1,G_1}^b,\dots,X_{N,G_N}^b$)), we estimate the central value of this quantity by taking $\langle f \rangle$ and its error $\Delta[f]$ by taking
\begin{flalign}
\Delta[f] = \sqrt{\frac{1}{N_\mathrm{b}}
\sum_{b=1}^{N_\mathrm{b}} \left(f^b-\langle f\rangle \right)^2
}\;.
\label{eq:bootstraperror}
\end{flalign}
By varying the number of bootstrap samples (from $N_\mathrm{b}=O(10^2)$ to $N_\mathrm{b}=O(10^4)$) and by building bins of different sizes of the raw simulation data, i.e. by averaging data obtained on consecutive (w.r.t.\ Monte Carlo time) gauge configurations, we checked the reliability of our estimates of the statistical errors.

We now explain how systematic errors are taken into account within this procedure. The data-driven procedure that we use to estimate systematic errors is explained in~\cref{sec:hlt} (see the paragraph around~\cref{eq:systematic_error}). Let's call $\Delta_{\mathrm{sys}}$ the data-driven estimate of a given systematic error affecting the observable $X$. In order to incorporate this error in our analysis procedure we generate $N_\mathrm{b}$ random bootstrap samples from a Gaussian distribution with zero mean and with variance $\Delta_{\mathrm{sys}}$. This allows us to represent this systematic error with the array $\Delta^b_{\mathrm{sys}}$ where $b$ is the bootstrap index.
Then, by starting from the original samples $X^b$, we build a new bootstrap representation of the observable $X$ by taking
\begin{flalign}
\hat X^b= X^b + \Delta^b_{\mathrm{sys}}\;.
\label{eq:Obotstrap1}
\end{flalign}
Given the fact that $\langle \Delta_{\mathrm{sys}}\rangle=0$ we have that the central value is unchanged,
\begin{flalign}
\langle \hat X\rangle= \langle X\rangle \;,
\end{flalign}
barring finite $N_\mathrm{b}$ effects.
Moreover, given the fact that $\langle X \Delta_{\mathrm{sys}}\rangle=0$, we have that $\Delta[\hat X]$ (computed using~\cref{{eq:bootstraperror}}) provides an estimate of the total error,
\begin{flalign}
\left(\Delta[\hat X]\right)^2= 
\left(\Delta[X]\right)^2 + 
\left(\Delta_{\mathrm{sys}}\right)^2\;,
\end{flalign}
barring again finite $N_\mathrm{b}$ effects. 

In order to combine different systematic errors $\Delta_{\mathrm{sys},i}$, where the index $i$ distinguishes the estimates of the different errors, we generate uncorrelated random bootstrap samples $\Delta^b_{\mathrm{sys},i}$, i.e.\ such that $\langle \Delta_{\mathrm{sys},i} \Delta_{\mathrm{sys},j}\rangle=0$ if $i\neq j$, and generalize \cref{eq:Obotstrap1} according to
\begin{flalign}
\hat X^b= X^b + \sum_i\Delta^b_{\mathrm{sys},i}\;.
\label{eq:Obotstrap2}
\end{flalign}

In the case of the systematic errors associated with the HLT spectral reconstruction we apply the previous procedure separately and independently for each ensemble. More precisely, by calling $X_G$ and $\Delta_{\mathrm{sys},G}$ the estimates of $d\Gamma^{(p)}_\star(\sigma)/d\vec \omega^2$ and of the associated systematic error $\Delta_\mathrm{HLT}^{(p)}(\vec \omega,\sigma)$ obtained on the ensemble $G$ (see~\cref{sec:hlt}), we define $\hat X^b_G=X^b_G+\Delta^b_{\mathrm{sys},G}$. In this way, the estimates associated with two different ensembles $G\neq G^\prime$ are totally uncorrelated, $\langle X_G X_{G^\prime}\rangle=0$. 
The systematic errors associated with the uncertainties of the renormalization constants (see~\cref{tab:configsrenormailzationconstants}) are treated with the same ensemble-dependent procedure just described\footnote{Although we use the same determination of renormalization constants for the B48 and B64 ensembles, given the fact that this systematic uncertainty is totally negligible w.r.t. the statistical errors of our physics results, we generate uncorrelated bootstrap samples for these two ensembles by starting though from the same estimates of $\Delta_{\mathrm{sys},Z_{V,A}}$.}.

In the case of the FSE systematic uncertainties we have two error estimates that have to be taken into account. 

The first, let's call it $\Delta_{L_\star}^{(p)}(\sigma,a_B)$, is the one coming from the difference between the linear and quadratic interpolations that we use to obtain our determination of $d\Gamma^{(p)}(\sigma;a_B,L_\star)/d\vec \omega^2$ (see \cref{fig:cs_FSE} and the related discussion in the main text). We stress that $\Delta_{L_\star}^{(p)}(\sigma,a_B)$ is only used to generate the bootstrap samples associated with the determination of $d\Gamma^{(p)}(\sigma;a_B,L_\star)/d\vec \omega^2$, i.e.\ at the coarsest value of the lattice spacing, and not to correct the determinations of $d\Gamma^{(p)}(\sigma;a,L_\star)/d\vec \omega^2$ on the other ensembles, i.e.\ the ones with $a< a_B$. Therefore, $\Delta_{L_\star}^{(p)}(\sigma,a_B)$ only affects the right-most point in~\cref{fig:cs_a} and the four determinations of $d\Gamma^{(p)}(\sigma;a,L_\star)/d\vec \omega^2$ that we use to perform our continuum extrapolations at $L=L_\star$ remain totally uncorrelated. 

After having performed the continuum extrapolations, we correct the resulting bootstrap samples according to~\cref{eq:Obotstrap2}. To this end we use our estimates of $\Delta_\mathrm{FSE}^{(p)}(\sigma, \vec \omega)$ (the second error associated with FSE that we estimate as explained in the paragraph around~\cref{eq:PFSE}) and our estimates $\Delta_a^{(p)}(\sigma, \vec \omega)$ of the systematic uncertainties associated with the continuum extrapolations.

We finally perform our $\sigma\mapsto 0$ extrapolations and correct the resulting bootstrap samples, according to~\cref{eq:Obotstrap1}, by using our estimates $\Delta_\sigma^{(p)}(\vec \omega)$ of the associated systematic uncertainties.

The bootstrap samples of our physics results, for each flavour contribution to the differential decay rate and to the differential lepton-energy moments, that is associated with the numbers quoted in \cref{eq:final_numbers}, \cref{tab:results}, \cref{tab:numbers_gamma}, \cref{tab:numbers_M1} and \cref{tab:numbers_M2} (see also next section), can be found in the  Supplemental Material~\cite{supplemental}. The bootstrap samples of our intermediate results, i.e.\ at $\sigma>0$ and at fixed lattice spacing, can be downloaded from~\cite{FK2/VQFYKW_2025}.

\FloatBarrier

\section{Tables and covariance matrices\label{sec:tables}}

\begin{table}[!t]
\begin{tabular}{lcccc}
$|\vec\omega|$ & $\qquad$ & $\quad\qquad\bar c s\quad\qquad$ & $\qquad$ & $\quad\qquad\bar c d\quad\qquad$ \\ [4pt] 
\hline
& & \\
0.05   &&  0.115(8)(7)[11]    &&   0.137(14)(10)[18]    \\ [4pt]
0.09   &&  0.238(15)(8)[17]    &&   0.291(26)(18)[31]    \\ [4pt]
0.14   &&  0.364(25)(9)[26]    &&   0.438(48)(34)[58]    \\ [4pt]
0.19   &&  0.483(29)(13)[32]    &&   0.620(47)(30)[56]    \\ [4pt]
0.24   &&  0.568(29)(22)[37]    &&   0.759(50)(10)[51]    \\ [4pt]
0.28   &&  0.637(39)(23)[45]    &&   0.901(64)(30)[70]    \\ [4pt]
0.33   &&  0.642(61)(37)[71]    &&   0.907(46)(42)[62]    \\ [4pt]
0.38   &&  0.535(51)(32)[60]    &&   0.740(81)(54)[97]    \\ [4pt]
0.42   &&  0.306(25)(36)[44]    &&   0.497(73)(57)[93]    \\ [4pt]
0.45   &&  -    &&   0.34(9)(5)[10]  
\end{tabular}
\caption{Our results for $\frac{1}{\bar\Gamma}\frac{\mathrm{d} \Gamma_{\bar f g}}{\mathrm{d}\vec \omega^2}$ corresponding to the black points in Figures~\ref{fig:cs_final} and \ref{fig:cd_final}. The first error is statistical, the second one is systematic and in the square-brackets we report the total error obtained as the combination in quadrature of the statistical and systematic uncertainties. The covariance between the points is given in Eq.~\ref{eq:cov_gamma_cs} and Eq.~\ref{eq:cov_gamma_cd}.
\label{tab:numbers_gamma}}
\end{table}

\begin{table}[!t]
\begin{tabular}{lcccc}
$|\vec\omega|$ & $\qquad$ & $\quad\qquad\bar c s\quad\qquad$ &$\qquad$& $\quad\qquad\bar c d\quad\qquad$ \\ [4pt] 
\hline
& & \\
0.05    &&   0.0521(47)(34)[58]    &&  0.0683(76)(63)[99]  \\ [4pt]
0.09    &&   0.108(7)(5)[9]       &&  0.141(18)(13)[22]  \\ [4pt]
0.14    &&   0.161(13)(8)[15]     &&  0.218(28)(24)[37]  \\ [4pt]
0.19    &&   0.216(14)(8)[16]     &&  0.302(36)(27)[45]  \\ [4pt]
0.24    &&   0.259(23)(9)[25]     &&  0.411(31)(7)[32]  \\ [4pt]
0.28    &&   0.292(24)(14)[27]     &&  0.438(38)(11)[39]  \\ [4pt]
0.33    &&   0.296(25)(17)[30]     &&  0.424(42)(28)[51]  \\ [4pt]
0.38    &&   0.245(30)(13)[33]     &&  0.350(60)(26)[65]  \\ [4pt]
0.42    &&   0.142(28)(12)[31]     &&  0.253(46)(28)[54]  \\ [4pt]
0.45    &&    -                   &&  0.205(39)(30)[50]  
\end{tabular}
\caption{Our results for $\frac{1}{\bar M_1}\frac{\mathrm{d} M_{1,\bar f g}}{\mathrm{d}\vec \omega^2}$ corresponding to the black points in the top and bottom panel of Figure~\ref{fig:DMDq2_final}. The first error is statistical, the second one is systematic and in the square-brackets we report the total error obtained as the combination in quadrature of the statistical and systematic uncertainties.  The covariance between the points is given in Eq.~\ref{eq:cov_M1_cs} and Eq.~\ref{eq:cov_M1_cd}.
\label{tab:numbers_M1}}
\end{table}

\begin{table}[!b]
\begin{tabular}{lcccc}
$|\vec\omega|$ &$\qquad$& $\quad\qquad\bar c s\quad\qquad$ &$\qquad$& $\quad\qquad\bar c d\quad\qquad$ \\ [4pt] 
\hline
& & \\
0.05     &&   0.112(16)(13)[20]  && 0.182(25)(20)[32]   \\ [4pt]
0.09     &&   0.249(26)(17)[31]  && 0.374(67)(42)[79]   \\ [4pt]
0.14     &&   0.387(35)(22)[42]  && 0.65(96)(55)[11]   \\ [4pt]
0.19     &&   0.517(43)(28)[51]  && 0.84(13)(8)[16]   \\ [4pt]
0.24     &&   0.658(37)(21)[43]  && 1.19(9)(3)[10]   \\ [4pt]
0.28     &&   0.730(56)(30)[64]  && 1.19(11)(5)[12]   \\ [4pt]
0.33     &&   0.734(49)(28)[56]  && 1.18(12)(8)[14]   \\ [4pt]
0.38     &&   0.615(62)(40)[74]  && 0.94(19)(10)[21]      \\ [4pt]
0.42     &&   0.343(79)(56)[97]  && 0.84(21)(12)[24]      \\ [4pt]
0.45     &&            -        && 0.71(18)(12)[22]       
\end{tabular}
\caption{Our results for $\frac{1}{\bar M_2}\frac{\mathrm{d} M_{2,\bar f g}}{\mathrm{d}\vec \omega^2}$ corresponding to the black points in the top and bottom panel of Figure~\ref{fig:DM2Dq2_final}. The first error is statistical, the second one is systematic and in the square-brackets we report the total error obtained as the combination in quadrature of the statistical and systematic uncertainties.  The covariance between the points is given in Eq.~\ref{eq:cov_M2_cs} and Eq.~\ref{eq:cov_M2_cd}.
\label{tab:numbers_M2}}
\end{table}

In this section we provide the tables containing the numerical values of our physical results for the differential decay rate and for the differential lepton-energy moments as well as the associated covariance matrices. 

The results for the differential decay rate $\frac{1}{\bar \Gamma} \frac{\mathrm{d}\Gamma_{\bar fg}}{\mathrm{d}\vec{\omega}^2}$, corresponding to all the simulated momenta $|\vec\omega|$, are listed in \cref{tab:numbers_gamma} for both the $\bar cs$ (black points in \cref{fig:cs_final} and $\bar cd$ channels (black points in \cref{fig:cd_final}. In each column and row, the first number corresponds to the central value, the first error is the statistical uncertainty, the second error the systematic uncertainty and the number in square-brackets is the total error obtained as the combination in quadrature of the statistical and systematic errors. The information about the correlations between the different momenta is provided in the covariance matrices reported in Eq.~\ref{eq:cov_gamma_cs} and Eq.~\ref{eq:cov_gamma_cd} for the channel $\bar c s$ and $\bar cd$, respectively. 

\begin{widetext}
\begin{flalign}
 \mathrm{Cov}\bigg[\frac{1}{\bar\Gamma}\frac{\mathrm{d} \Gamma_{\bar c s}}{\mathrm{d}\vec \omega^2}\bigg] = \begin{pmatrix}
  1.19 &   1.66 &   2.23 &   2.11 &   1.88 &   1.54 &   1.65 &   0.83 &   0.06 \\
  1.66 &   2.84 &   3.98 &   4.06 &   3.46 &   2.86 &   2.61 &   0.86 &   0.06 \\
  2.23 &   3.98 &   6.90 &   7.47 &   6.90 &   5.78 &   5.37 &   1.73 &   0.24 \\
  2.11 &   4.06 &   7.47 &   9.93 &   9.61 &   8.86 &   8.18 &   2.54 &   0.21 \\
  1.88 &   3.46 &   6.90 &   9.61 &  13.38 &  12.88 &  14.05 &   5.66 &  -0.52 \\
  1.54 &   2.86 &   5.78 &   8.86 &  12.88 &  20.30 &  24.60 &  10.74 &  -1.45 \\
  1.65 &   2.61 &   5.37 &   8.18 &  14.05 &  24.60 &  50.18 &  26.40 &  -1.87 \\
  0.83 &   0.86 &   1.73 &   2.54 &   5.66 &  10.74 &  26.40 &  36.05 &   9.46 \\
  0.06 &   0.06 &   0.24 &   0.21 &  -0.52 &  -1.45 &  -1.87 &   9.46 &  19.20 \\
 \end{pmatrix}\times 10^{-4}\;.
\label{eq:cov_gamma_cs}
\end{flalign}

\begin{flalign}
 \mathrm{Cov}\bigg[\frac{1}{\bar\Gamma}\frac{\mathrm{d} \Gamma_{\bar c d}}{\mathrm{d}\vec \omega^2}\bigg] = \begin{pmatrix}
  3.10 &   4.05 &   4.72 &   2.85 &   0.49 &   2.16 &   1.21 &   0.33 &  -0.53 &  -0.72 \\
  4.05 &   9.84 &  13.29 &   7.94 &   0.59 &   5.79 &   4.02 &   2.35 &   0.86 &   0.61 \\
  4.72 &  13.29 &  34.00 &  22.73 &   0.93 &  11.73 &   8.78 &   4.87 &   2.72 &   1.68 \\
  2.85 &   7.94 &  22.73 &  31.13 &   2.91 &  15.41 &   9.22 &   7.33 &   0.74 &  -0.74 \\
  0.49 &   0.59 &   0.93 &   2.91 &  25.88 &   3.25 &   3.72 &   0.69 &   1.98 &   2.54 \\
  2.16 &   5.79 &  11.73 &  15.41 &   3.25 &  49.28 &  28.97 &  17.47 &   6.66 &   5.39 \\
  1.21 &   4.02 &   8.78 &   9.22 &   3.72 &  28.97 &  38.45 &  35.30 &  13.78 &  10.02 \\
  0.33 &   2.35 &   4.87 &   7.33 &   0.69 &  17.47 &  35.30 &  95.03 &  45.74 &  40.04 \\
 -0.53 &   0.86 &   2.72 &   0.74 &   1.98 &   6.66 &  13.78 &  45.74 &  86.52 &  82.75 \\
 -0.72 &   0.61 &   1.68 &  -0.74 &   2.54 &   5.39 &  10.02 &  40.04 &  82.75 & 101.57 \\
 \end{pmatrix}\times 10^{-4}\;.
\label{eq:cov_gamma_cd}
\end{flalign}
\end{widetext}

The covariance matrices are calculated in the standard way, i.e.\ the values on the diagonal correspond to the variance (squared error). Rows and columns are ordered according to the ordering of the momenta provided in the corresponding tables. For instance, \cref{eq:cov_gamma_cs} is a symmetric $9\times 9$ matrix for the 9 numbers provided in the column denoted by $\bar cs$ in \cref{tab:numbers_gamma}. The square root of the top-left number, $\sqrt{1.19\times 10^{-4}}\simeq 0.011$, is the total error of $\frac{1}{\bar \Gamma}\frac{\mathrm{d}\Gamma_{\bar c s}}{\mathrm{d}\vec\omega^2}$ for $|\vec{\omega}|=0.05$ (0.115(8)(7)[11] is the corresponding value shown in \cref{tab:numbers_gamma}). Similarly, the square root of bottom-right number is the total error of $\frac{1}{\bar \Gamma}\frac{\mathrm{d}\Gamma_{\bar c s}}{\mathrm{d}\vec\omega^2}$ for $|\vec{\omega}|=0.42$.

The corresponding results for $\frac{1}{\bar M_1}\frac{\mathrm{d}M_{1,\bar f g}}{\mathrm{d}\vec\omega^2}$ (black points in Figure~\ref{fig:DMDq2_final}) and $\frac{1}{\bar M_2}\frac{\mathrm{d}M_{2,\bar f g}}{\mathrm{d}\vec\omega^2}$ (black points in \cref{fig:DM2Dq2_final}) for the two channels are given respectively in \cref{tab:numbers_M1} and \cref{tab:numbers_M2}. The associated covariance matrices, organized as those for the decay rate, are given in \cref{eq:cov_M1_cs}, \cref{eq:cov_M1_cd}, \cref{eq:cov_M2_cs} and \cref{eq:cov_M2_cd}.

\begin{widetext}

\begin{flalign}
 \mathrm{Cov}\bigg[\frac{1}{\bar M_1}\frac{\mathrm{d} M_{1,\bar c s}}{\mathrm{d}\vec \omega^2}\bigg] = \begin{pmatrix}
  0.34 &   0.47 &   0.71 &   0.56 &   0.61 &   0.47 &   0.36 &   0.25 &   0.16 \\
  0.47 &   0.84 &   1.16 &   1.10 &   1.14 &   1.00 &   0.83 &   0.57 &   0.30 \\
  0.71 &   1.16 &   2.38 &   2.11 &   2.52 &   1.77 &   1.38 &   0.92 &   0.38 \\
  0.56 &   1.10 &   2.11 &   2.60 &   3.23 &   2.57 &   2.09 &   1.35 &   0.35 \\
  0.61 &   1.14 &   2.52 &   3.23 &   6.05 &   4.49 &   3.82 &   2.78 &   0.29 \\
  0.47 &   1.00 &   1.77 &   2.57 &   4.49 &   7.42 &   7.46 &   4.97 &   1.90 \\
  0.36 &   0.83 &   1.38 &   2.09 &   3.82 &   7.46 &   9.24 &   7.63 &   3.95 \\
  0.25 &   0.57 &   0.92 &   1.35 &   2.78 &   4.97 &   7.63 &  11.02 &   6.78 \\
  0.16 &   0.30 &   0.38 &   0.35 &   0.29 &   1.90 &   3.95 &   6.78 &   9.58 \\
 \end{pmatrix}\times 10^{-4}\;.
\label{eq:cov_M1_cs}
\end{flalign}

\begin{flalign}
 \mathrm{Cov}\bigg[\frac{1}{\bar M_1}\frac{\mathrm{d} M_{1,\bar c d}}{\mathrm{d}\vec \omega^2}\bigg] = \begin{pmatrix}
  0.98 &   1.60 &   1.45 &   1.23 &   0.24 &   0.69 &   0.40 &   0.04 &   0.13 &   0.22 \\
  1.60 &   4.82 &   5.74 &   4.16 &   0.45 &   2.21 &   1.50 &   0.21 &   0.71 &   0.71 \\
  1.45 &   5.74 &  13.97 &  11.97 &   0.59 &   4.42 &   3.79 &   1.27 &   1.94 &   1.11 \\
  1.23 &   4.16 &  11.97 &  20.38 &   0.45 &   6.99 &   5.46 &   1.70 &   2.48 &   0.39 \\
  0.24 &   0.45 &   0.59 &   0.45 &   9.94 &   1.10 &   1.64 &  -0.58 &   0.51 &   1.06 \\
  0.69 &   2.21 &   4.42 &   6.99 &   1.10 &  15.19 &  10.98 &   4.71 &   3.10 &   2.13 \\
  0.40 &   1.50 &   3.79 &   5.46 &   1.64 &  10.98 &  25.82 &  16.89 &   6.59 &   4.17 \\
  0.04 &   0.21 &   1.27 &   1.70 &  -0.58 &   4.71 &  16.89 &  42.60 &  20.60 &  14.72 \\
  0.13 &   0.71 &   1.94 &   2.48 &   0.51 &   3.10 &   6.59 &  20.60 &  29.31 &  23.28 \\
  0.22 &   0.71 &   1.11 &   0.39 &   1.06 &   2.13 &   4.17 &  14.72 &  23.28 &  24.74 \\
 \end{pmatrix}\times 10^{-4}\;.
\label{eq:cov_M1_cd}
\end{flalign}

\begin{flalign}
 \mathrm{Cov}\bigg[\frac{1}{\bar M_2}\frac{\mathrm{d} M_{2,\bar c s}}{\mathrm{d}\vec \omega^2}\bigg] = \begin{pmatrix}
  4.15 &   5.97 &   6.69 &   6.59 &   3.82 &   4.05 &   2.74 &   2.74 &   1.58 \\
  5.97 &   9.71 &  12.13 &  12.56 &   7.98 &   7.90 &   5.22 &   4.59 &   2.58 \\
  6.69 &  12.13 &  17.43 &  19.92 &  13.64 &  13.23 &   8.82 &   7.33 &   3.21 \\
  6.59 &  12.56 &  19.92 &  26.04 &  18.22 &  18.46 &  12.50 &   9.98 &   2.04 \\
  3.82 &   7.98 &  13.64 &  18.22 &  18.17 &  20.64 &  16.12 &  12.69 &   2.65 \\
  4.05 &   7.90 &  13.23 &  18.46 &  20.64 &  40.43 &  31.16 &  24.78 &   2.71 \\
  2.74 &   5.22 &   8.82 &  12.50 &  16.12 &  31.16 &  31.87 &  32.34 &  11.14 \\
  2.74 &   4.59 &   7.33 &   9.98 &  12.69 &  24.78 &  32.34 &  54.57 &  37.47 \\
  1.58 &   2.58 &   3.21 &   2.04 &   2.65 &   2.71 &  11.14 &  37.47 &  94.33 \\
 \end{pmatrix}\times 10^{-4}\;.
\label{eq:cov_M2_cs}
\end{flalign}

\begin{flalign}
 \mathrm{Cov}\bigg[\frac{1}{\bar M_2}\frac{\mathrm{d} M_{2,\bar c d}}{\mathrm{d}\vec \omega^2}\bigg] = \begin{pmatrix}
 10.54 &  18.24 &  13.16 &  12.23 &   1.27 &   5.91 &   3.45 &   2.11 &  -1.87 &   2.88 \\
 18.24 &  63.02 &  62.62 &  51.55 &   0.85 &  21.00 &  18.88 &   6.97 &  -5.72 &   3.31 \\
 13.16 &  62.62 & 122.36 & 118.41 &  -0.48 &  37.06 &  38.40 &  11.27 &  -1.64 &  -1.09 \\
 12.23 &  51.55 & 118.41 & 239.12 & -15.54 &  65.55 &  53.91 &   9.99 &  -0.19 &  -7.81 \\
  1.27 &   0.85 &  -0.48 & -15.54 &  95.01 &  -1.01 &   2.23 &  -8.69 &  -3.79 &   9.23 \\
  5.91 &  21.00 &  37.06 &  65.55 &  -1.01 & 144.74 & 107.54 &  40.05 &  30.10 &  30.10 \\
  3.45 &  18.88 &  38.40 &  53.91 &   2.23 & 107.54 & 205.37 & 122.29 &  51.49 &  21.35 \\
  2.11 &   6.97 &  11.27 &   9.99 &  -8.69 &  40.05 & 122.29 & 452.53 & 295.03 & 199.95 \\
 -1.87 &  -5.72 &  -1.64 &  -0.19 &  -3.79 &  30.10 &  51.49 & 295.03 & 577.72 & 446.52 \\
  2.88 &   3.31 &  -1.09 &  -7.81 &   9.23 &  30.10 &  21.35 & 199.95 & 446.52 & 473.50 \\
 \end{pmatrix}\times 10^{-4}\;.
\label{eq:cov_M2_cd}
\end{flalign}
\end{widetext}

Finally, the covariance matrix between the six observables listed in \cref{tab:results} is provided in \cref{tab:covariance_results}. In this case the association between the matrix entries and the corresponding observables is made manifest and the elements on the diagonal still correspond to the squared total errors.

\begin{widetext}

\begin{table}[t]
\begin{tabular}{lcccccc}
  & \quad $10^{14}\times\Gamma_{\bar c s}$ \quad 
  & \quad $10^{14}\times\Gamma_{\bar c d}$ \quad 
  & \quad $\Gamma M_{1,\bar c s}/\Gamma_{\bar cs}$ \quad 
  & \quad $\Gamma M_{1,\bar c d}/\Gamma_{\bar cs}$ \quad 
  & \quad $\Gamma M_{2,\bar c s}/\Gamma_{\bar cs}$ \quad 
  & \quad $\Gamma M_{2,\bar c d}/\Gamma_{\bar cs}$\\ [4pt] 
\hline
\\
\quad $10^{14}\times\Gamma_{\bar c s}$         & 30.783  &  23.542  &   0.139  &  -0.890  &  -0.004  &  -0.505      \\ [8pt]
\quad $10^{14}\times\Gamma_{\bar c d}$         & 23.542  &  85.857  &  -0.037  &   3.044  &  -0.067  &   1.689      \\ [8pt]
\quad $\Gamma M_{1,\bar c s}/\Gamma_{\bar cs}$ & 0.139  &  -0.037  &   0.055  &  -0.002  &   0.019  &   0.003      \\ [8pt]
\quad $\Gamma M_{1,\bar c d}/\Gamma_{\bar cs}$ & -0.890  &   3.044  &  -0.002  &   0.378  &   0.003  &   0.233      \\ [8pt]
\quad $\Gamma M_{2,\bar c s}/\Gamma_{\bar cs}$ & -0.004  &  -0.067  &   0.019  &   0.003  &   0.011  &   0.006      \\ [8pt]
\quad $\Gamma M_{2,\bar c d}/\Gamma_{\bar cs}$ & -0.505  &   1.689  &   0.003  &   0.233  &   0.006  &   0.181      \\ [2pt]
\end{tabular}
\caption{Covariance matrix of the final results given in \cref{tab:results}. All the numbers in the table are multiplied by $10^2$.
\label{tab:covariance_results}}
\end{table}

\end{widetext}

\section{Analysis of the lepton-energy moments
\label{sec:anamom}}

In this appendix we present aggregated information, analogous to that discussed for the decay rate, from the analysis of the first and second lepton-energy moment. The analysis is carried out in a equivalent way to that extensively discussed for the decay rate. We have computed the lepton-energy moments for the (quark-connected) $\bar cs$ and $\bar c d$ channels and neglected the further contributions. Where not specified, the pull variables are obtained by collecting together the data from the two channels. 

\begin{figure}
    \centering
    \includegraphics[width=\columnwidth]{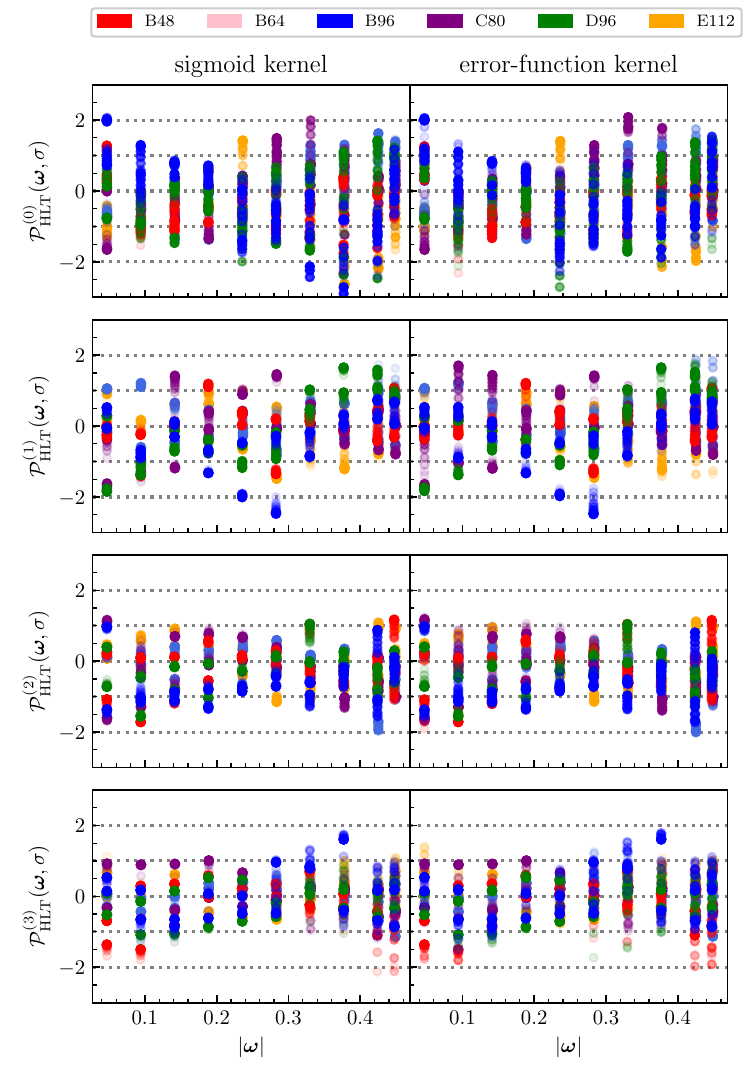}
    \caption{The same as \cref{fig:cs_pull_HLT} for the first lepton-energy moment. Data for the $\bar c s$ and $\bar c d$ are displayed together.}
    \label{fig:DMDq2_pull_HLT}
\end{figure}

\begin{figure}
    \centering
    \includegraphics[width=\columnwidth]{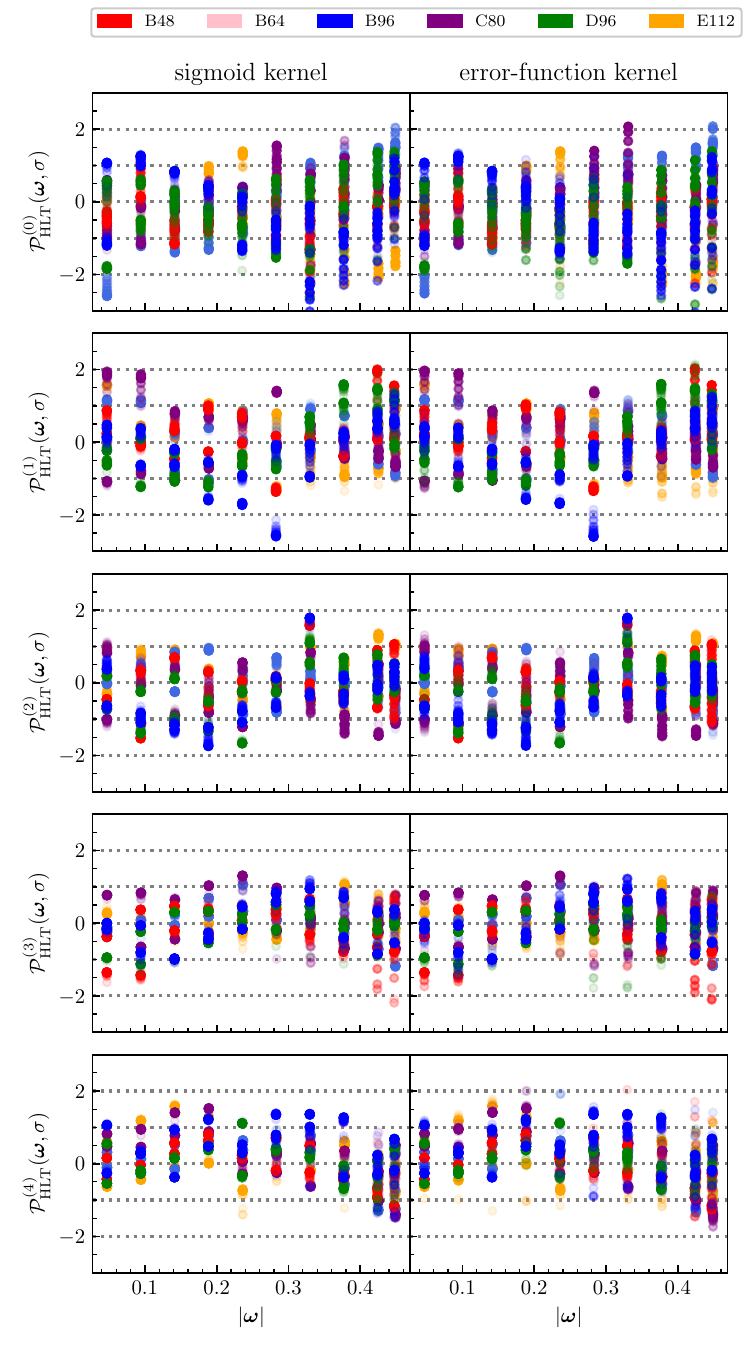}
    \caption{The same as \cref{fig:cs_pull_HLT} for the second lepton-energy moment. Data for the $\bar c s$ and $\bar c d$ are displayed together.}
    \label{fig:DM2Dq2_pull_HLT}
\end{figure}

\begin{figure}
    \centering
    \includegraphics[width=\columnwidth]{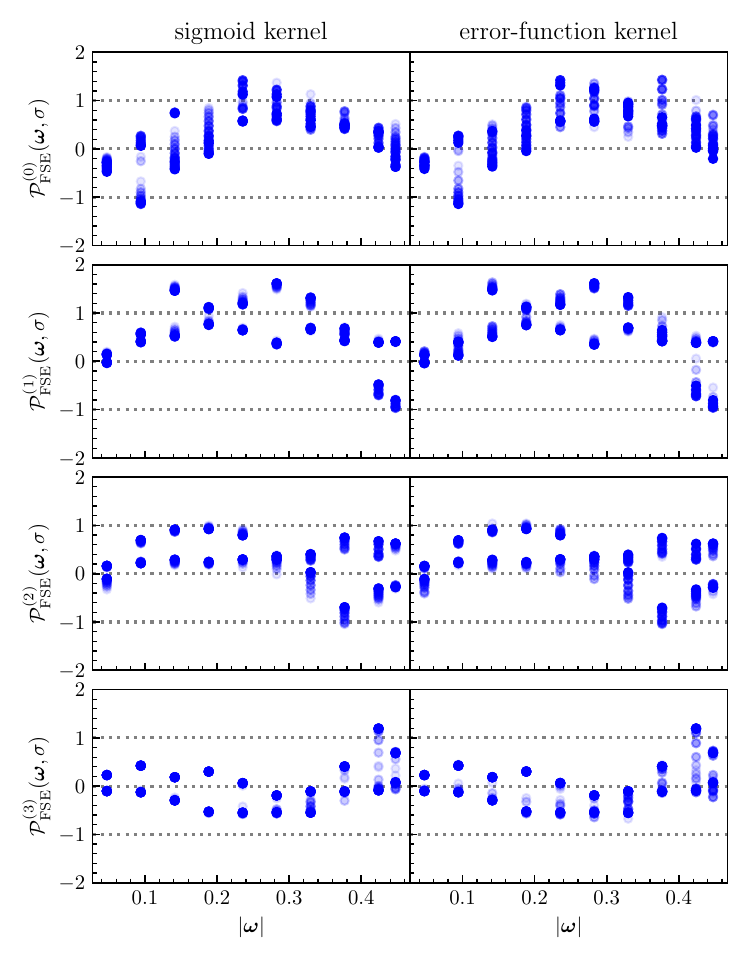}
    \caption{The same as \cref{fig:cs_FSEpull} for the first lepton-energy moment. Data for the $\bar c s$ and $\bar c d$ are displayed together.}
    \label{fig:DMDq2_pull_FSE}
\end{figure}

\begin{figure}
    \centering
    \includegraphics[width=\columnwidth]{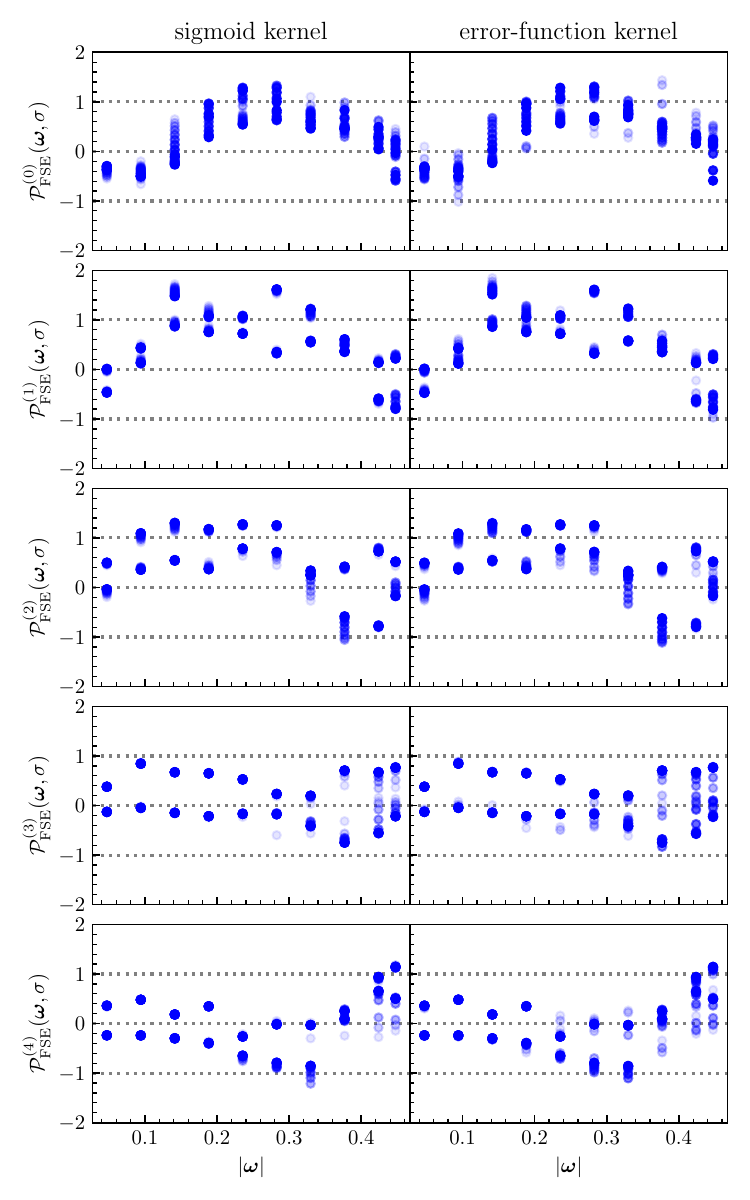}
    \caption{The same as \cref{fig:cs_FSEpull} for the second lepton-energy moment. Data for the $\bar c s$ and $\bar c d$ are displayed together.}
    \label{fig:DM2Dq2_pull_FSE}
\end{figure}

\begin{figure}
    \centering
    \includegraphics[width=\columnwidth]{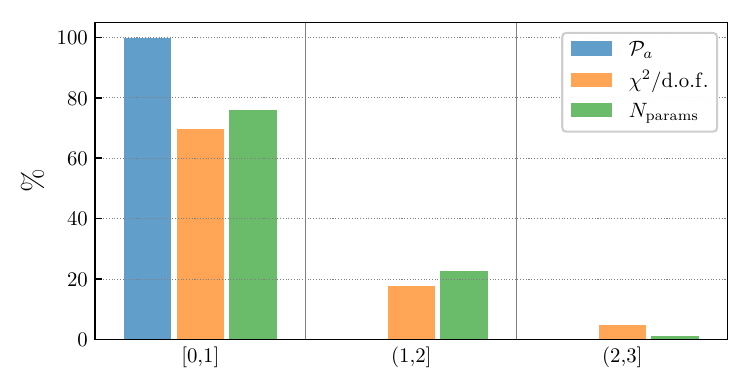}
    \includegraphics[width=\columnwidth]{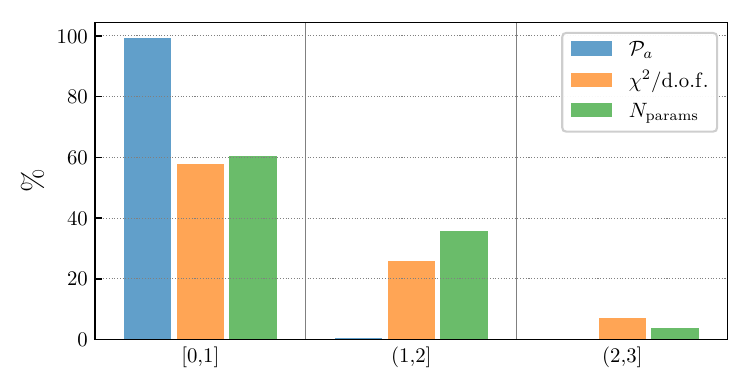}
    \caption{The same as \cref{fig:cs_pull_a} for the first (top-panel) and second (bottom-panel) lepton-energy moment. The histograms gather together the $\bar cs$ and $\bar cd$ channels.}
    \label{fig:DMxDq2_pull_a_to_0}
\end{figure}

The pull variable $\mathcal{P}_\mathrm{HLT}^{(p)}(\vec \omega, \sigma)$ for the first and second lepton-energy moment is shown in \cref{fig:DMDq2_pull_HLT} and \cref{fig:DM2Dq2_pull_HLT} respectively. The plots show that $|\mathcal{P}_\mathrm{HLT}^{(p)}(\vec \omega, \sigma)|<2$ in the majority of the cases and always $|\mathcal{P}_\mathrm{HLT}^{(p)}(\vec \omega, \sigma)|<3$ meaning that the stability analysis are dominated by statistics. The pull variables $\mathcal{P}_\mathrm{FSE}^{(p)}(\vec \omega,\sigma)$ are shown respectively in \cref{fig:DMDq2_pull_FSE} and \cref{fig:DM2Dq2_pull_FSE}. In almost all the cases $|\mathcal{P}_\mathrm{FSE}^{(p)}(\vec \omega,\sigma)|<1$ and always $|\mathcal{P}_\mathrm{FSE}^{(p)}(\vec \omega,\sigma)|<2$. The finite size effects are therefore subdominant for the two lepton-energy moments as well. The histograms for the variables $\mathcal{P}_a$, $\chi^2/\mathrm{d.o.f.}$ and $N_\mathrm{params}$, providing a global quantitative measure of the goodness for the continuum limits,  are shown in the top- and bottom-panel of \cref{fig:DMxDq2_pull_a_to_0} respectively for the first and second lepton-energy moment. Similarly to the decay rate, the figure shows that lattice artifacts are almost completely absent ($\mathcal{P}_a<1$), the quality of the fits are good ($\chi^2/\mathrm{d.o.f.}<1$ in more than half of the cases and $\chi^2/\mathrm{d.o.f.}>2$ only in less than 10\% of the cases for the second lepton-energy moment) and dominated by constant and linear ansatze.

\begin{figure}
    \centering
    \includegraphics[width=\columnwidth]{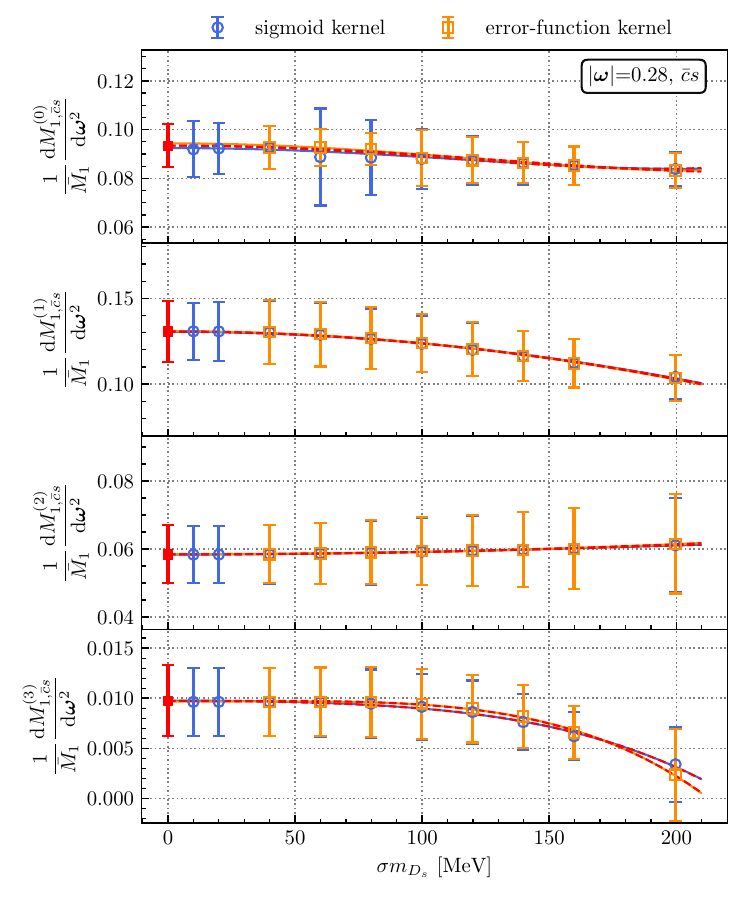}
\caption{$\sigma \mapsto 0$ extrapolation of the  $d M^{(p)}_{1,\bar c s}/d\vec \omega^2$ contribution to the differential first lepton-energy moment for $|\vec \omega|=0.28$. See the analogous \cref{fig:cs_sigma_1}.}  
\label{fig:DMDq2_th6_sigma_to_0}
\end{figure}

\begin{figure}
    \centering
    \includegraphics[width=\columnwidth]{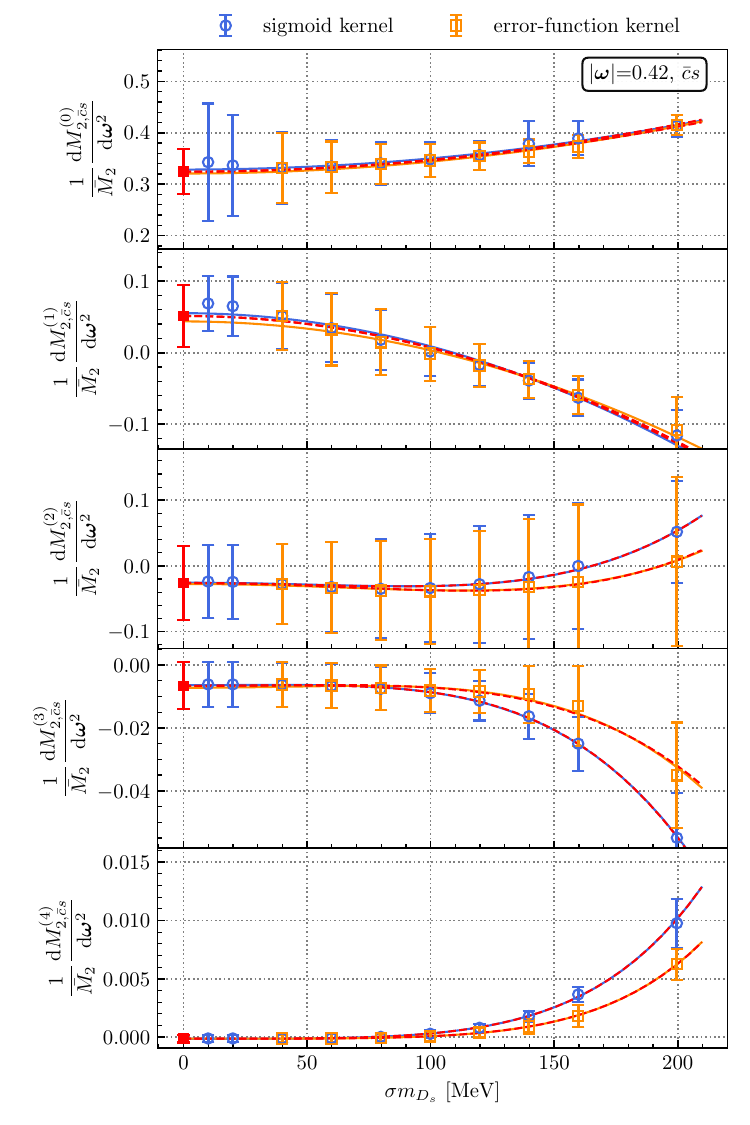}
\caption{$\sigma \mapsto 0$ extrapolation of the  $d M^{(p)}_{2,\bar c s}/d\vec \omega^2$ contribution to the differential second lepton-energy moment for $|\vec \omega|=0.42$. See the analogous \cref{fig:cs_sigma_1}.}   
\label{fig:DM2Dq2_th9_sigma_to_0}
\end{figure}

\begin{figure}
    \centering
    \includegraphics[width=\columnwidth]{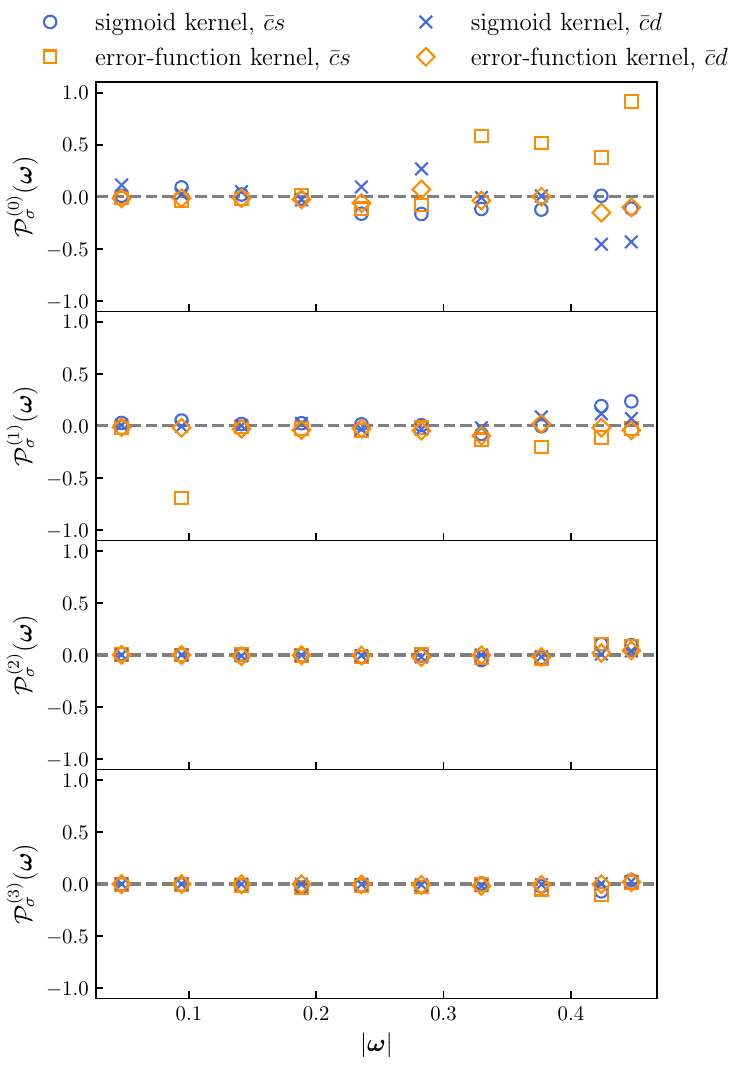}
    \caption{The same as \cref{fig:cs_sigma_pull} for the first lepton-energy moment. Data for the $\bar c s$ and $\bar c d$ are displayed together.}
    \label{fig:DMDq2_pull_sigma_to_0}
\end{figure}

\begin{figure}
    \centering
    \includegraphics[width=\columnwidth]{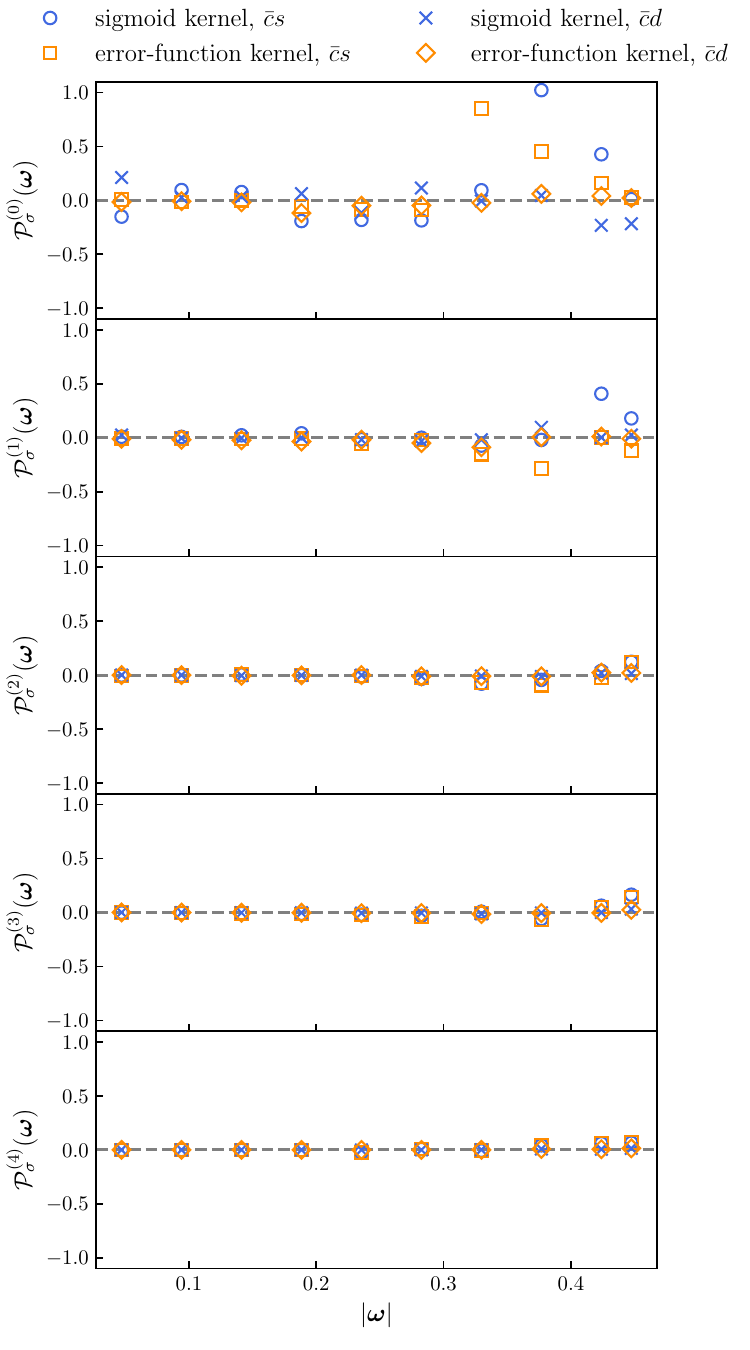}
    \caption{The same as \cref{fig:cs_sigma_pull} for the second lepton-energy moment. Data for the $\bar c s$ and $\bar c d$ are displayed together.}
    \label{fig:DM2Dq2_pull_sigma_to_0}
\end{figure}

Concerning the $\sigma \mapsto 0$ limit, the first lepton-energy moment has an additional contribution labeled by $p=3$ and the second lepton-energy moment has two more labeled by $p=3,4$. According to the asymptotic expansion for small $\sigma$ done in \cref{sec:sigmato0}, for these new contributions we consider the following polynomial fits,
\begin{flalign}\label{eq:sigma_ansatz1}
\frac{d M^{(3),\mathrm{I}}_{1,2}(\sigma)}{d \vec \omega^2}
&=
C_0^{(3),\mathrm{I}}+C_1^{(3),\mathrm{I}}\sigma^4+C_2^{(3),\mathrm{I}}\sigma^6
\end{flalign}
and
\begin{flalign}\label{eq:sigma_ansatz2}
\frac{d M^{(4),\mathrm{I}}_{2}(\sigma)}{d \vec \omega^2}
&=
C_0^{(4),\mathrm{I}}+C_1^{(4),\mathrm{I}}\sigma^6+C_2^{(4),\mathrm{I}}\sigma^8.
\end{flalign}
\Cref{fig:DMDq2_th6_sigma_to_0} shows the $\sigma \mapsto 0$ extrapolation for the quantity $dM^{(p)}_{1,\bar cs}/d\vec{\omega}^2$ in correspondence of $|\vec \omega|=0.28$. \Cref{fig:DM2Dq2_th9_sigma_to_0} shows instead the $\sigma \mapsto 0$ extrapolation for the quantity $dM^{(p)}_{2,\bar cs}/d\vec{\omega}^2$ for $|\vec \omega|=0.42$.  As can be appreciated in both the figures, the fit ansatz proposed above excellently reproduces the trend of the data points. The pull variables $\mathcal{P}_\sigma^{(p)}$ are shown in \cref{fig:DMDq2_pull_sigma_to_0,fig:DM2Dq2_pull_sigma_to_0} for the first and second lepton-energy moment respectively. Again, $\mathcal{P}_\sigma^{(p)}$ is very small and, besides few exceptions,$|\mathcal{P}_\sigma^{(p)}|<0.5$ in all the cases showing the goodness of the $\sigma \mapsto 0$ extrapolations.

\begin{figure}
    \centering
    \includegraphics[width=\columnwidth]{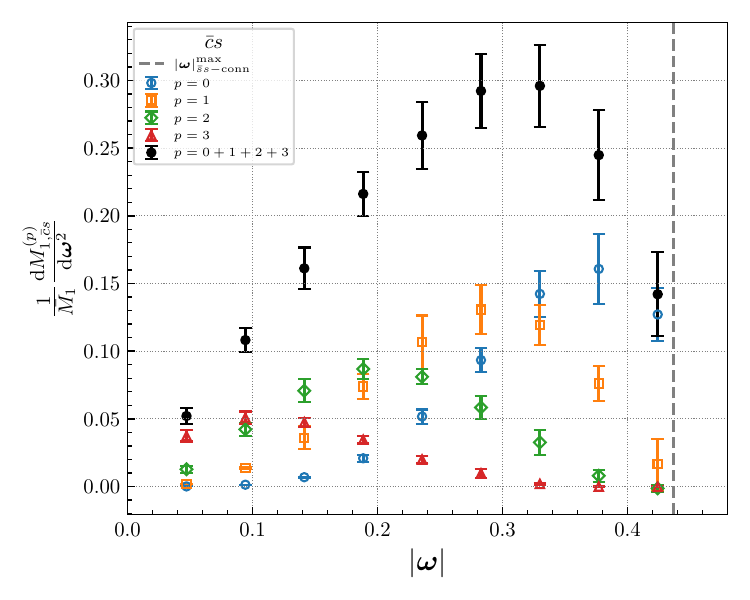}
    \includegraphics[width=\columnwidth]{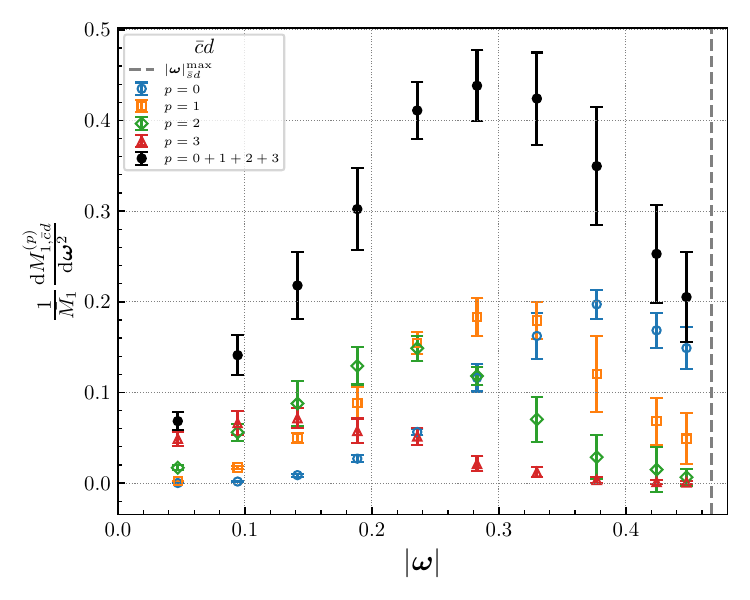}
    \caption{$d M_{1,\bar c s}/d\vec \omega^2$ (top-panel) and $d M_{1,\bar c d}/d\vec \omega^2$ (bottom-panel) contributions to the first lepton-energy moment. See the analogous \cref{fig:cs_final}. The central values and the associated errors of the black points are listed in the column denoted by $\bar c s$ (top-panel) and $\bar c d$ (bottom-panel) of~\cref{tab:numbers_M1}.}
    \label{fig:DMDq2_final}
\end{figure}

\begin{figure}
    \centering
    \includegraphics[width=\columnwidth]{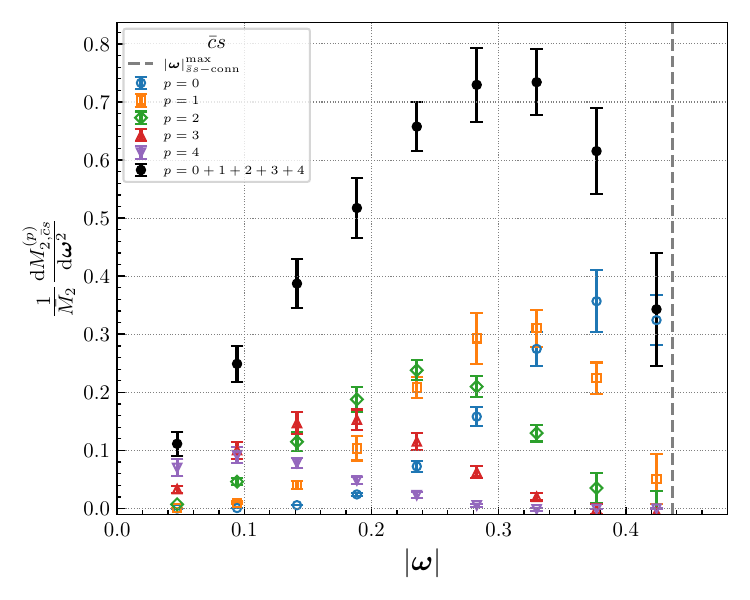}
    \includegraphics[width=\columnwidth]{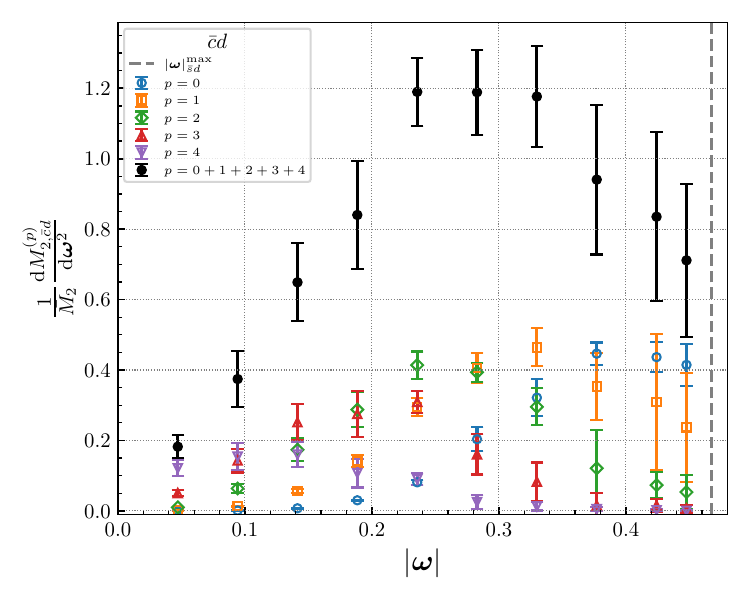}
    \caption{$d M_{2,\bar c s}/d\vec \omega^2$ (top-panel) and $d M_{2,\bar c d}/d\vec \omega^2$ (bottom-panel) contributions to the second lepton-energy moment. See the analogous \cref{fig:cs_final}.  The central values and the associated errors of the black points are listed in the column denoted by $\bar c s$ (top-panel) and $\bar c d$ (bottom-panel) of~\cref{tab:numbers_M2}.}
    \label{fig:DM2Dq2_final}
\end{figure}

\begin{figure}
    \centering
    \includegraphics[width=\columnwidth]{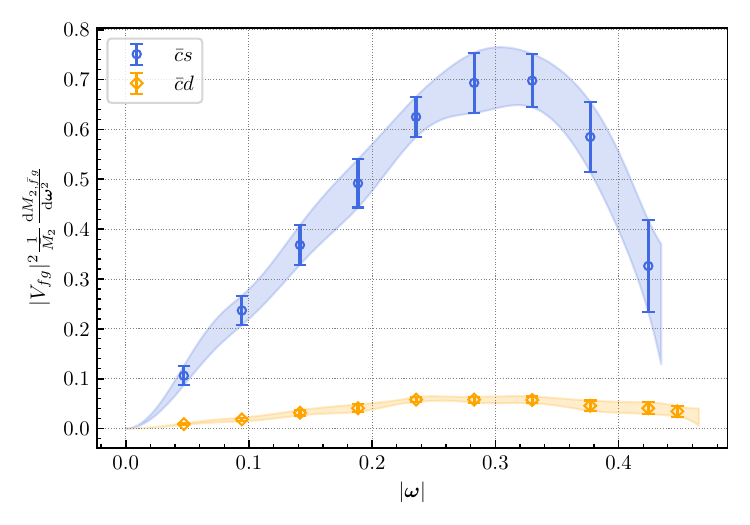}
    \caption{
    The same as \cref{fig:final_differential_DMDq2} but for the second lepton-energy moment.
    \label{fig:final_differential_DM2Dq2}
    }
\end{figure}

\begin{figure}
    \centering
    \includegraphics[width=\columnwidth]{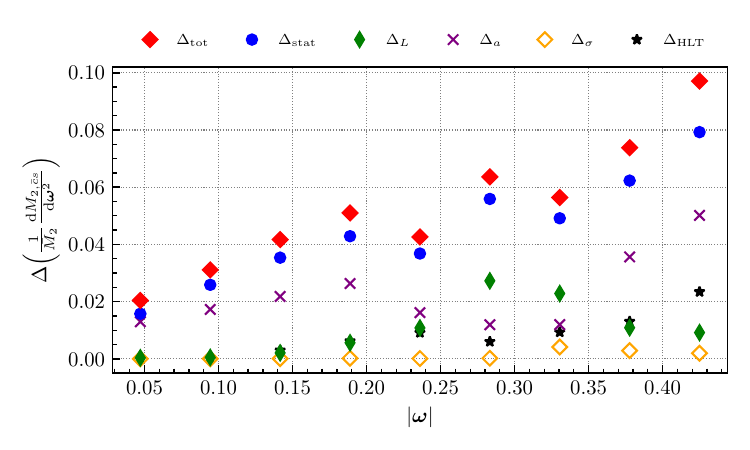}
    \includegraphics[width=\columnwidth]{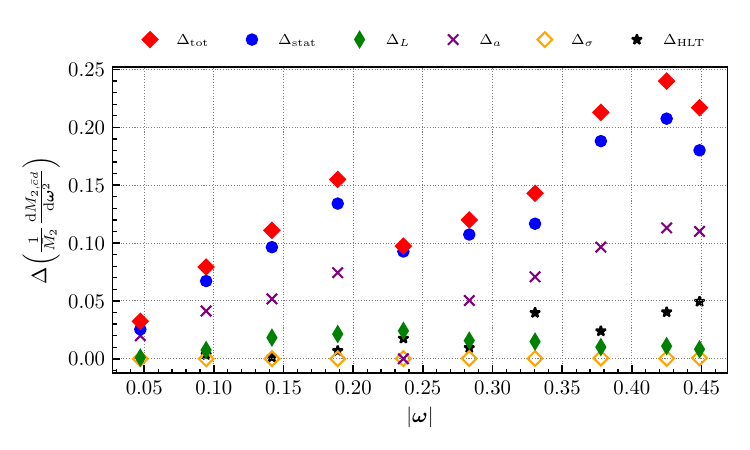}
    \caption{Error budgets of the second lepton-energy moment for the channels $\bar c s$ (top-panel) and $\bar c d$ (bottom-panel), analogous to \cref{fig:DMDq2_error_budget}.}
    \label{fig:DM2Dq2_error_budget}
\end{figure}

\Cref{fig:DMDq2_final} shows the quantity $d M_{1,\bar f g}/d\vec \omega^2$ for the $\bar cs$ (top-panel) and $\bar c d$ (bottom-panel) channels. The analogous plot for the second lepton-energy moment is displayed in \cref{fig:DM2Dq2_final}. Our final results for the differential lepton energy moments are shown in \cref{fig:final_differential_DMDq2,fig:final_differential_DM2Dq2}, respectively. Finally, \cref{fig:DMDq2_error_budget} shows the error budget of $d M_{1,\bar f g}/d\vec \omega^2$ for the $\bar c s$ (top-panel) and $\bar cd$ (bottom-panel) channels. The corresponding plots for the second lepton-energy moment are shown in \cref{fig:DM2Dq2_error_budget}. Analogously to what we found in the case of the decay rate, also for lepton-energy moments the main source of uncertainty is the statistical one.


\section{Experimental measurements of the branching-ratios and the decay rate
\label{sec:exp}}

The experimental decay rate is given by
\begin{flalign}
    \Gamma^\mathrm{exp}=\Gamma^\mathrm{exp}_\mathrm{tot}\cdot \mathcal{B}^\mathrm{exp}(D_s^+\mapsto X e^+ \nu_e),
\end{flalign}
where $\mathcal{B}^\mathrm{exp}(D_s^+\mapsto X e^+ \nu_e)$ is the experimental branching-ratio for the semileptonic mode and $\Gamma^\mathrm{exp}_\mathrm{tot}$ is the total decay rate of the $D_s$ meson. By using for the mean lifetime the of $D_s$ meson  the value $\tau=501.2(2.2)\times 10^{-15}$~s from Ref.~\cite{ParticleDataGroup:2024cfk}, we obtain
\begin{flalign}
    \Gamma^\mathrm{exp}_\mathrm{tot}=\frac{\hbar c}{\tau} = 131.33(58) \times 10^{-14}\,\mathrm{GeV},
\end{flalign}
with $\hbar c=6.5821\times 10^{-25}\,\mathrm{GeV}\times \mathrm{s}$. Currently, the available experimental branching-ratios are measured by the CLEO collaboration Ref.~\cite{CLEO:2009uah} and by the BES-III collaboration Ref~\cite{BESIII:2021duu}
\begin{flalign}
    &  \mathcal{B}_\mathrm{CLEO}^\mathrm{exp}(D_s^+\mapsto X e^+ \nu_e) = 6.52(39)(15)\%\,,
    \\[4pt]
    &  \mathcal{B}_\mathrm{BES-III}^\mathrm{exp}(D_s^+\mapsto X e^+ \nu_e) = 6.30(13)(10)\%\,,
    \\[4pt]
    &  \mathcal{B}_\mathrm{average}^\mathrm{exp}(D_s^+\mapsto X e^+ \nu_e) = 6.33(15)\%\,,
\end{flalign}
where the average has been taken from Ref.~\cite{ParticleDataGroup:2024cfk}.
The corresponding decay rates for the inclusive semileptonic channel are given in \cref{eq:expresults}.

\FloatBarrier

\bibliography{incds}% Produces the bibliography via BibTeX.

\begin{thebibliography}{62}%
\makeatletter
\providecommand \@ifxundefined [1]{%
 \@ifx{#1\undefined}
}%
\providecommand \@ifnum [1]{%
 \ifnum #1\expandafter \@firstoftwo
 \else \expandafter \@secondoftwo
 \fi
}%
\providecommand \@ifx [1]{%
 \ifx #1\expandafter \@firstoftwo
 \else \expandafter \@secondoftwo
 \fi
}%
\providecommand \natexlab [1]{#1}%
\providecommand \enquote  [1]{``#1''}%
\providecommand \bibnamefont  [1]{#1}%
\providecommand \bibfnamefont [1]{#1}%
\providecommand \citenamefont [1]{#1}%
\providecommand \href@noop [0]{\@secondoftwo}%
\providecommand \href [0]{\begingroup \@sanitize@url \@href}%
\providecommand \@href[1]{\@@startlink{#1}\@@href}%
\providecommand \@@href[1]{\endgroup#1\@@endlink}%
\providecommand \@sanitize@url [0]{\catcode `\\12\catcode `\$12\catcode `\&12\catcode `\#12\catcode `\^12\catcode `\_12\catcode `\%12\relax}%
\providecommand \@@startlink[1]{}%
\providecommand \@@endlink[0]{}%
\providecommand \url  [0]{\begingroup\@sanitize@url \@url }%
\providecommand \@url [1]{\endgroup\@href {#1}{\urlprefix }}%
\providecommand \urlprefix  [0]{URL }%
\providecommand \Eprint [0]{\href }%
\providecommand \doibase [0]{https://doi.org/}%
\providecommand \selectlanguage [0]{\@gobble}%
\providecommand \bibinfo  [0]{\@secondoftwo}%
\providecommand \bibfield  [0]{\@secondoftwo}%
\providecommand \translation [1]{[#1]}%
\providecommand \BibitemOpen [0]{}%
\providecommand \bibitemStop [0]{}%
\providecommand \bibitemNoStop [0]{.\EOS\space}%
\providecommand \EOS [0]{\spacefactor3000\relax}%
\providecommand \BibitemShut  [1]{\csname bibitem#1\endcsname}%
\let\auto@bib@innerbib\@empty
%</preamble>
\bibitem [{\citenamefont {De~Santis}\ \emph {et~al.}(2025{\natexlab{a}})\citenamefont {De~Santis} \emph {et~al.}}]{DeSantis:2025yfm}%
  \BibitemOpen
  \bibfield  {author} {\bibinfo {author} {\bibfnamefont {A.}~\bibnamefont {De~Santis}} \emph {et~al.},\ }\bibfield  {title} {\bibinfo {title} {{Inclusive Semileptonic Decays of the Ds Meson: Lattice QCD Confronts Experiments}},\ }\href {https://doi.org/10.1103/snc6-cpz6} {\bibfield  {journal} {\bibinfo  {journal} {Phys. Rev. Lett.}\ }\textbf {\bibinfo {volume} {135}},\ \bibinfo {pages} {121901} (\bibinfo {year} {2025}{\natexlab{a}})},\ \Eprint {https://arxiv.org/abs/2504.06064} {arXiv:2504.06064 [hep-lat]} \BibitemShut {NoStop}%
\bibitem [{\citenamefont {Aoki}\ \emph {et~al.}(2024)\citenamefont {Aoki} \emph {et~al.}}]{FlavourLatticeAveragingGroupFLAG:2024oxs}%
  \BibitemOpen
  \bibfield  {author} {\bibinfo {author} {\bibfnamefont {Y.}~\bibnamefont {Aoki}} \emph {et~al.} (\bibinfo {collaboration} {Flavour Lattice Averaging Group (FLAG)}),\ }\bibfield  {title} {\bibinfo {title} {{FLAG Review 2024}},\ }\href@noop {} {\  (\bibinfo {year} {2024})},\ \Eprint {https://arxiv.org/abs/2411.04268} {arXiv:2411.04268 [hep-lat]} \BibitemShut {NoStop}%
\bibitem [{\citenamefont {Asner}\ \emph {et~al.}(2010)\citenamefont {Asner} \emph {et~al.}}]{CLEO:2009uah}%
  \BibitemOpen
  \bibfield  {author} {\bibinfo {author} {\bibfnamefont {D.~M.}\ \bibnamefont {Asner}} \emph {et~al.} (\bibinfo {collaboration} {CLEO}),\ }\bibfield  {title} {\bibinfo {title} {{Measurement of absolute branching fractions of inclusive semileptonic decays of charm and charmed-strange mesons}},\ }\href {https://doi.org/10.1103/PhysRevD.81.052007} {\bibfield  {journal} {\bibinfo  {journal} {Phys. Rev. D}\ }\textbf {\bibinfo {volume} {81}},\ \bibinfo {pages} {052007} (\bibinfo {year} {2010})},\ \Eprint {https://arxiv.org/abs/0912.4232} {arXiv:0912.4232 [hep-ex]} \BibitemShut {NoStop}%
\bibitem [{\citenamefont {Ablikim}\ \emph {et~al.}(2021)\citenamefont {Ablikim} \emph {et~al.}}]{BESIII:2021duu}%
  \BibitemOpen
  \bibfield  {author} {\bibinfo {author} {\bibfnamefont {M.}~\bibnamefont {Ablikim}} \emph {et~al.} (\bibinfo {collaboration} {BESIII}),\ }\bibfield  {title} {\bibinfo {title} {{Measurement of the absolute branching fraction of inclusive semielectronic $D_s^+$ decays}},\ }\href {https://doi.org/10.1103/PhysRevD.104.012003} {\bibfield  {journal} {\bibinfo  {journal} {Phys. Rev. D}\ }\textbf {\bibinfo {volume} {104}},\ \bibinfo {pages} {012003} (\bibinfo {year} {2021})},\ \Eprint {https://arxiv.org/abs/2104.07311} {arXiv:2104.07311 [hep-ex]} \BibitemShut {NoStop}%
\bibitem [{\citenamefont {Barata}\ and\ \citenamefont {Fredenhagen}(1991)}]{Barata:1990rn}%
  \BibitemOpen
  \bibfield  {author} {\bibinfo {author} {\bibfnamefont {J.~C.~A.}\ \bibnamefont {Barata}}\ and\ \bibinfo {author} {\bibfnamefont {K.}~\bibnamefont {Fredenhagen}},\ }\bibfield  {title} {\bibinfo {title} {{Particle scattering in Euclidean lattice field theories}},\ }\href {https://doi.org/10.1007/BF02102039} {\bibfield  {journal} {\bibinfo  {journal} {Commun. Math. Phys.}\ }\textbf {\bibinfo {volume} {138}},\ \bibinfo {pages} {507} (\bibinfo {year} {1991})}\BibitemShut {NoStop}%
\bibitem [{\citenamefont {Patella}\ and\ \citenamefont {Tantalo}(2025)}]{Patella:2024cto}%
  \BibitemOpen
  \bibfield  {author} {\bibinfo {author} {\bibfnamefont {A.}~\bibnamefont {Patella}}\ and\ \bibinfo {author} {\bibfnamefont {N.}~\bibnamefont {Tantalo}},\ }\bibfield  {title} {\bibinfo {title} {{Scattering amplitudes from Euclidean correlators: Haag-Ruelle theory and approximation formulae}},\ }\href {https://doi.org/10.1007/JHEP01(2025)091} {\bibfield  {journal} {\bibinfo  {journal} {JHEP}\ }\textbf {\bibinfo {volume} {01}},\ \bibinfo {pages} {091}},\ \Eprint {https://arxiv.org/abs/2407.02069} {arXiv:2407.02069 [hep-lat]} \BibitemShut {NoStop}%
\bibitem [{\citenamefont {Hansen}\ \emph {et~al.}(2017)\citenamefont {Hansen}, \citenamefont {Meyer},\ and\ \citenamefont {Robaina}}]{Hansen:2017mnd}%
  \BibitemOpen
  \bibfield  {author} {\bibinfo {author} {\bibfnamefont {M.~T.}\ \bibnamefont {Hansen}}, \bibinfo {author} {\bibfnamefont {H.~B.}\ \bibnamefont {Meyer}},\ and\ \bibinfo {author} {\bibfnamefont {D.}~\bibnamefont {Robaina}},\ }\bibfield  {title} {\bibinfo {title} {{From deep inelastic scattering to heavy-flavor semileptonic decays: Total rates into multihadron final states from lattice QCD}},\ }\href {https://doi.org/10.1103/PhysRevD.96.094513} {\bibfield  {journal} {\bibinfo  {journal} {Phys. Rev. D}\ }\textbf {\bibinfo {volume} {96}},\ \bibinfo {pages} {094513} (\bibinfo {year} {2017})},\ \Eprint {https://arxiv.org/abs/1704.08993} {arXiv:1704.08993 [hep-lat]} \BibitemShut {NoStop}%
\bibitem [{\citenamefont {Hashimoto}(2017)}]{Hashimoto:2017wqo}%
  \BibitemOpen
  \bibfield  {author} {\bibinfo {author} {\bibfnamefont {S.}~\bibnamefont {Hashimoto}},\ }\bibfield  {title} {\bibinfo {title} {{Inclusive semi-leptonic B meson decay structure functions from lattice QCD}},\ }\href {https://doi.org/10.1093/ptep/ptx052} {\bibfield  {journal} {\bibinfo  {journal} {PTEP}\ }\textbf {\bibinfo {volume} {2017}},\ \bibinfo {pages} {053B03} (\bibinfo {year} {2017})},\ \Eprint {https://arxiv.org/abs/1703.01881} {arXiv:1703.01881 [hep-lat]} \BibitemShut {NoStop}%
\bibitem [{\citenamefont {Hansen}\ \emph {et~al.}(2019)\citenamefont {Hansen}, \citenamefont {Lupo},\ and\ \citenamefont {Tantalo}}]{Hansen:2019idp}%
  \BibitemOpen
  \bibfield  {author} {\bibinfo {author} {\bibfnamefont {M.}~\bibnamefont {Hansen}}, \bibinfo {author} {\bibfnamefont {A.}~\bibnamefont {Lupo}},\ and\ \bibinfo {author} {\bibfnamefont {N.}~\bibnamefont {Tantalo}},\ }\bibfield  {title} {\bibinfo {title} {{Extraction of spectral densities from lattice correlators}},\ }\href {https://doi.org/10.1103/PhysRevD.99.094508} {\bibfield  {journal} {\bibinfo  {journal} {Phys. Rev. D}\ }\textbf {\bibinfo {volume} {99}},\ \bibinfo {pages} {094508} (\bibinfo {year} {2019})},\ \Eprint {https://arxiv.org/abs/1903.06476} {arXiv:1903.06476 [hep-lat]} \BibitemShut {NoStop}%
\bibitem [{\citenamefont {Gambino}\ and\ \citenamefont {Hashimoto}(2020)}]{Gambino:2020crt}%
  \BibitemOpen
  \bibfield  {author} {\bibinfo {author} {\bibfnamefont {P.}~\bibnamefont {Gambino}}\ and\ \bibinfo {author} {\bibfnamefont {S.}~\bibnamefont {Hashimoto}},\ }\bibfield  {title} {\bibinfo {title} {{Inclusive Semileptonic Decays from Lattice QCD}},\ }\href {https://doi.org/10.1103/PhysRevLett.125.032001} {\bibfield  {journal} {\bibinfo  {journal} {Phys. Rev. Lett.}\ }\textbf {\bibinfo {volume} {125}},\ \bibinfo {pages} {032001} (\bibinfo {year} {2020})},\ \Eprint {https://arxiv.org/abs/2005.13730} {arXiv:2005.13730 [hep-lat]} \BibitemShut {NoStop}%
\bibitem [{\citenamefont {Gambino}\ \emph {et~al.}(2022)\citenamefont {Gambino}, \citenamefont {Hashimoto}, \citenamefont {M{\"a}chler}, \citenamefont {Panero}, \citenamefont {Sanfilippo}, \citenamefont {Simula}, \citenamefont {Smecca},\ and\ \citenamefont {Tantalo}}]{Gambino:2022dvu}%
  \BibitemOpen
  \bibfield  {author} {\bibinfo {author} {\bibfnamefont {P.}~\bibnamefont {Gambino}}, \bibinfo {author} {\bibfnamefont {S.}~\bibnamefont {Hashimoto}}, \bibinfo {author} {\bibfnamefont {S.}~\bibnamefont {M{\"a}chler}}, \bibinfo {author} {\bibfnamefont {M.}~\bibnamefont {Panero}}, \bibinfo {author} {\bibfnamefont {F.}~\bibnamefont {Sanfilippo}}, \bibinfo {author} {\bibfnamefont {S.}~\bibnamefont {Simula}}, \bibinfo {author} {\bibfnamefont {A.}~\bibnamefont {Smecca}},\ and\ \bibinfo {author} {\bibfnamefont {N.}~\bibnamefont {Tantalo}},\ }\bibfield  {title} {\bibinfo {title} {{Lattice QCD study of inclusive semileptonic decays of heavy mesons}},\ }\href {https://doi.org/10.1007/JHEP07(2022)083} {\bibfield  {journal} {\bibinfo  {journal} {JHEP}\ }\textbf {\bibinfo {volume} {07}},\ \bibinfo {pages} {083}},\ \Eprint {https://arxiv.org/abs/2203.11762} {arXiv:2203.11762 [hep-lat]} \BibitemShut {NoStop}%
\bibitem [{\citenamefont {Manohar}\ and\ \citenamefont {Wise}(1994)}]{Manohar:1993qn}%
  \BibitemOpen
  \bibfield  {author} {\bibinfo {author} {\bibfnamefont {A.~V.}\ \bibnamefont {Manohar}}\ and\ \bibinfo {author} {\bibfnamefont {M.~B.}\ \bibnamefont {Wise}},\ }\bibfield  {title} {\bibinfo {title} {{Inclusive semileptonic B and polarized Lambda(b) decays from QCD}},\ }\href {https://doi.org/10.1103/PhysRevD.49.1310} {\bibfield  {journal} {\bibinfo  {journal} {Phys. Rev. D}\ }\textbf {\bibinfo {volume} {49}},\ \bibinfo {pages} {1310} (\bibinfo {year} {1994})},\ \Eprint {https://arxiv.org/abs/hep-ph/9308246} {arXiv:hep-ph/9308246} \BibitemShut {NoStop}%
\bibitem [{\citenamefont {Blok}\ \emph {et~al.}(1994)\citenamefont {Blok}, \citenamefont {Koyrakh}, \citenamefont {Shifman},\ and\ \citenamefont {Vainshtein}}]{Blok:1993va}%
  \BibitemOpen
  \bibfield  {author} {\bibinfo {author} {\bibfnamefont {B.}~\bibnamefont {Blok}}, \bibinfo {author} {\bibfnamefont {L.}~\bibnamefont {Koyrakh}}, \bibinfo {author} {\bibfnamefont {M.~A.}\ \bibnamefont {Shifman}},\ and\ \bibinfo {author} {\bibfnamefont {A.~I.}\ \bibnamefont {Vainshtein}},\ }\bibfield  {title} {\bibinfo {title} {{Differential distributions in semileptonic decays of the heavy flavors in QCD}},\ }\href {https://doi.org/10.1103/PhysRevD.50.3572} {\bibfield  {journal} {\bibinfo  {journal} {Phys. Rev. D}\ }\textbf {\bibinfo {volume} {49}},\ \bibinfo {pages} {3356} (\bibinfo {year} {1994})},\ \bibinfo {note} {[Erratum: Phys.Rev.D 50, 3572 (1994)]},\ \Eprint {https://arxiv.org/abs/hep-ph/9307247} {arXiv:hep-ph/9307247} \BibitemShut {NoStop}%
\bibitem [{\citenamefont {Bigi}\ \emph {et~al.}(1992)\citenamefont {Bigi}, \citenamefont {Uraltsev},\ and\ \citenamefont {Vainshtein}}]{Bigi:1992su}%
  \BibitemOpen
  \bibfield  {author} {\bibinfo {author} {\bibfnamefont {I.~I.~Y.}\ \bibnamefont {Bigi}}, \bibinfo {author} {\bibfnamefont {N.~G.}\ \bibnamefont {Uraltsev}},\ and\ \bibinfo {author} {\bibfnamefont {A.~I.}\ \bibnamefont {Vainshtein}},\ }\bibfield  {title} {\bibinfo {title} {{Nonperturbative corrections to inclusive beauty and charm decays: QCD versus phenomenological models}},\ }\href {https://doi.org/10.1016/0370-2693(92)90908-M} {\bibfield  {journal} {\bibinfo  {journal} {Phys. Lett. B}\ }\textbf {\bibinfo {volume} {293}},\ \bibinfo {pages} {430} (\bibinfo {year} {1992})},\ \bibinfo {note} {[Erratum: Phys.Lett.B 297, 477--477 (1992)]},\ \Eprint {https://arxiv.org/abs/hep-ph/9207214} {arXiv:hep-ph/9207214} \BibitemShut {NoStop}%
\bibitem [{\citenamefont {Bigi}\ \emph {et~al.}(1993)\citenamefont {Bigi}, \citenamefont {Shifman}, \citenamefont {Uraltsev},\ and\ \citenamefont {Vainshtein}}]{Bigi:1993fe}%
  \BibitemOpen
  \bibfield  {author} {\bibinfo {author} {\bibfnamefont {I.~I.~Y.}\ \bibnamefont {Bigi}}, \bibinfo {author} {\bibfnamefont {M.~A.}\ \bibnamefont {Shifman}}, \bibinfo {author} {\bibfnamefont {N.~G.}\ \bibnamefont {Uraltsev}},\ and\ \bibinfo {author} {\bibfnamefont {A.~I.}\ \bibnamefont {Vainshtein}},\ }\bibfield  {title} {\bibinfo {title} {{QCD predictions for lepton spectra in inclusive heavy flavor decays}},\ }\href {https://doi.org/10.1103/PhysRevLett.71.496} {\bibfield  {journal} {\bibinfo  {journal} {Phys. Rev. Lett.}\ }\textbf {\bibinfo {volume} {71}},\ \bibinfo {pages} {496} (\bibinfo {year} {1993})},\ \Eprint {https://arxiv.org/abs/hep-ph/9304225} {arXiv:hep-ph/9304225} \BibitemShut {NoStop}%
\bibitem [{\citenamefont {Chay}\ \emph {et~al.}(1990)\citenamefont {Chay}, \citenamefont {Georgi},\ and\ \citenamefont {Grinstein}}]{Chay:1990da}%
  \BibitemOpen
  \bibfield  {author} {\bibinfo {author} {\bibfnamefont {J.}~\bibnamefont {Chay}}, \bibinfo {author} {\bibfnamefont {H.}~\bibnamefont {Georgi}},\ and\ \bibinfo {author} {\bibfnamefont {B.}~\bibnamefont {Grinstein}},\ }\bibfield  {title} {\bibinfo {title} {{Lepton energy distributions in heavy meson decays from QCD}},\ }\href {https://doi.org/10.1016/0370-2693(90)90916-T} {\bibfield  {journal} {\bibinfo  {journal} {Phys. Lett. B}\ }\textbf {\bibinfo {volume} {247}},\ \bibinfo {pages} {399} (\bibinfo {year} {1990})}\BibitemShut {NoStop}%
\bibitem [{\citenamefont {Alexandrou}\ \emph {et~al.}(2023{\natexlab{a}})\citenamefont {Alexandrou} \emph {et~al.}}]{ExtendedTwistedMassCollaborationETMC:2022sta}%
  \BibitemOpen
  \bibfield  {author} {\bibinfo {author} {\bibfnamefont {C.}~\bibnamefont {Alexandrou}} \emph {et~al.} (\bibinfo {collaboration} {Extended Twisted Mass Collaboration (ETMC)}),\ }\bibfield  {title} {\bibinfo {title} {{Probing the Energy-Smeared R Ratio Using Lattice QCD}},\ }\href {https://doi.org/10.1103/PhysRevLett.130.241901} {\bibfield  {journal} {\bibinfo  {journal} {Phys. Rev. Lett.}\ }\textbf {\bibinfo {volume} {130}},\ \bibinfo {pages} {241901} (\bibinfo {year} {2023}{\natexlab{a}})},\ \Eprint {https://arxiv.org/abs/2212.08467} {arXiv:2212.08467 [hep-lat]} \BibitemShut {NoStop}%
\bibitem [{\citenamefont {Evangelista}\ \emph {et~al.}(2023)\citenamefont {Evangelista}, \citenamefont {Frezzotti}, \citenamefont {Tantalo}, \citenamefont {Gagliardi}, \citenamefont {Sanfilippo}, \citenamefont {Simula},\ and\ \citenamefont {Lubicz}}]{Evangelista:2023fmt}%
  \BibitemOpen
  \bibfield  {author} {\bibinfo {author} {\bibfnamefont {A.}~\bibnamefont {Evangelista}}, \bibinfo {author} {\bibfnamefont {R.}~\bibnamefont {Frezzotti}}, \bibinfo {author} {\bibfnamefont {N.}~\bibnamefont {Tantalo}}, \bibinfo {author} {\bibfnamefont {G.}~\bibnamefont {Gagliardi}}, \bibinfo {author} {\bibfnamefont {F.}~\bibnamefont {Sanfilippo}}, \bibinfo {author} {\bibfnamefont {S.}~\bibnamefont {Simula}},\ and\ \bibinfo {author} {\bibfnamefont {V.}~\bibnamefont {Lubicz}} (\bibinfo {collaboration} {Extended Twisted Mass}),\ }\bibfield  {title} {\bibinfo {title} {{Inclusive hadronic decay rate of the {\ensuremath{\tau}} lepton from lattice QCD}},\ }\href {https://doi.org/10.1103/PhysRevD.108.074513} {\bibfield  {journal} {\bibinfo  {journal} {Phys. Rev. D}\ }\textbf {\bibinfo {volume} {108}},\ \bibinfo {pages} {074513} (\bibinfo {year} {2023})},\ \Eprint {https://arxiv.org/abs/2308.03125} {arXiv:2308.03125 [hep-lat]} \BibitemShut {NoStop}%
\bibitem [{\citenamefont {Alexandrou}\ \emph {et~al.}(2024)\citenamefont {Alexandrou} \emph {et~al.}}]{ExtendedTwistedMass:2024myu}%
  \BibitemOpen
  \bibfield  {author} {\bibinfo {author} {\bibfnamefont {C.}~\bibnamefont {Alexandrou}} \emph {et~al.} (\bibinfo {collaboration} {Extended Twisted Mass}),\ }\bibfield  {title} {\bibinfo {title} {{Inclusive Hadronic Decay Rate of the {\ensuremath{\tau}} Lepton from Lattice QCD: The u{\textasciimacron}s Flavor Channel and the Cabibbo Angle}},\ }\href {https://doi.org/10.1103/PhysRevLett.132.261901} {\bibfield  {journal} {\bibinfo  {journal} {Phys. Rev. Lett.}\ }\textbf {\bibinfo {volume} {132}},\ \bibinfo {pages} {261901} (\bibinfo {year} {2024})},\ \Eprint {https://arxiv.org/abs/2403.05404} {arXiv:2403.05404 [hep-lat]} \BibitemShut {NoStop}%
\bibitem [{\citenamefont {Barone}\ \emph {et~al.}(2023{\natexlab{a}})\citenamefont {Barone}, \citenamefont {J{\"u}ttner}, \citenamefont {Hashimoto}, \citenamefont {Kaneko},\ and\ \citenamefont {Kellermann}}]{Barone:2022gkn}%
  \BibitemOpen
  \bibfield  {author} {\bibinfo {author} {\bibfnamefont {A.}~\bibnamefont {Barone}}, \bibinfo {author} {\bibfnamefont {A.}~\bibnamefont {J{\"u}ttner}}, \bibinfo {author} {\bibfnamefont {S.}~\bibnamefont {Hashimoto}}, \bibinfo {author} {\bibfnamefont {T.}~\bibnamefont {Kaneko}},\ and\ \bibinfo {author} {\bibfnamefont {R.}~\bibnamefont {Kellermann}},\ }\bibfield  {title} {\bibinfo {title} {{Inclusive semi-leptonic $B_{(s)}$ mesons decay at the physical $b$ quark mass}},\ }\href {https://doi.org/10.22323/1.430.0403} {\bibfield  {journal} {\bibinfo  {journal} {PoS}\ }\textbf {\bibinfo {volume} {LATTICE2022}},\ \bibinfo {pages} {403} (\bibinfo {year} {2023}{\natexlab{a}})},\ \Eprint {https://arxiv.org/abs/2211.15623} {arXiv:2211.15623 [hep-lat]} \BibitemShut {NoStop}%
\bibitem [{\citenamefont {Kellermann}\ \emph {et~al.}(2023)\citenamefont {Kellermann}, \citenamefont {Barone}, \citenamefont {Hashimoto}, \citenamefont {J{\"u}ttner},\ and\ \citenamefont {Kaneko}}]{Kellermann:2022mms}%
  \BibitemOpen
  \bibfield  {author} {\bibinfo {author} {\bibfnamefont {R.}~\bibnamefont {Kellermann}}, \bibinfo {author} {\bibfnamefont {A.}~\bibnamefont {Barone}}, \bibinfo {author} {\bibfnamefont {S.}~\bibnamefont {Hashimoto}}, \bibinfo {author} {\bibfnamefont {A.}~\bibnamefont {J{\"u}ttner}},\ and\ \bibinfo {author} {\bibfnamefont {T.}~\bibnamefont {Kaneko}},\ }\bibfield  {title} {\bibinfo {title} {{Inclusive semi-leptonic decays of charmed mesons with M{\"o}bius domain wall fermions}},\ }\href {https://doi.org/10.22323/1.430.0414} {\bibfield  {journal} {\bibinfo  {journal} {PoS}\ }\textbf {\bibinfo {volume} {LATTICE2022}},\ \bibinfo {pages} {414} (\bibinfo {year} {2023})},\ \Eprint {https://arxiv.org/abs/2211.16830} {arXiv:2211.16830 [hep-lat]} \BibitemShut {NoStop}%
\bibitem [{\citenamefont {Barone}\ \emph {et~al.}(2023{\natexlab{b}})\citenamefont {Barone}, \citenamefont {Hashimoto}, \citenamefont {J{\"u}ttner}, \citenamefont {Kaneko},\ and\ \citenamefont {Kellermann}}]{Barone:2023tbl}%
  \BibitemOpen
  \bibfield  {author} {\bibinfo {author} {\bibfnamefont {A.}~\bibnamefont {Barone}}, \bibinfo {author} {\bibfnamefont {S.}~\bibnamefont {Hashimoto}}, \bibinfo {author} {\bibfnamefont {A.}~\bibnamefont {J{\"u}ttner}}, \bibinfo {author} {\bibfnamefont {T.}~\bibnamefont {Kaneko}},\ and\ \bibinfo {author} {\bibfnamefont {R.}~\bibnamefont {Kellermann}},\ }\bibfield  {title} {\bibinfo {title} {{Approaches to inclusive semileptonic B$_{(s)}$-meson decays from Lattice QCD}},\ }\href {https://doi.org/10.1007/JHEP07(2023)145} {\bibfield  {journal} {\bibinfo  {journal} {JHEP}\ }\textbf {\bibinfo {volume} {07}},\ \bibinfo {pages} {145}},\ \Eprint {https://arxiv.org/abs/2305.14092} {arXiv:2305.14092 [hep-lat]} \BibitemShut {NoStop}%
\bibitem [{\citenamefont {Kellermann}\ \emph {et~al.}(2024{\natexlab{a}})\citenamefont {Kellermann}, \citenamefont {Barone}, \citenamefont {Hashimoto}, \citenamefont {J{\"u}ttner{\ensuremath{\mathit{c}}}},\ and\ \citenamefont {Kaneko{\ensuremath{\mathit{a}}}}}]{Kellermann:2023yec}%
  \BibitemOpen
  \bibfield  {author} {\bibinfo {author} {\bibfnamefont {R.}~\bibnamefont {Kellermann}}, \bibinfo {author} {\bibfnamefont {A.}~\bibnamefont {Barone}}, \bibinfo {author} {\bibfnamefont {S.}~\bibnamefont {Hashimoto}}, \bibinfo {author} {\bibfnamefont {A.}~\bibnamefont {J{\"u}ttner{\ensuremath{\mathit{c}}}}},\ and\ \bibinfo {author} {\bibfnamefont {T.}~\bibnamefont {Kaneko{\ensuremath{\mathit{a}}}}},\ }\bibfield  {title} {\bibinfo {title} {{Studies on finite-volume effects in the inclusive semileptonic decays of charmed mesons}},\ }\href {https://doi.org/10.22323/1.453.0272} {\bibfield  {journal} {\bibinfo  {journal} {PoS}\ }\textbf {\bibinfo {volume} {LATTICE2023}},\ \bibinfo {pages} {272} (\bibinfo {year} {2024}{\natexlab{a}})},\ \Eprint {https://arxiv.org/abs/2312.16442} {arXiv:2312.16442 [hep-lat]} \BibitemShut {NoStop}%
\bibitem [{\citenamefont {Barone}\ \emph {et~al.}(2024)\citenamefont {Barone}, \citenamefont {Hashimoto}, \citenamefont {J{\"u}ttner}, \citenamefont {Kaneko},\ and\ \citenamefont {Kellermann}}]{Barone:2023iat}%
  \BibitemOpen
  \bibfield  {author} {\bibinfo {author} {\bibfnamefont {A.}~\bibnamefont {Barone}}, \bibinfo {author} {\bibfnamefont {S.}~\bibnamefont {Hashimoto}}, \bibinfo {author} {\bibfnamefont {A.}~\bibnamefont {J{\"u}ttner}}, \bibinfo {author} {\bibfnamefont {T.}~\bibnamefont {Kaneko}},\ and\ \bibinfo {author} {\bibfnamefont {R.}~\bibnamefont {Kellermann}},\ }\bibfield  {title} {\bibinfo {title} {{Chebyshev and Backus-Gilbert reconstruction for inclusive semileptonic $B_{(s)}$-meson decays from Lattice QCD}},\ }\href {https://doi.org/10.22323/1.453.0236} {\bibfield  {journal} {\bibinfo  {journal} {PoS}\ }\textbf {\bibinfo {volume} {LATTICE2023}},\ \bibinfo {pages} {236} (\bibinfo {year} {2024})},\ \Eprint {https://arxiv.org/abs/2312.17401} {arXiv:2312.17401 [hep-lat]} \BibitemShut {NoStop}%
\bibitem [{\citenamefont {Kellermann}\ \emph {et~al.}(2024{\natexlab{b}})\citenamefont {Kellermann}, \citenamefont {Barone}, \citenamefont {Hashimoto}, \citenamefont {J{\"u}ttner},\ and\ \citenamefont {Kaneko}}]{Kellermann:2024zfy}%
  \BibitemOpen
  \bibfield  {author} {\bibinfo {author} {\bibfnamefont {R.}~\bibnamefont {Kellermann}}, \bibinfo {author} {\bibfnamefont {A.}~\bibnamefont {Barone}}, \bibinfo {author} {\bibfnamefont {S.}~\bibnamefont {Hashimoto}}, \bibinfo {author} {\bibfnamefont {A.}~\bibnamefont {J{\"u}ttner}},\ and\ \bibinfo {author} {\bibfnamefont {T.}~\bibnamefont {Kaneko}},\ }\bibfield  {title} {\bibinfo {title} {{Updates on inclusive charmed and bottomed meson decays from the lattice}},\ }in\ \href@noop {} {\emph {\bibinfo {booktitle} {{12th International Workshop on the CKM Unitarity Triangle}}}}\ (\bibinfo {year} {2024})\ \Eprint {https://arxiv.org/abs/2405.06152} {arXiv:2405.06152 [hep-lat]} \BibitemShut {NoStop}%
\bibitem [{\citenamefont {Hashimoto}(2024)}]{Hashimoto:2024pnd}%
  \BibitemOpen
  \bibfield  {author} {\bibinfo {author} {\bibfnamefont {S.}~\bibnamefont {Hashimoto}},\ }\bibfield  {title} {\bibinfo {title} {{Towards the understanding of the inclusive vs exclusive puzzles in the |Vcb| determinations}},\ }\href {https://doi.org/10.22323/1.451.0012} {\bibfield  {journal} {\bibinfo  {journal} {PoS}\ }\textbf {\bibinfo {volume} {EuroPLEx2023}},\ \bibinfo {pages} {012} (\bibinfo {year} {2024})},\ \Eprint {https://arxiv.org/abs/2406.04579} {arXiv:2406.04579 [hep-lat]} \BibitemShut {NoStop}%
\bibitem [{\citenamefont {Kellermann}\ \emph {et~al.}(2025)\citenamefont {Kellermann}, \citenamefont {Hu}, \citenamefont {Barone}, \citenamefont {Elgaziari}, \citenamefont {Hashimoto}, \citenamefont {Kaneko},\ and\ \citenamefont {J{\"u}ttner}}]{Kellermann:2025pzt}%
  \BibitemOpen
  \bibfield  {author} {\bibinfo {author} {\bibfnamefont {R.}~\bibnamefont {Kellermann}}, \bibinfo {author} {\bibfnamefont {Z.}~\bibnamefont {Hu}}, \bibinfo {author} {\bibfnamefont {A.}~\bibnamefont {Barone}}, \bibinfo {author} {\bibfnamefont {A.}~\bibnamefont {Elgaziari}}, \bibinfo {author} {\bibfnamefont {S.}~\bibnamefont {Hashimoto}}, \bibinfo {author} {\bibfnamefont {T.}~\bibnamefont {Kaneko}},\ and\ \bibinfo {author} {\bibfnamefont {A.}~\bibnamefont {J{\"u}ttner}},\ }\bibfield  {title} {\bibinfo {title} {{Inclusive semileptonic decays from lattice QCD: Analysis of systematic effects}},\ }\href {https://doi.org/10.1103/kltp-1p1f} {\bibfield  {journal} {\bibinfo  {journal} {Phys. Rev. D}\ }\textbf {\bibinfo {volume} {112}},\ \bibinfo {pages} {014501} (\bibinfo {year} {2025})},\ \Eprint {https://arxiv.org/abs/2504.03358} {arXiv:2504.03358 [hep-lat]} \BibitemShut {NoStop}%
\bibitem [{\citenamefont {Sirlin}(1982)}]{Sirlin:1981ie}%
  \BibitemOpen
  \bibfield  {author} {\bibinfo {author} {\bibfnamefont {A.}~\bibnamefont {Sirlin}},\ }\bibfield  {title} {\bibinfo {title} {{Large m(W), m(Z) Behavior of the O(alpha) Corrections to Semileptonic Processes Mediated by W}},\ }\href {https://doi.org/10.1016/0550-3213(82)90303-0} {\bibfield  {journal} {\bibinfo  {journal} {Nucl. Phys. B}\ }\textbf {\bibinfo {volume} {196}},\ \bibinfo {pages} {83} (\bibinfo {year} {1982})}\BibitemShut {NoStop}%
\bibitem [{\citenamefont {Bigi}\ \emph {et~al.}(2023)\citenamefont {Bigi}, \citenamefont {Bordone}, \citenamefont {Gambino}, \citenamefont {Haisch},\ and\ \citenamefont {Piccione}}]{Bigi:2023cbv}%
  \BibitemOpen
  \bibfield  {author} {\bibinfo {author} {\bibfnamefont {D.}~\bibnamefont {Bigi}}, \bibinfo {author} {\bibfnamefont {M.}~\bibnamefont {Bordone}}, \bibinfo {author} {\bibfnamefont {P.}~\bibnamefont {Gambino}}, \bibinfo {author} {\bibfnamefont {U.}~\bibnamefont {Haisch}},\ and\ \bibinfo {author} {\bibfnamefont {A.}~\bibnamefont {Piccione}},\ }\bibfield  {title} {\bibinfo {title} {{QED effects in inclusive semi-leptonic B decays}},\ }\href {https://doi.org/10.1007/JHEP11(2023)163} {\bibfield  {journal} {\bibinfo  {journal} {JHEP}\ }\textbf {\bibinfo {volume} {11}},\ \bibinfo {pages} {163}},\ \bibinfo {note} {[Erratum: JHEP 03, 078 (2025)]},\ \Eprint {https://arxiv.org/abs/2309.02849} {arXiv:2309.02849 [hep-ph]} \BibitemShut {NoStop}%
\bibitem [{\citenamefont {Alexandrou}\ \emph {et~al.}(2025{\natexlab{a}})\citenamefont {Alexandrou} \emph {et~al.}}]{ExtendedTwistedMass:2024nyi}%
  \BibitemOpen
  \bibfield  {author} {\bibinfo {author} {\bibfnamefont {C.}~\bibnamefont {Alexandrou}} \emph {et~al.} (\bibinfo {collaboration} {Extended Twisted Mass}),\ }\bibfield  {title} {\bibinfo {title} {{Strange and charm quark contributions to the muon anomalous magnetic moment in lattice QCD with twisted-mass fermions}},\ }\href {https://doi.org/10.1103/PhysRevD.111.054502} {\bibfield  {journal} {\bibinfo  {journal} {Phys. Rev. D}\ }\textbf {\bibinfo {volume} {111}},\ \bibinfo {pages} {054502} (\bibinfo {year} {2025}{\natexlab{a}})},\ \Eprint {https://arxiv.org/abs/2411.08852} {arXiv:2411.08852 [hep-lat]} \BibitemShut {NoStop}%
\bibitem [{\citenamefont {Alexandrou}\ \emph {et~al.}(2018)\citenamefont {Alexandrou} \emph {et~al.}}]{Alexandrou:2018egz}%
  \BibitemOpen
  \bibfield  {author} {\bibinfo {author} {\bibfnamefont {C.}~\bibnamefont {Alexandrou}} \emph {et~al.},\ }\bibfield  {title} {\bibinfo {title} {{Simulating twisted mass fermions at physical light, strange and charm quark masses}},\ }\href {https://doi.org/10.1103/PhysRevD.98.054518} {\bibfield  {journal} {\bibinfo  {journal} {Phys. Rev. D}\ }\textbf {\bibinfo {volume} {98}},\ \bibinfo {pages} {054518} (\bibinfo {year} {2018})},\ \Eprint {https://arxiv.org/abs/1807.00495} {arXiv:1807.00495 [hep-lat]} \BibitemShut {NoStop}%
\bibitem [{\citenamefont {Bergner}\ \emph {et~al.}(2020)\citenamefont {Bergner}, \citenamefont {Dimopoulos}, \citenamefont {Finkenrath}, \citenamefont {Fiorenza}, \citenamefont {Frezzotti}, \citenamefont {Garofalo}, \citenamefont {Kostrzewa}, \citenamefont {Sanfilippo}, \citenamefont {Simula},\ and\ \citenamefont {Wenger}}]{ExtendedTwistedMass:2020tvp}%
  \BibitemOpen
  \bibfield  {author} {\bibinfo {author} {\bibfnamefont {G.}~\bibnamefont {Bergner}}, \bibinfo {author} {\bibfnamefont {P.}~\bibnamefont {Dimopoulos}}, \bibinfo {author} {\bibfnamefont {J.}~\bibnamefont {Finkenrath}}, \bibinfo {author} {\bibfnamefont {E.}~\bibnamefont {Fiorenza}}, \bibinfo {author} {\bibfnamefont {R.}~\bibnamefont {Frezzotti}}, \bibinfo {author} {\bibfnamefont {M.}~\bibnamefont {Garofalo}}, \bibinfo {author} {\bibfnamefont {B.}~\bibnamefont {Kostrzewa}}, \bibinfo {author} {\bibfnamefont {F.}~\bibnamefont {Sanfilippo}}, \bibinfo {author} {\bibfnamefont {S.}~\bibnamefont {Simula}},\ and\ \bibinfo {author} {\bibfnamefont {U.}~\bibnamefont {Wenger}} (\bibinfo {collaboration} {Extended Twisted Mass}),\ }\bibfield  {title} {\bibinfo {title} {{Quark masses and decay constants in $N_f=2+1+1$ isoQCD with Wilson clover twisted mass fermions}},\ }\href {https://doi.org/10.22323/1.363.0181} {\bibfield  {journal} {\bibinfo  {journal} {PoS}\ }\textbf {\bibinfo {volume} {LATTICE2019}},\ \bibinfo {pages}
  {181} (\bibinfo {year} {2020})},\ \Eprint {https://arxiv.org/abs/2001.09116} {arXiv:2001.09116 [hep-lat]} \BibitemShut {NoStop}%
\bibitem [{\citenamefont {Alexandrou}\ \emph {et~al.}(2021)\citenamefont {Alexandrou} \emph {et~al.}}]{ExtendedTwistedMass:2021qui}%
  \BibitemOpen
  \bibfield  {author} {\bibinfo {author} {\bibfnamefont {C.}~\bibnamefont {Alexandrou}} \emph {et~al.} (\bibinfo {collaboration} {Extended Twisted Mass}),\ }\bibfield  {title} {\bibinfo {title} {{Ratio of kaon and pion leptonic decay constants with Nf=2+1+1 Wilson-clover twisted-mass fermions}},\ }\href {https://doi.org/10.1103/PhysRevD.104.074520} {\bibfield  {journal} {\bibinfo  {journal} {Phys. Rev. D}\ }\textbf {\bibinfo {volume} {104}},\ \bibinfo {pages} {074520} (\bibinfo {year} {2021})},\ \Eprint {https://arxiv.org/abs/2104.06747} {arXiv:2104.06747 [hep-lat]} \BibitemShut {NoStop}%
\bibitem [{\citenamefont {Finkenrath}\ \emph {et~al.}(2022)\citenamefont {Finkenrath} \emph {et~al.}}]{Finkenrath:2022eon}%
  \BibitemOpen
  \bibfield  {author} {\bibinfo {author} {\bibfnamefont {J.}~\bibnamefont {Finkenrath}} \emph {et~al.},\ }\bibfield  {title} {\bibinfo {title} {{Twisted mass gauge ensembles at physical values of the light, strange and charm quark masses}},\ }\href {https://doi.org/10.22323/1.396.0284} {\bibfield  {journal} {\bibinfo  {journal} {PoS}\ }\textbf {\bibinfo {volume} {LATTICE2021}},\ \bibinfo {pages} {284} (\bibinfo {year} {2022})},\ \Eprint {https://arxiv.org/abs/2201.02551} {arXiv:2201.02551 [hep-lat]} \BibitemShut {NoStop}%
\bibitem [{\citenamefont {Frezzotti}\ \emph {et~al.}(2001)\citenamefont {Frezzotti}, \citenamefont {Grassi}, \citenamefont {Sint},\ and\ \citenamefont {Weisz}}]{Frezzotti:2000nk}%
  \BibitemOpen
  \bibfield  {author} {\bibinfo {author} {\bibfnamefont {R.}~\bibnamefont {Frezzotti}}, \bibinfo {author} {\bibfnamefont {P.~A.}\ \bibnamefont {Grassi}}, \bibinfo {author} {\bibfnamefont {S.}~\bibnamefont {Sint}},\ and\ \bibinfo {author} {\bibfnamefont {P.}~\bibnamefont {Weisz}} (\bibinfo {collaboration} {Alpha}),\ }\bibfield  {title} {\bibinfo {title} {{Lattice QCD with a chirally twisted mass term}},\ }\href {https://doi.org/10.1088/1126-6708/2001/08/058} {\bibfield  {journal} {\bibinfo  {journal} {JHEP}\ }\textbf {\bibinfo {volume} {08}},\ \bibinfo {pages} {058}},\ \Eprint {https://arxiv.org/abs/hep-lat/0101001} {arXiv:hep-lat/0101001} \BibitemShut {NoStop}%
\bibitem [{\citenamefont {Frezzotti}\ and\ \citenamefont {Rossi}(2004{\natexlab{a}})}]{Frezzotti:2003xj}%
  \BibitemOpen
  \bibfield  {author} {\bibinfo {author} {\bibfnamefont {R.}~\bibnamefont {Frezzotti}}\ and\ \bibinfo {author} {\bibfnamefont {G.~C.}\ \bibnamefont {Rossi}},\ }\bibfield  {title} {\bibinfo {title} {{Twisted mass lattice QCD with mass nondegenerate quarks}},\ }\href {https://doi.org/10.1016/S0920-5632(03)02477-0} {\bibfield  {journal} {\bibinfo  {journal} {Nucl. Phys. B Proc. Suppl.}\ }\textbf {\bibinfo {volume} {128}},\ \bibinfo {pages} {193} (\bibinfo {year} {2004}{\natexlab{a}})},\ \Eprint {https://arxiv.org/abs/hep-lat/0311008} {arXiv:hep-lat/0311008} \BibitemShut {NoStop}%
\bibitem [{\citenamefont {Frezzotti}\ and\ \citenamefont {Rossi}(2004{\natexlab{b}})}]{Frezzotti:2004wz}%
  \BibitemOpen
  \bibfield  {author} {\bibinfo {author} {\bibfnamefont {R.}~\bibnamefont {Frezzotti}}\ and\ \bibinfo {author} {\bibfnamefont {G.~C.}\ \bibnamefont {Rossi}},\ }\bibfield  {title} {\bibinfo {title} {{Chirally improving Wilson fermions. II. Four-quark operators}},\ }\href {https://doi.org/10.1088/1126-6708/2004/10/070} {\bibfield  {journal} {\bibinfo  {journal} {JHEP}\ }\textbf {\bibinfo {volume} {10}},\ \bibinfo {pages} {070}},\ \Eprint {https://arxiv.org/abs/hep-lat/0407002} {arXiv:hep-lat/0407002} \BibitemShut {NoStop}%
\bibitem [{\citenamefont {Alexandrou}\ \emph {et~al.}(2025{\natexlab{b}})\citenamefont {Alexandrou} \emph {et~al.}}]{ExtendedTwistedMassCollaborationETMC:2024xdf}%
  \BibitemOpen
  \bibfield  {author} {\bibinfo {author} {\bibfnamefont {C.}~\bibnamefont {Alexandrou}} \emph {et~al.} (\bibinfo {collaboration} {Extended Twisted Mass}),\ }\bibfield  {title} {\bibinfo {title} {{Strange and charm quark contributions to the muon anomalous magnetic moment in lattice QCD with twisted-mass fermions}},\ }\href {https://doi.org/10.1103/PhysRevD.111.054502} {\bibfield  {journal} {\bibinfo  {journal} {Phys. Rev. D}\ }\textbf {\bibinfo {volume} {111}},\ \bibinfo {pages} {054502} (\bibinfo {year} {2025}{\natexlab{b}})},\ \Eprint {https://arxiv.org/abs/2411.08852} {arXiv:2411.08852 [hep-lat]} \BibitemShut {NoStop}%
\bibitem [{\citenamefont {Albanese}\ \emph {et~al.}(1987)\citenamefont {Albanese} \emph {et~al.}}]{APE:1987ehd}%
  \BibitemOpen
  \bibfield  {author} {\bibinfo {author} {\bibfnamefont {M.}~\bibnamefont {Albanese}} \emph {et~al.} (\bibinfo {collaboration} {APE}),\ }\bibfield  {title} {\bibinfo {title} {{Glueball Masses and String Tension in Lattice QCD}},\ }\href {https://doi.org/10.1016/0370-2693(87)91160-9} {\bibfield  {journal} {\bibinfo  {journal} {Phys. Lett. B}\ }\textbf {\bibinfo {volume} {192}},\ \bibinfo {pages} {163} (\bibinfo {year} {1987})}\BibitemShut {NoStop}%
\bibitem [{\citenamefont {de~Divitiis}\ \emph {et~al.}(2004)\citenamefont {de~Divitiis}, \citenamefont {Petronzio},\ and\ \citenamefont {Tantalo}}]{deDivitiis:2004kq}%
  \BibitemOpen
  \bibfield  {author} {\bibinfo {author} {\bibfnamefont {G.~M.}\ \bibnamefont {de~Divitiis}}, \bibinfo {author} {\bibfnamefont {R.}~\bibnamefont {Petronzio}},\ and\ \bibinfo {author} {\bibfnamefont {N.}~\bibnamefont {Tantalo}},\ }\bibfield  {title} {\bibinfo {title} {{On the discretization of physical momenta in lattice QCD}},\ }\href {https://doi.org/10.1016/j.physletb.2004.06.035} {\bibfield  {journal} {\bibinfo  {journal} {Phys. Lett. B}\ }\textbf {\bibinfo {volume} {595}},\ \bibinfo {pages} {408} (\bibinfo {year} {2004})},\ \Eprint {https://arxiv.org/abs/hep-lat/0405002} {arXiv:hep-lat/0405002} \BibitemShut {NoStop}%
\bibitem [{\citenamefont {Alexandrou}\ \emph {et~al.}(2023{\natexlab{b}})\citenamefont {Alexandrou} \emph {et~al.}}]{ExtendedTwistedMass:2022jpw}%
  \BibitemOpen
  \bibfield  {author} {\bibinfo {author} {\bibfnamefont {C.}~\bibnamefont {Alexandrou}} \emph {et~al.} (\bibinfo {collaboration} {Extended Twisted Mass}),\ }\bibfield  {title} {\bibinfo {title} {{Lattice calculation of the short and intermediate time-distance hadronic vacuum polarization contributions to the muon magnetic moment using twisted-mass fermions}},\ }\href {https://doi.org/10.1103/PhysRevD.107.074506} {\bibfield  {journal} {\bibinfo  {journal} {Phys. Rev. D}\ }\textbf {\bibinfo {volume} {107}},\ \bibinfo {pages} {074506} (\bibinfo {year} {2023}{\natexlab{b}})},\ \Eprint {https://arxiv.org/abs/2206.15084} {arXiv:2206.15084 [hep-lat]} \BibitemShut {NoStop}%
\bibitem [{\citenamefont {Backus}\ and\ \citenamefont {Gilbert}(1968)}]{SMBackus}%
  \BibitemOpen
  \bibfield  {author} {\bibinfo {author} {\bibfnamefont {G.}~\bibnamefont {Backus}}\ and\ \bibinfo {author} {\bibfnamefont {F.}~\bibnamefont {Gilbert}},\ }\bibfield  {title} {\bibinfo {title} {{The resolving power of gross earth data}},\ }\href {https://doi.org/10.1111/j.1365-246X.1968.tb00216.x} {\bibfield  {journal} {\bibinfo  {journal} {Geophys. J. Int.}\ }\textbf {\bibinfo {volume} {16}},\ \bibinfo {pages} {169} (\bibinfo {year} {1968})}\BibitemShut {NoStop}%
\bibitem [{\citenamefont {Bulava}\ \emph {et~al.}(2022)\citenamefont {Bulava}, \citenamefont {Hansen}, \citenamefont {Hansen}, \citenamefont {Patella},\ and\ \citenamefont {Tantalo}}]{Bulava:2021fre}%
  \BibitemOpen
  \bibfield  {author} {\bibinfo {author} {\bibfnamefont {J.}~\bibnamefont {Bulava}}, \bibinfo {author} {\bibfnamefont {M.~T.}\ \bibnamefont {Hansen}}, \bibinfo {author} {\bibfnamefont {M.~W.}\ \bibnamefont {Hansen}}, \bibinfo {author} {\bibfnamefont {A.}~\bibnamefont {Patella}},\ and\ \bibinfo {author} {\bibfnamefont {N.}~\bibnamefont {Tantalo}},\ }\bibfield  {title} {\bibinfo {title} {{Inclusive rates from smeared spectral densities in the two-dimensional O(3) non-linear {\ensuremath{\sigma}}-model}},\ }\href {https://doi.org/10.1007/JHEP07(2022)034} {\bibfield  {journal} {\bibinfo  {journal} {JHEP}\ }\textbf {\bibinfo {volume} {07}},\ \bibinfo {pages} {034}},\ \Eprint {https://arxiv.org/abs/2111.12774} {arXiv:2111.12774 [hep-lat]} \BibitemShut {NoStop}%
\bibitem [{\citenamefont {Frezzotti}\ \emph {et~al.}(2023)\citenamefont {Frezzotti}, \citenamefont {Tantalo}, \citenamefont {Gagliardi}, \citenamefont {Sanfilippo}, \citenamefont {Simula},\ and\ \citenamefont {Lubicz}}]{Frezzotti:2023nun}%
  \BibitemOpen
  \bibfield  {author} {\bibinfo {author} {\bibfnamefont {R.}~\bibnamefont {Frezzotti}}, \bibinfo {author} {\bibfnamefont {N.}~\bibnamefont {Tantalo}}, \bibinfo {author} {\bibfnamefont {G.}~\bibnamefont {Gagliardi}}, \bibinfo {author} {\bibfnamefont {F.}~\bibnamefont {Sanfilippo}}, \bibinfo {author} {\bibfnamefont {S.}~\bibnamefont {Simula}},\ and\ \bibinfo {author} {\bibfnamefont {V.}~\bibnamefont {Lubicz}},\ }\bibfield  {title} {\bibinfo {title} {{Spectral-function determination of complex electroweak amplitudes with lattice QCD}},\ }\href {https://doi.org/10.1103/PhysRevD.108.074510} {\bibfield  {journal} {\bibinfo  {journal} {Phys. Rev. D}\ }\textbf {\bibinfo {volume} {108}},\ \bibinfo {pages} {074510} (\bibinfo {year} {2023})},\ \Eprint {https://arxiv.org/abs/2306.07228} {arXiv:2306.07228 [hep-lat]} \BibitemShut {NoStop}%
\bibitem [{\citenamefont {Bonanno}\ \emph {et~al.}(2023)\citenamefont {Bonanno}, \citenamefont {D'Angelo}, \citenamefont {D'Elia}, \citenamefont {Maio},\ and\ \citenamefont {Naviglio}}]{Bonanno:2023ljc}%
  \BibitemOpen
  \bibfield  {author} {\bibinfo {author} {\bibfnamefont {C.}~\bibnamefont {Bonanno}}, \bibinfo {author} {\bibfnamefont {F.}~\bibnamefont {D'Angelo}}, \bibinfo {author} {\bibfnamefont {M.}~\bibnamefont {D'Elia}}, \bibinfo {author} {\bibfnamefont {L.}~\bibnamefont {Maio}},\ and\ \bibinfo {author} {\bibfnamefont {M.}~\bibnamefont {Naviglio}},\ }\bibfield  {title} {\bibinfo {title} {{Sphaleron rate from a modified Backus-Gilbert inversion method}},\ }\href {https://doi.org/10.1103/PhysRevD.108.074515} {\bibfield  {journal} {\bibinfo  {journal} {Phys. Rev. D}\ }\textbf {\bibinfo {volume} {108}},\ \bibinfo {pages} {074515} (\bibinfo {year} {2023})},\ \Eprint {https://arxiv.org/abs/2305.17120} {arXiv:2305.17120 [hep-lat]} \BibitemShut {NoStop}%
\bibitem [{\citenamefont {Bonanno}\ \emph {et~al.}(2024)\citenamefont {Bonanno}, \citenamefont {D'Angelo}, \citenamefont {D'Elia}, \citenamefont {Maio},\ and\ \citenamefont {Naviglio}}]{Bonanno:2023thi}%
  \BibitemOpen
  \bibfield  {author} {\bibinfo {author} {\bibfnamefont {C.}~\bibnamefont {Bonanno}}, \bibinfo {author} {\bibfnamefont {F.}~\bibnamefont {D'Angelo}}, \bibinfo {author} {\bibfnamefont {M.}~\bibnamefont {D'Elia}}, \bibinfo {author} {\bibfnamefont {L.}~\bibnamefont {Maio}},\ and\ \bibinfo {author} {\bibfnamefont {M.}~\bibnamefont {Naviglio}},\ }\bibfield  {title} {\bibinfo {title} {{Sphaleron Rate of Nf=2+1 QCD}},\ }\href {https://doi.org/10.1103/PhysRevLett.132.051903} {\bibfield  {journal} {\bibinfo  {journal} {Phys. Rev. Lett.}\ }\textbf {\bibinfo {volume} {132}},\ \bibinfo {pages} {051903} (\bibinfo {year} {2024})},\ \Eprint {https://arxiv.org/abs/2308.01287} {arXiv:2308.01287 [hep-lat]} \BibitemShut {NoStop}%
\bibitem [{\citenamefont {Bulava}(2024)}]{Bulava:2023brj}%
  \BibitemOpen
  \bibfield  {author} {\bibinfo {author} {\bibfnamefont {J.}~\bibnamefont {Bulava}},\ }\bibfield  {title} {\bibinfo {title} {{Spectral reconstruction of Euclidean correlator moments in lattice QCD}},\ }\href {https://doi.org/10.1393/ncc/i2024-24199-3} {\bibfield  {journal} {\bibinfo  {journal} {Nuovo Cim. C}\ }\textbf {\bibinfo {volume} {47}},\ \bibinfo {pages} {199} (\bibinfo {year} {2024})},\ \Eprint {https://arxiv.org/abs/2311.06027} {arXiv:2311.06027 [hep-lat]} \BibitemShut {NoStop}%
\bibitem [{\citenamefont {Del~Debbio}\ \emph {et~al.}(2025)\citenamefont {Del~Debbio}, \citenamefont {Lupo}, \citenamefont {Panero},\ and\ \citenamefont {Tantalo}}]{DelDebbio:2024lwm}%
  \BibitemOpen
  \bibfield  {author} {\bibinfo {author} {\bibfnamefont {L.}~\bibnamefont {Del~Debbio}}, \bibinfo {author} {\bibfnamefont {A.}~\bibnamefont {Lupo}}, \bibinfo {author} {\bibfnamefont {M.}~\bibnamefont {Panero}},\ and\ \bibinfo {author} {\bibfnamefont {N.}~\bibnamefont {Tantalo}},\ }\bibfield  {title} {\bibinfo {title} {{Bayesian solution to the inverse problem and its relation to Backus{\textendash}Gilbert methods}},\ }\href {https://doi.org/10.1140/epjc/s10052-025-13885-9} {\bibfield  {journal} {\bibinfo  {journal} {Eur. Phys. J. C}\ }\textbf {\bibinfo {volume} {85}},\ \bibinfo {pages} {185} (\bibinfo {year} {2025})},\ \Eprint {https://arxiv.org/abs/2409.04413} {arXiv:2409.04413 [hep-lat]} \BibitemShut {NoStop}%
\bibitem [{\citenamefont {Bennett}\ \emph {et~al.}(2024)\citenamefont {Bennett} \emph {et~al.}}]{Bennett:2024cqv}%
  \BibitemOpen
  \bibfield  {author} {\bibinfo {author} {\bibfnamefont {E.}~\bibnamefont {Bennett}} \emph {et~al.},\ }\bibfield  {title} {\bibinfo {title} {{Meson spectroscopy from spectral densities in lattice gauge theories}},\ }\href {https://doi.org/10.1103/PhysRevD.110.074509} {\bibfield  {journal} {\bibinfo  {journal} {Phys. Rev. D}\ }\textbf {\bibinfo {volume} {110}},\ \bibinfo {pages} {074509} (\bibinfo {year} {2024})},\ \Eprint {https://arxiv.org/abs/2405.01388} {arXiv:2405.01388 [hep-lat]} \BibitemShut {NoStop}%
\bibitem [{\citenamefont {Blum}\ \emph {et~al.}(2025)\citenamefont {Blum}, \citenamefont {Jay}, \citenamefont {Jin}, \citenamefont {Kronfeld},\ and\ \citenamefont {Stewart}}]{Blum:2024hyr}%
  \BibitemOpen
  \bibfield  {author} {\bibinfo {author} {\bibfnamefont {T.}~\bibnamefont {Blum}}, \bibinfo {author} {\bibfnamefont {W.~I.}\ \bibnamefont {Jay}}, \bibinfo {author} {\bibfnamefont {L.}~\bibnamefont {Jin}}, \bibinfo {author} {\bibfnamefont {A.~S.}\ \bibnamefont {Kronfeld}},\ and\ \bibinfo {author} {\bibfnamefont {D.~B.~A.}\ \bibnamefont {Stewart}},\ }\bibfield  {title} {\bibinfo {title} {{Toward inclusive observables with staggered quarks: the smeared $R${\textasciitilde}ratio}},\ }\href {https://doi.org/10.22323/1.466.0126} {\bibfield  {journal} {\bibinfo  {journal} {PoS}\ }\textbf {\bibinfo {volume} {LATTICE2024}},\ \bibinfo {pages} {126} (\bibinfo {year} {2025})},\ \Eprint {https://arxiv.org/abs/2411.14300} {arXiv:2411.14300 [hep-lat]} \BibitemShut {NoStop}%
\bibitem [{\citenamefont {Almirante}\ \emph {et~al.}(2025)\citenamefont {Almirante}, \citenamefont {Astrakhantsev}, \citenamefont {Braguta}, \citenamefont {D'Elia}, \citenamefont {Maio}, \citenamefont {Naviglio}, \citenamefont {Sanfilippo},\ and\ \citenamefont {Trunin}}]{Almirante:2024lqn}%
  \BibitemOpen
  \bibfield  {author} {\bibinfo {author} {\bibfnamefont {G.}~\bibnamefont {Almirante}}, \bibinfo {author} {\bibfnamefont {N.}~\bibnamefont {Astrakhantsev}}, \bibinfo {author} {\bibfnamefont {V.~V.}\ \bibnamefont {Braguta}}, \bibinfo {author} {\bibfnamefont {M.}~\bibnamefont {D'Elia}}, \bibinfo {author} {\bibfnamefont {L.}~\bibnamefont {Maio}}, \bibinfo {author} {\bibfnamefont {M.}~\bibnamefont {Naviglio}}, \bibinfo {author} {\bibfnamefont {F.}~\bibnamefont {Sanfilippo}},\ and\ \bibinfo {author} {\bibfnamefont {A.}~\bibnamefont {Trunin}},\ }\bibfield  {title} {\bibinfo {title} {{Electrical conductivity of the quark-gluon plasma in the presence of strong magnetic fields}},\ }\href {https://doi.org/10.1103/PhysRevD.111.034505} {\bibfield  {journal} {\bibinfo  {journal} {Phys. Rev. D}\ }\textbf {\bibinfo {volume} {111}},\ \bibinfo {pages} {034505} (\bibinfo {year} {2025})},\ \Eprint {https://arxiv.org/abs/2406.18504} {arXiv:2406.18504 [hep-lat]} \BibitemShut {NoStop}%
\bibitem [{\citenamefont {Smecca}\ \emph {et~al.}(2025)\citenamefont {Smecca} \emph {et~al.}}]{Smecca:2025hfw}%
  \BibitemOpen
  \bibfield  {author} {\bibinfo {author} {\bibfnamefont {A.}~\bibnamefont {Smecca}} \emph {et~al.},\ }\bibfield  {title} {\bibinfo {title} {{The NRQCD $\Upsilon$ spectrum at non-zero temperatures using Backus-Gilbert regularisations}},\ }\href {https://doi.org/10.22323/1.466.0197} {\bibfield  {journal} {\bibinfo  {journal} {PoS}\ }\textbf {\bibinfo {volume} {LATTICE2024}},\ \bibinfo {pages} {197} (\bibinfo {year} {2025})},\ \Eprint {https://arxiv.org/abs/2502.03060} {arXiv:2502.03060 [hep-lat]} \BibitemShut {NoStop}%
\bibitem [{\citenamefont {Akaike}(1974)}]{Akaike:1974vps}%
  \BibitemOpen
  \bibfield  {author} {\bibinfo {author} {\bibfnamefont {H.}~\bibnamefont {Akaike}},\ }\bibfield  {title} {\bibinfo {title} {{A new look at the statistical model identification}},\ }\href {https://doi.org/10.1109/TAC.1974.1100705} {\bibfield  {journal} {\bibinfo  {journal} {IEEE Trans. Automatic Control}\ }\textbf {\bibinfo {volume} {19}},\ \bibinfo {pages} {716} (\bibinfo {year} {1974})}\BibitemShut {NoStop}%
\bibitem [{\citenamefont {Bigi}\ and\ \citenamefont {Uraltsev}(1994)}]{Bigi:1993bh}%
  \BibitemOpen
  \bibfield  {author} {\bibinfo {author} {\bibfnamefont {I.~I.~Y.}\ \bibnamefont {Bigi}}\ and\ \bibinfo {author} {\bibfnamefont {N.~G.}\ \bibnamefont {Uraltsev}},\ }\bibfield  {title} {\bibinfo {title} {{Weak annihilation and the endpoint spectrum in semileptonic B decays}},\ }\href {https://doi.org/10.1016/0550-3213(94)90564-9} {\bibfield  {journal} {\bibinfo  {journal} {Nucl. Phys. B}\ }\textbf {\bibinfo {volume} {423}},\ \bibinfo {pages} {33} (\bibinfo {year} {1994})},\ \Eprint {https://arxiv.org/abs/hep-ph/9310285} {arXiv:hep-ph/9310285} \BibitemShut {NoStop}%
\bibitem [{\citenamefont {Ligeti}\ \emph {et~al.}(2010)\citenamefont {Ligeti}, \citenamefont {Luke},\ and\ \citenamefont {Manohar}}]{Ligeti:2010vd}%
  \BibitemOpen
  \bibfield  {author} {\bibinfo {author} {\bibfnamefont {Z.}~\bibnamefont {Ligeti}}, \bibinfo {author} {\bibfnamefont {M.}~\bibnamefont {Luke}},\ and\ \bibinfo {author} {\bibfnamefont {A.~V.}\ \bibnamefont {Manohar}},\ }\bibfield  {title} {\bibinfo {title} {{Constraining weak annihilation using semileptonic D decays}},\ }\href {https://doi.org/10.1103/PhysRevD.82.033003} {\bibfield  {journal} {\bibinfo  {journal} {Phys. Rev. D}\ }\textbf {\bibinfo {volume} {82}},\ \bibinfo {pages} {033003} (\bibinfo {year} {2010})},\ \Eprint {https://arxiv.org/abs/1003.1351} {arXiv:1003.1351 [hep-ph]} \BibitemShut {NoStop}%
\bibitem [{\citenamefont {Gambino}\ and\ \citenamefont {Kamenik}(2010)}]{Gambino:2010jz}%
  \BibitemOpen
  \bibfield  {author} {\bibinfo {author} {\bibfnamefont {P.}~\bibnamefont {Gambino}}\ and\ \bibinfo {author} {\bibfnamefont {J.~F.}\ \bibnamefont {Kamenik}},\ }\bibfield  {title} {\bibinfo {title} {{Lepton energy moments in semileptonic charm decays}},\ }\href {https://doi.org/10.1016/j.nuclphysb.2010.07.019} {\bibfield  {journal} {\bibinfo  {journal} {Nucl. Phys. B}\ }\textbf {\bibinfo {volume} {840}},\ \bibinfo {pages} {424} (\bibinfo {year} {2010})},\ \Eprint {https://arxiv.org/abs/1004.0114} {arXiv:1004.0114 [hep-ph]} \BibitemShut {NoStop}%
\bibitem [{\citenamefont {Bigi}\ \emph {et~al.}(2010)\citenamefont {Bigi}, \citenamefont {Mannel}, \citenamefont {Turczyk},\ and\ \citenamefont {Uraltsev}}]{Bigi:2009ym}%
  \BibitemOpen
  \bibfield  {author} {\bibinfo {author} {\bibfnamefont {I.}~\bibnamefont {Bigi}}, \bibinfo {author} {\bibfnamefont {T.}~\bibnamefont {Mannel}}, \bibinfo {author} {\bibfnamefont {S.}~\bibnamefont {Turczyk}},\ and\ \bibinfo {author} {\bibfnamefont {N.}~\bibnamefont {Uraltsev}},\ }\bibfield  {title} {\bibinfo {title} {{The Two Roads to 'Intrinsic Charm' in B Decays}},\ }\href {https://doi.org/10.1007/JHEP04(2010)073} {\bibfield  {journal} {\bibinfo  {journal} {JHEP}\ }\textbf {\bibinfo {volume} {04}},\ \bibinfo {pages} {073}},\ \Eprint {https://arxiv.org/abs/0911.3322} {arXiv:0911.3322 [hep-ph]} \BibitemShut {NoStop}%
\bibitem [{\citenamefont {Navas}\ \emph {et~al.}(0 24)\citenamefont {Navas} \emph {et~al.}}]{ParticleDataGroup:2024cfk}%
  \BibitemOpen
  \bibfield  {author} {\bibinfo {author} {\bibfnamefont {S.}~\bibnamefont {Navas}} \emph {et~al.} (\bibinfo {collaboration} {Particle Data Group}),\ }\bibfield  {title} {\bibinfo {title} {{Review of particle physics}},\ }\href {https://doi.org/10.1103/PhysRevD.110.030001} {\bibfield  {journal} {\bibinfo  {journal} {Phys. Rev. D}\ }\textbf {\bibinfo {volume} {110}},\ \bibinfo {pages} {030001} (\bibinfo {year} {20 24})}\BibitemShut {NoStop}%
\bibitem [{Cub()}]{CubicSpline}%
  \BibitemOpen
  \href {https://docs.scipy.org/doc/scipy/reference/generated/scipy.interpolate.CubicSpline.html} {\bibinfo {title} {Scipy - cubicspline package}}\BibitemShut {NoStop}%
\bibitem [{\citenamefont {{J\"{u}lich Supercomputing Centre}}(2021)}]{JUWELS}%
  \BibitemOpen
  \bibfield  {author} {\bibinfo {author} {\bibnamefont {{J\"{u}lich Supercomputing Centre}}},\ }\bibfield  {title} {\bibinfo {title} {{JUWELS Cluster and Booster: Exascale Pathfinder with Modular Supercomputing Architecture at Juelich Supercomputing Centre}},\ }\bibfield  {journal} {\bibinfo  {journal} {Journal of large-scale research facilities}\ }\textbf {\bibinfo {volume} {7}},\ \href {https://doi.org/10.17815/jlsrf-7-183} {10.17815/jlsrf-7-183} (\bibinfo {year} {2021})\BibitemShut {NoStop}%
\bibitem [{sup()}]{supplemental}%
  \BibitemOpen
  \href {to-be-added-from-aps} {\bibinfo {title} {{Supplemental Material for this article}}}\BibitemShut {NoStop}%
\bibitem [{\citenamefont {De~Santis}\ \emph {et~al.}(2025{\natexlab{b}})\citenamefont {De~Santis}, \citenamefont {Evangelista}, \citenamefont {Frezzotti}, \citenamefont {Gigliardi}, \citenamefont {Gambino}, \citenamefont {Garofalo}, \citenamefont {Groß}, \citenamefont {Kostrzewa}, \citenamefont {Lubicz}, \citenamefont {Margari}, \citenamefont {Panero}, \citenamefont {Sanfilippo}, \citenamefont {Simula}, \citenamefont {Smecca}, \citenamefont {Tantalo},\ and\ \citenamefont {Urbach}}]{FK2/VQFYKW_2025}%
  \BibitemOpen
  \bibfield  {author} {\bibinfo {author} {\bibfnamefont {A.}~\bibnamefont {De~Santis}}, \bibinfo {author} {\bibfnamefont {A.}~\bibnamefont {Evangelista}}, \bibinfo {author} {\bibfnamefont {R.}~\bibnamefont {Frezzotti}}, \bibinfo {author} {\bibfnamefont {G.}~\bibnamefont {Gigliardi}}, \bibinfo {author} {\bibfnamefont {P.}~\bibnamefont {Gambino}}, \bibinfo {author} {\bibfnamefont {M.}~\bibnamefont {Garofalo}}, \bibinfo {author} {\bibfnamefont {C.~F.}\ \bibnamefont {Groß}}, \bibinfo {author} {\bibfnamefont {B.}~\bibnamefont {Kostrzewa}}, \bibinfo {author} {\bibfnamefont {V.}~\bibnamefont {Lubicz}}, \bibinfo {author} {\bibfnamefont {F.}~\bibnamefont {Margari}}, \bibinfo {author} {\bibfnamefont {M.}~\bibnamefont {Panero}}, \bibinfo {author} {\bibfnamefont {F.}~\bibnamefont {Sanfilippo}}, \bibinfo {author} {\bibfnamefont {S.}~\bibnamefont {Simula}}, \bibinfo {author} {\bibfnamefont {A.}~\bibnamefont {Smecca}}, \bibinfo {author} {\bibfnamefont {N.}~\bibnamefont {Tantalo}},\ and\ \bibinfo {author} {\bibfnamefont
  {C.}~\bibnamefont {Urbach}},\ }\href {https://doi.org/10.60507/FK2/VQFYKW} {\bibinfo {title} {{Supplementary data for ``Inclusive semileptonic decays of the $D_s$ meson''}}} (\bibinfo {year} {2025}{\natexlab{b}})\BibitemShut {NoStop}%
\end{thebibliography}%

\end{document}